\documentclass[twocolumn,letterpaper,amsmath,amssymb,amsfonts,floatfix,pre,superscriptaddress,showkeys,showpacs]{revtex4-1}

\usepackage{dcolumn}  
\usepackage{multirow}
\usepackage{hyperref} 
\usepackage{verbatim}
\usepackage{bm} 
\usepackage{graphicx}
\usepackage{epstopdf}
\usepackage{ifpdf} 
\usepackage{epsfig}
\usepackage{color}
\usepackage[usenames,dvipsnames]{xcolor}
\usepackage{float}

\begin{document}

\title{Population splitting of rodlike swimmers in Couette flow}

\author{Hossein Nili}
\affiliation{School of Physics, Institute for Research in Fundamental Sciences (IPM), P.O. Box 19395-5531, Tehran, Iran}
\author{Masoud Kheyri}
\affiliation{Department of Physics, Sharif University of Technology,  P.O. Box 11155-9161, Tehran, Iran}
\author{Javad Abazari}
\affiliation{School of Mechanical Engineering, University of Tehran, P.O. Box 11155-4563, Tehran, Iran}
\author{Ali Fahimniya}
\affiliation{Department of Physics, Sharif University of Technology,  P.O. Box 11155-9161, Tehran, Iran}
\affiliation{Department of Computer Engineering, Sharif University of Technology, P.O. Box 11155-9517, Tehran, Iran}
\author{Ali Naji} \thanks{Corresponding author. Email: a.naji@ipm.ir}
\affiliation{School of Physics, Institute for Research in Fundamental Sciences (IPM), P.O. Box 19395-5531, Tehran, Iran}


\begin{abstract}
We present a quantitative analysis on the response of a dilute active suspension of self-propelled rods (swimmers) in a planar channel subjected to an imposed shear flow. To best capture the salient features of shear-induced effects, we consider the case of an imposed Couette flow, providing a constant shear rate across the channel. We argue that the steady-state behavior of swimmers can be understood in the light of a population splitting phenomenon, occurring as the shear rate exceeds a certain threshold, initiating the reversal of swimming direction for a finite fraction of swimmers from down- to upstream or vice versa, depending on swimmer position within the channel. Swimmers thus split into two distinct, statistically significant and oppositely swimming majority and minority populations. The onset of population splitting translates into a transition from a self-propulsion-dominated regime to a shear-dominated regime, corresponding to a unimodal-to-bimodal change in the probability distribution function of the swimmer orientation. We present a phase diagram in terms of the swim and flow P\'eclet numbers showing the separation of these two regimes by a discontinuous transition line. Our results shed further light on the behavior of swimmers in a shear flow and provide an explanation for the previously reported non-monotonic behavior of the mean, near-wall, parallel-to-flow orientation of swimmers with increasing shear strength.
\end{abstract}


\maketitle

\section{Introduction}

Self-propelled particles, or swimmers, and other active matter systems \cite{ramaswamyreview,marchetti_review,gompper_review,goldstein_review,son_review,cates_review,rusconi-stocker-review,golestanian_review,paxton-review,Romanczuk:EPJ2012,bechinger_review} have been at the focus of extensive multidisciplinary research over the last few decades due to their important applications from micro- down to nanoscale \cite{cargo,dreyfus,paxton,sperm-carrying,robotic,micro-bio-robot,klumpp1,klumpp2} and also  their remarkable theoretical and fundamental aspects \cite{purcellmain,najafiramin,laugascallop1,zottlcollective,sokolovcollective,golestanian_njp,gompper_review,marchetti_review,ramaswamyreview,golestanian_review,goldstein_review,bechinger_review,becker,najafiramin05,cates_review,son_review,berg1973bacteria,bergbook,budrene,planktonic,rusconi-stocker-review,ramin_molmach,blair1995bacteria,Romanczuk:EPJ2012,paxton-review}. Swimmers are commonplace in biology and biotechnological applications \cite{ramaswamyreview,marchetti_review,gompper_review,goldstein_review,son_review,cates_review,rusconi-stocker-review,golestanian_review,paxton-review,Romanczuk:EPJ2012,bechinger_review,rusconi-stocker-review,berg1973bacteria,bergbook,budrene,planktonic} with examples ranging from uniflagellated sperm cells  \cite{graysperm,shacksperm,woolleysperm,julichersperm,julicherspermchemx}, the biflagellate alga {\em Chlamydomonas reinhardtii} \cite{ringochlamy,goldstein_review,goldstein_science}, the colonial alga {\em Volvox carteri} \cite{volvox1,volvox2} and the multiflagellate bacterium {\em Escherichia coli} \cite{bergbook,berg1973bacteria,blair1995bacteria} to artificial micro-/nanoswimmers, fabricated using patchy colloids and Janus micro-/nanoparticles \cite{janusmain,valadares2010catalytic}. This latter class of swimmers exhibit well-defined, stochastic, self-propelled motion due to asymmetry in their catalytic surface reactions in specific solvent mixtures \cite{sanojanus,ramin07,volpe1,paxton-review,ramin_molmach,golestanian_review,gompper_review,marchetti_review,ramaswamyreview,Romanczuk:EPJ2012,bechinger_review,takagi2013,Saintillan2007,Hong2007}; their motion closely resembles microorganismal swimming, which involves the hugely more complex mechanisms such as flagellar or ciliary movements.   

Fluid suspensions of active particles are typically confined by rigid boundaries. The near-wall behavior of swimmers has received a lot of attention (see, e.g.,  Refs. \cite{nwirezhang,sabass,elgeti2016,wallattraction,wallattraction2,wallattraction3,li2011accumulation,li2011accumulation2,ardekani,Hernandez-Ortiz1,Hernandez-Ortiz2,Hernandez-Ortiz3,stark-wall,gompper-wall,elgeti2009,elgeti2013,sperm-rheotaxis,upstream-goldstein,rusconi,elgeti2015run,costanzo,catesupstream,Mathijssen:2016a,Mathijssen:2016b,Mathijssen:2016c,ezhilan,upstream2015prl,zottl_Poiseuille,Underhill2015} and the references therein) and turns out to have a profound role in many biological processes, such as biofilm formation  \cite{biofilmformation,biofilmstoodley} and sperm motility \cite{spermsurface1,spermsurface2,spermsurface3,sperm-rheotaxis}, as well as microfluidic setups used for manipulation and separation of active agents, such as bacteria \cite{ufluidics1,ufluidics2,son_review}. 

Very commonly, external flow is present within the confined regions containing active suspensions  \cite{flowbiofilm1,flowbiofilm2}. Specifically, near-surface fluid flow is known to cause swimming against the flow, i.e., upstream swimming, which has been observed in experiments \cite{son_review,ecoliupstream,ecoliupstream2,ecoliupstream3,upstreamold,upstream3,rusconi,zottl_Poiseuille} and the effect attributed to a number of factors, including hydrodynamic surface interactions and/or shear reorientation of swimmers  \cite{ecoliupstream2,ecoliupstream3,ezhilan,catesupstream,ecoliupstream,upstream2015prl,sperm-rheotaxis,upstream-goldstein,rusconi,Mathijssen:2016a,Mathijssen:2016b,Mathijssen:2016c,zottl_Poiseuille}.

In the context of continuum models, Ezhilan and Saintillan \cite{ezhilan} have conducted a thorough study on the response of a confined suspension of active rodlike particles to imposed Poiseuille flow, using a kinetic model in which the active suspension is described by a joint position-orientation probability distribution function (PDF) governed by a non-interacting Smoluchowski equation. Without considering hydrodynamic inter-particle and surface-particle interactions, they have been able to observe many of the important effects arising from confinement and imposed flow coupled with self-propulsion. These include wall accumulation, shear-trapping and upstream swimming (see Ref. \cite{ezhilan} for a comprehensive review of the literature on these effects). It has thus been suggested that hydrodynamic interactions may not be essential to the major phenomena occurring in active suspensions under confinement. 

Basing our work on the approach adopted by the above authors, we address the steady-state behavior of a dilute suspension of self-propelled, rodlike, Brownian particles in a planar channel, subjected to an imposed Couette flow. The characteristic feature of the Couette flow is that it presents a constant shear rate  across the channel width, modeled here using a linear fluid velocity profile increasing from zero on one of the surfaces (bottom wall) to a finite value on the other surface (top wall). Thus, unlike the pressure-driven Poiseuille flow, where the shear rate varies gradually from zero at the centerline of the channel to maximum values at its boundaries, the Couette model provides a situation where the swimmers experience the same shear-induced torque, independent of their position within the channel. This helps create a more transparent arrangement to investigate the effects of imposed shear on swimmer density and orientation. Three characteristic regions were shown to exist under  an imposed Poiseuille flow \cite{ezhilan}. These were (a) a centerline depletion region, where self-propulsion and rotational diffusion dominate, (b) an intermediate shear-trapping region, which develops at sufficiently low (high) swim (flow) P\'eclet numbers due to the interplay between self-propulsion and imposed shear, and (c)   swimmer accumulation regions near the walls, where all previous factors play a role. From these three, only the latter two exist when the flow is of the Couette type, as the centerline behavior does not manifest itself in the absence of changes in the shear rate across the channel width. 

We shall demonstrate that the steady-state behavior of swimmers in the channel for different strengths of self-propulsion and imposed shear, represented by swim and flow P\'eclet numbers, respectively, can be described in the light of a {\em population splitting phenomenon}. The phenomenon is caused by shear alignment at flows strong enough to flip the swimming direction of a statistically significant (macroscopic) number of active particles from upstream to downstream, or  vice versa. In other words, 
upon increasing the flow P\'eclet number, a growing fraction of swimmers from a {\em majority population} of up- or downstream swimmers will be converting to develop a {\em minority population} of swimmers moving in the opposite direction. This behavior can happen near each of the walls, provided that the flow P\'eclet number is increased beyond a threshold value. It corresponds to a transition from a {\em unimodal} to a  {\em bimodal} PDF of swimmers, and underlies some of the salient features arising from the interplay between shear-induced torque 
and active self-propulsion, which can be studied and elucidated most clearly in the case of Couette flow. These include the non-monotonic behavior of the mean parallel-to-flow component of the orientation vector of swimmers, showing first an increasing and, then, a decreasing trend toward zero with increasing flow P\'eclet number. Such a behavior (which has also been found, but remained unexplored, in the case of Poiseuille flow \cite{ezhilan}) portrays the onset of the population splitting phenomenon. 

Our results thus indicate that the standard picture for the near-wall behavior of rodlike swimmers in an imposed flow, in which the active particles are perceived to swim either up- or downstream near the walls, may be inaccurate as it is determined based on the {\em mean} parallel-to-flow orientation of swimmers. 
This latter quantity ceases to act as a suitable measure describing the `typical' near-wall behavior of swimmers, when population splitting takes place and gives rise to a minority, but significant and measurable, population of particles swimming in the direction opposite to the majority population; hence, a situation whose characterization requires a knowledge of the higher-order moments of the PDF. 

The organization of this paper is as follows: In Section \ref{sec:model}, we introduce the model used in our study of the shear-induced orientational behavior of rodlike swimmers subject to Couette flow. In Section \ref{results},  we present the results obtained from our numerical investigations, which we conclude in Section \ref{conclusion}.

\section{Model}
\label{sec:model}

\subsection{Physical specifications}
\label{phys-system}

\begin{figure}[t!]
\begin{center}
\includegraphics[width=8.cm]{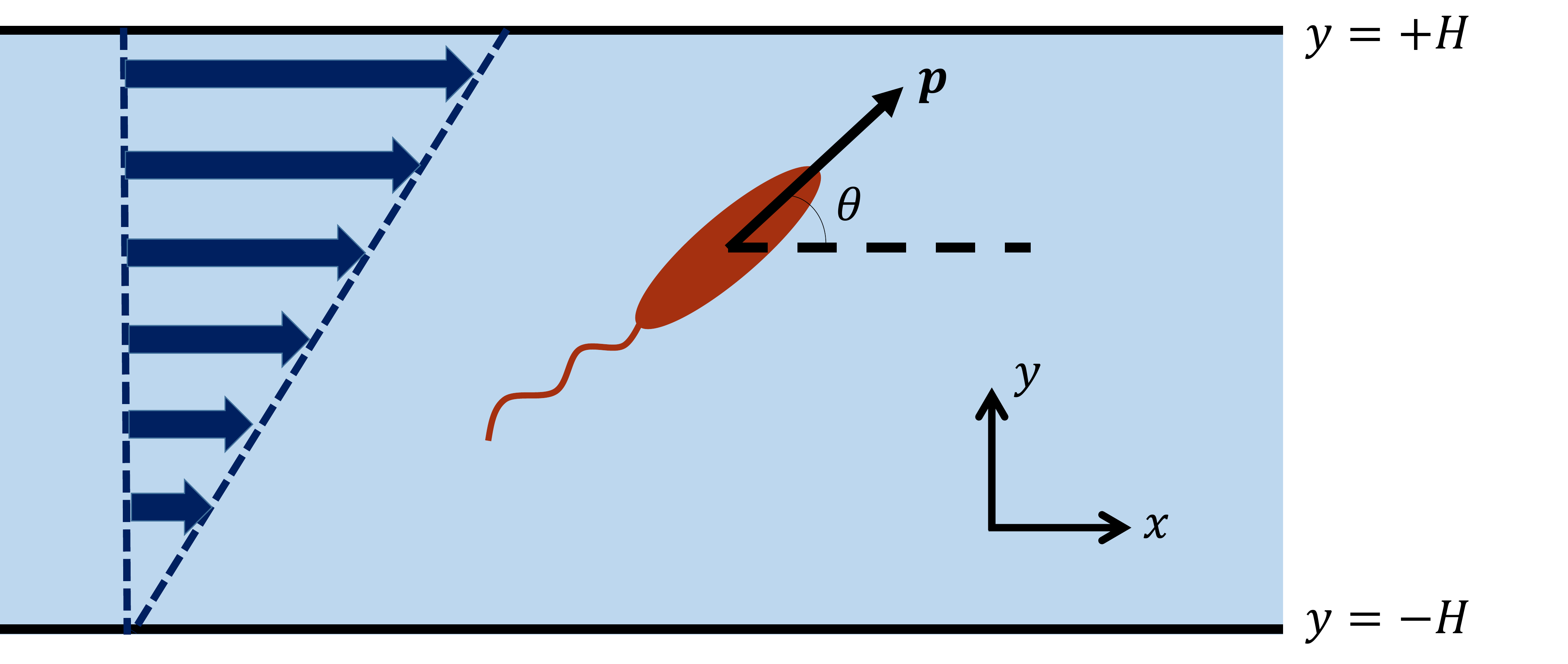}
\caption{Schematic view of a long planar channel with rigid no-slip walls at $y=\pm H$, confining a dilute suspension of rodlike swimmers subjected to Couette flow (thick, dark blue arrows). Swimmers self-propel at a constant speed $V_s$ in the direction $\mathbf{p}=(\cos \theta, \sin \theta)$, while they are also subject to translational and rotational Brownian motion and shear-induced torque.} 
\label{figure_no1}
\end{center}
\end{figure}

We consider a dilute suspension of self-propelled, rodlike, Brownian particles (swimmers) confined by two plane-parallel, rigid, no-slip walls forming a long, planar channel of width $2H$,  subjected to an imposed Couette flow with constant shear rate, $\dot{\gamma}$, and  laminar flow profile 
\begin{equation} \label{eq_couette}
{\mathbf u}(\mathbf{r})=\frac{U_{0}}{2}\cdot \frac{y+H}{H} {\mathbf {\hat x}} =\dot{\gamma} \,(y+H){\mathbf {\hat x}}, 
\end{equation} 
where $U_{0}$ is the (constant) speed of the moving top plate, ${\mathbf {\hat x}}$ the unit vector in the  flow (or channel) direction, $y$ the perpendicular coordinate (see Fig. \ref{figure_no1}), and 
\begin{equation} \label{eq_couette_shear}
\dot{\gamma}=\frac{U_{0}}{2H}. 
\end{equation} 
 
Swimmers are modeled as thin rods (needles) of high aspect ratio giving a Bretherton constant \cite{bretherton} approaching unity. They are described by the orientation vector $\mathbf{p}$, denoting the direction of active swimming at a constant speed $V_s$ (in the case of microorganisms, $\mathbf{p}$ may correspond to the tail-head direction as shown schematically in Fig. \ref{figure_no1}; note, however, that our model is generically applicable also to synthetic self-propelled micro- and nanorods \cite{paxton,takagi2013,Saintillan2007,Hong2007}). We shall be interested primarily in the steady-state properties of the system that exhibit translational symmetry in $x$-direction as well as in the direction perpendicular to the $x-y$ plane and, therefore, depend only on the coordinate $y$ and the orientation angle $\theta$. We neglect hydrodynamic as well as excluded-volume interactions between individual swimmers and between swimmers and the channel walls. This assumption is expected to remain valid in the dilute regime \cite{ezhilan}. 
 
\subsection{The continuum model}
\label{eqns}

In the continuum description, the joint PDF $\Psi(\mathbf{r},\mathbf{p},t)$ of finding a swimmer at position $\mathbf{r}$ with orientation $\mathbf{p}$ at time $t$ is governed by the Smoluchowski equation \cite{doiedwards,ezhilan}
\begin{equation} \label{smoluch}
\frac{\partial \Psi}{\partial t}+\nabla_{\mathbf{r}}\cdot (\mathbf{v}_\mathbf{r} \Psi)+\nabla_{\mathbf{p}}\cdot (\mathbf{v}_\mathbf{p} \Psi)=D_T\nabla_{\mathbf{r}}^2 \Psi + D_R\nabla_{\mathbf{p}}^2 \Psi,
\end{equation} 
where $D_T$ and $D_R$ are translational and rotational diffusivities, and $\mathbf{v}_\mathbf{r}$ and $\mathbf{v}_\mathbf{p}$ the deterministic translational and rotational  flux velocities, respectively. For rodlike swimmers under the action of the Couette flow, one has  
\begin{subequations} \label{eq4}
\begin{align}
\mathbf{v}_\mathbf{r}&=V_s\mathbf{p}+\mathbf{u}(\mathbf{r}),\\ 
\mathbf{v}_\mathbf{p}&=\dot{\gamma}(\mathbf{\hat{y}\cdot p})(\mathbf{I-pp})\cdot \mathbf{\hat{x}},  
\label{eq4b}
\end{align}
\end{subequations}
where $\mathbf{I}$ is the identity tensor. The deterministic translational flux takes contributions from self-propulsion and advection by the flow, while the rotational one takes contributions from the shear rate only. It is the combined action of self-propulsion, advection and shear by the imposed flow, as well as translational and rotational diffusion, that leads to an eventual steady-state distribution function for swimmers within the channel   $\Psi=\Psi(y,\theta)$. Using $\mathbf{p}=(p_x, p_y)=(\cos \theta, \sin \theta)$, the Smoluchowski equation at steady state reduces to
\begin{equation} \label{pde_dim}
V_s\frac{\partial }{\partial y}(\Psi \sin \theta)-\dot{\gamma}\frac{\partial }{\partial \theta}(\Psi \sin^{2}\theta)=D_T\frac{\partial^2 \Psi}{\partial y^2} + D_R\frac{\partial^2 \Psi}{\partial \theta^2}.
\end{equation} 

Since the walls are rigid and impermeable to fluid flow and active particles, the normal component of the deterministic translational flux must balance the normal component of the stochastic (diffusion) flux  at $y=\pm H$, supplementing Eq. (\ref{pde_dim}) with the boundary conditions
\begin{equation} \label{eq_bc_dim}
\left(D_T\frac{\partial }{\partial y} - V_s\sin\theta\right) \Psi(y, \theta)\bigg |_{y=\pm H} =0.
\end{equation} 

\subsection{Non-dimensionalization}
\label{non-dim}

Using the channel half-width $H$, and the rotational diffusion timescale $1/D_R$, as characteristic scales for length and time, respectively, the non-dimensionalization of Eq. (\ref{pde_dim}) proceeds immediately as  
\begin{equation} 
\label{pde_nondim}
2Pe_s\frac{\partial }{\partial \tilde y} (\tilde{\Psi} \sin\theta)-Pe_f\frac{\partial }{\partial \theta}\left (\tilde{\Psi}\sin^{2}\theta  \right)=\xi^{2}\frac{\partial^2 \tilde{\Psi}}{\partial \tilde y^2}+\frac{\partial^2 \tilde{\Psi}}{\partial \theta^2},
\end{equation}  
subject to the boundary conditions
\begin{equation}  
\label{bc_nondim}
\left(\xi^2 \frac{\partial }{\partial \tilde y} - 2Pe_s\sin\theta \right)\tilde{\Psi}(\tilde y, \theta)\bigg |_{\tilde y=\pm 1} = 0, 
\end{equation}
where $\tilde y=y/H$ is the dimensionless coordinate across the channel width and $\tilde{\Psi}=\Psi/c_0$ is the dimensionless PDF, with $c_0$ being the mean number density of swimmers in the channel.  
In Eqs.  (\ref{pde_nondim}) and (\ref{bc_nondim}), we have adopted a notation close to that of Ref. \cite{ezhilan} and defined the dimensionless system parameters as follows: (i) the swim P\'eclet number,
\begin{equation} \label{Pes_def}
	Pe_s=
		\frac{1/D_R}{2H/V_s}=\frac{V_s}{2HD_R},
\end{equation} 
which gives the ratio of the rotational diffusion timescale $1/D_R$, to the swim (across channel) timescale $2H/V_s$, (or, equivalently, the ratio of  the run length $V_s/D_R$, to the channel width), (ii) the flow P\'eclet number, 
\begin{equation} \label{Pef_def}
	Pe_f=\frac{1/D_R}{1/\dot{\gamma}} = \frac{U_0}{2HD_R},
\end{equation} 
which gives the ratio of the rotational diffusion timescale  to the timescale for swimmer alignment with the axis of flow $1/\dot{\gamma}$, as the flow would attempt to align the swimmers in/against the direction of flow (i.e., the positive/negative $x$-axis, corresponding to the self-propelled particles swimming downstream/upstream, respectively),
and (iii)  the dimensionless parameter 
\begin{equation} \label{xi_def}
		\xi^{2}=\frac{1/D_R}{H^{2}/D_T}=\frac{D_T}{H^{2}D_R},
\end{equation} 
which gives the ratio of the rotational diffusion timescale over the translational diffusion timescale. 

The number density of swimmers,
\begin{equation} \label{c_def}
c(y) = \int_0^{2\pi} {\Psi}({y},\theta)\,\mathrm{d}\theta,
\end{equation}  
 is also normalized by $c_0$, giving the following constraint on the rescaled swimmer density, $\tilde{c}({\tilde y}) \equiv c(y)/c_0$:
\begin{equation}
\label{c_int_cons}
\int_{-1}^{1}\tilde{c}(\tilde y)\, \mathrm{d}\tilde y=2.
\end{equation}  

The steady-state properties of swimmers in the channel follow from the solution of Eqs. (\ref{pde_nondim}), (\ref{bc_nondim}) and  (\ref{c_int_cons}) in the computational domain $-1\leq \tilde y \leq 1$ and $0\leq \theta < 2\pi$ (see Appendix \ref{methods} for details). The parameter space of the system is spanned by the three dimensionless parameters $Pe_f$, $Pe_s$ and $\xi$. The focus of our study will be on the interplay between shear alignment and self-propulsion of swimmers and, as such, we will vary $Pe_f$ and $Pe_s$ and, unless otherwise noted (Section \ref{phasediag}), we fix the third dimensionless parameter as $\xi=\sqrt{2/3}\simeq 0.82$. This particular value is chosen to match the base parameter values assumed in Ref. \cite{ezhilan} to enable like-for-like (quantitative) comparison between the cases of imposed Poiseuille flow considered in that work and the Couette flow considered here, as will be seen throughout the text (note that the third dimensionless parameter in Ref. \cite{ezhilan} is denoted by  $\Lambda$ and its chosen base value $\Lambda=1/6$ corresponds to the mentioned base value for $\xi$, when the relation $\Lambda=\xi^2/(4Pe_s^2)$ is used with $Pe_s=1$). We should, however, note that these base values imply a channel half-width comparable to particle size, if typical thermal values for the translational and rotational diffusion coefficients are assumed. Therefore, we shall later vary $\xi$ over a wider range, including much smaller values of $\xi$ (much larger channel widths), showing that the key outcomes of our analysis for the shear-induced behavior of swimmers indeed hold over a wide range of dimensionless parameter values (which can themselves be mapped to a broad range of values for the actual system parameters).

\section{Results and Discussion}
\label{results}

\begin{figure}[t!]
\begin{center}
	\begin{minipage}[t]{0.238\textwidth}\begin{center}
		\includegraphics[width=\textwidth]{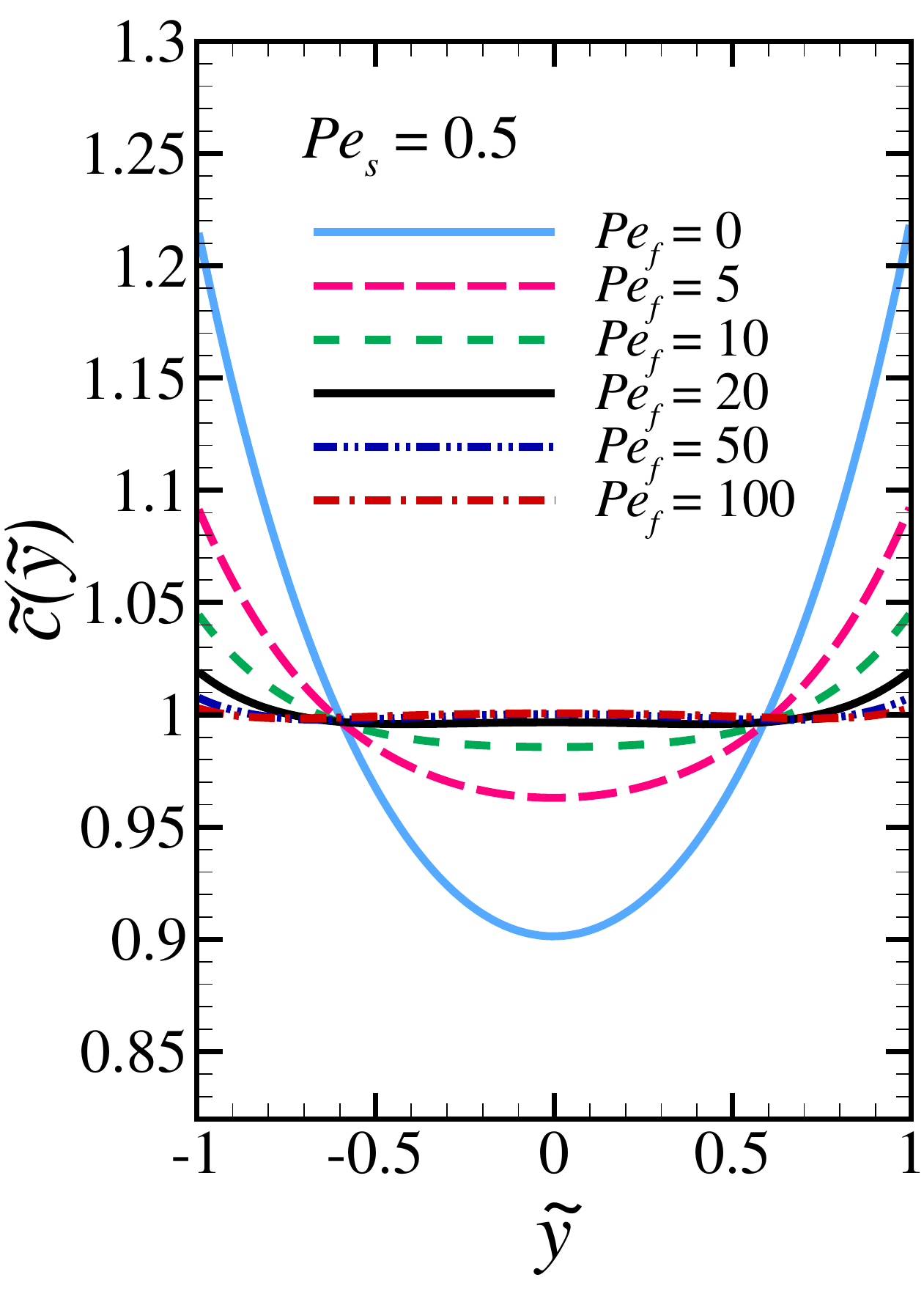} (a)
		\label{c_s0p5_fvar}
	\end{center}\end{minipage} \hskip0.cm	
	\begin{minipage}[t]{0.238\textwidth}\begin{center}
		\includegraphics[width=\textwidth]{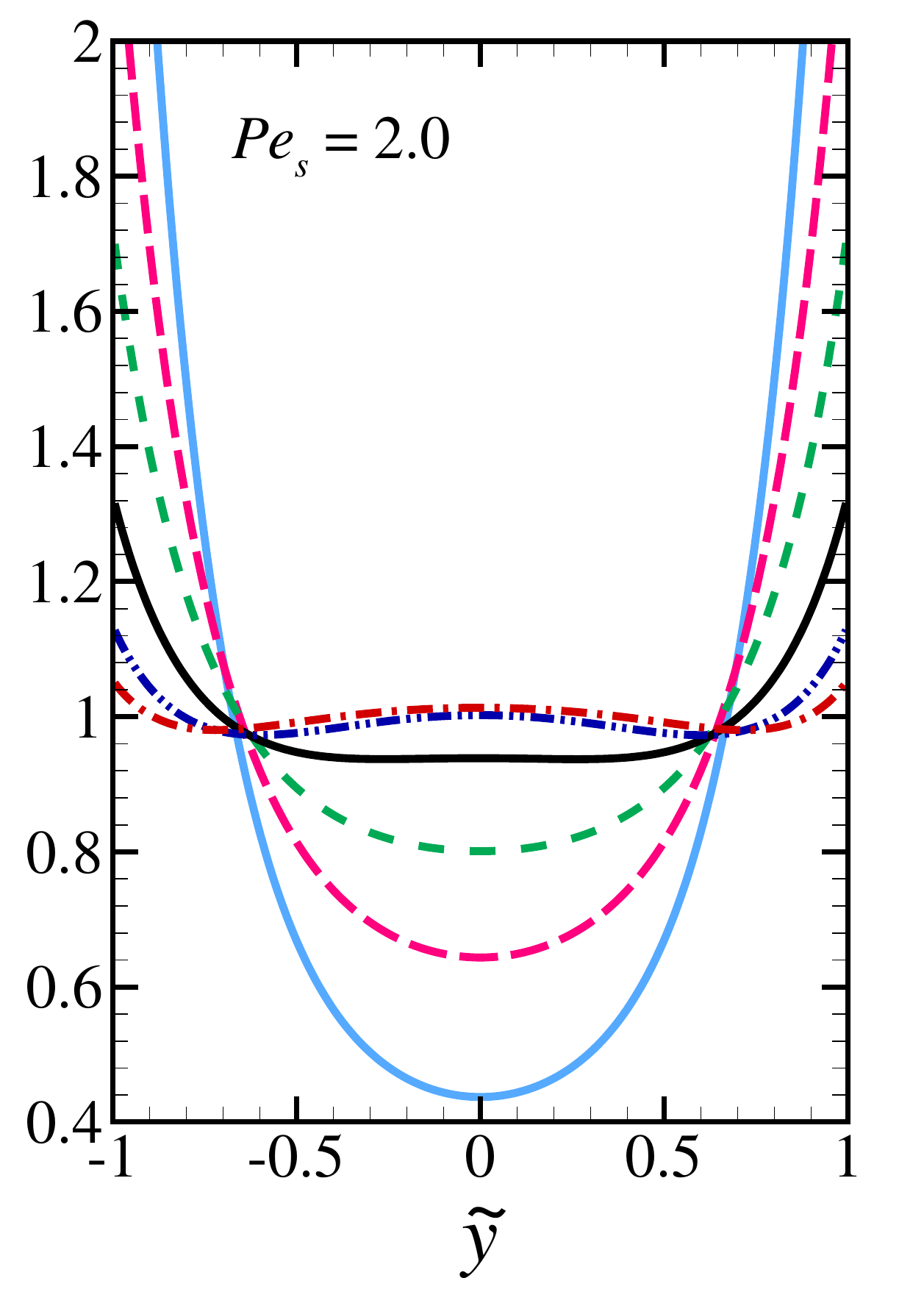} (b)
		\label{c_s2_fvar}
	\end{center}\end{minipage} 
\caption{Rescaled swimmer density profile as a function of the rescaled coordinate $\tilde y$ across the channel for swim P\'eclet numbers (a) $Pe_s=0.5$ and (b) $Pe_s=2$, fixed $\xi=0.82$ and different flow P\'eclet numbers $Pe_f$, as shown on the graphs. } 
\label{c_profiles}
\end{center}
\end{figure}

\subsection{Swimmer density profiles}

Figure \ref{c_profiles} shows the numerically calculated swimmer density profiles across the channel for different swim and flow P\'eclet numbers ($Pe_s=0.5$ and 2 in panels a and b, respectively, with $Pe_f$ increased from zero to 100). The results confirm the well-established observation that swimmers tend to accumulate at the boundaries, in this case the channel walls at $\tilde y=\pm 1$. While active swimming (increased $Pe_s$) tends to drive the particles toward the walls, making the profiles more steep near the walls, the imposed shear (increased $Pe_f$) tends to align the particles in the horizontal direction, making the profiles more uniform (see, e.g., the nearly uniform density profiles in Fig. \ref{c_profiles}a,  where active self-propulsion is relatively weak).  For sufficiently large values of $Pe_s$ and $Pe_f$ (Fig. \ref{c_profiles}b), the interplay between wall accumulation and shear alignment effects leads to a plateau-like region around the centerline of the channel and regions with lowered density between the centerline and the walls. The behavior in these latter regions is thus intrinsically different from the centerline depletion found in the case of an imposed Poiseuille flow \cite{ezhilan}, where swimmer depletion around the centerline (with vanishing shear rate) is driven by self-propulsion and rotational diffusion. (Note also that, although for the chosen parameter values the curves in Fig. \ref{c_profiles} appear to go through nearly the same points close to $\tilde y= \pm0.5$, this feature does not necessarily hold in other ranges of parameters such as larger values of $Pe_s$.) 

\begin{figure}[t!]
\begin{center}
	\begin{minipage}[t]{0.238\textwidth}\begin{center}
		\includegraphics[width=\textwidth]{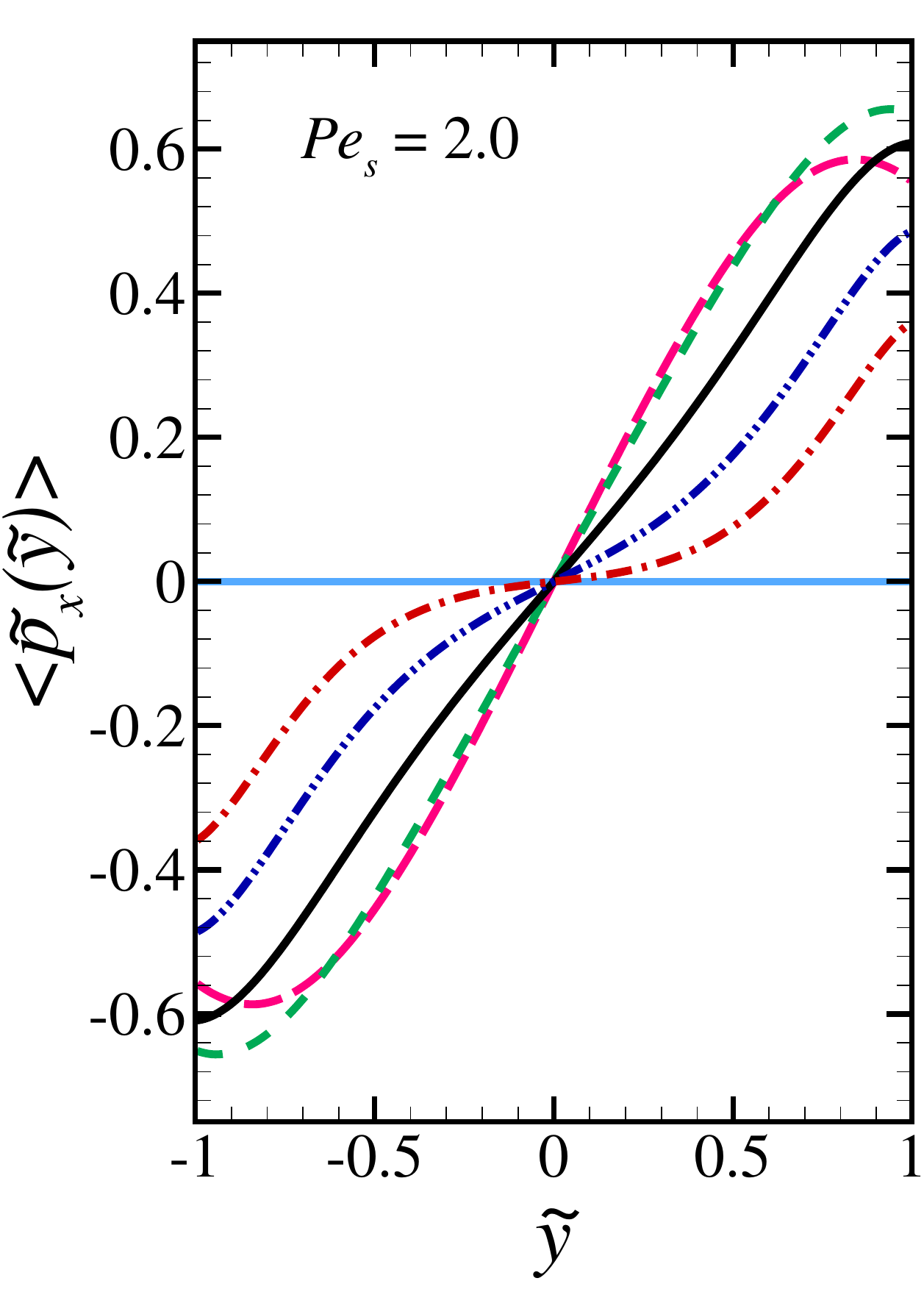} (a)
		\label{psi_a}
	\end{center}\end{minipage} \hskip0.cm	
	\begin{minipage}[t]{0.238\textwidth}\begin{center}
		\includegraphics[width=\textwidth]{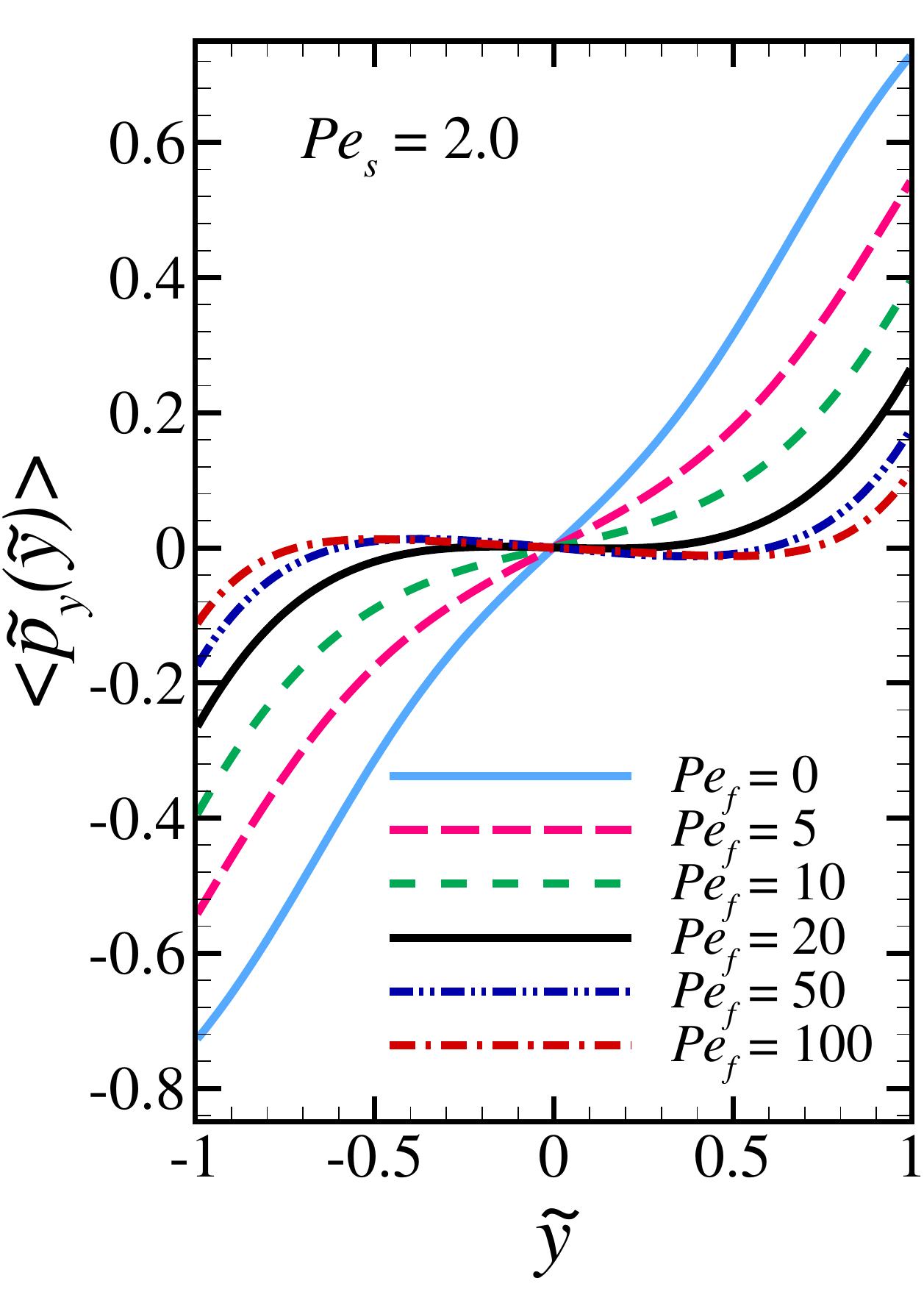} (b)
		\label{psi_b}
	\end{center}\end{minipage}
\caption{Components  of the mean swimmer orientation vector in the direction of the flow, $\langle \tilde p_x\rangle$ (a), and perpendicular to it, $\langle \tilde p_y\rangle$ (b), as a function of the rescaled coordinate $\tilde y$ across the channel for $Pe_s=2$, $\xi=0.82$ and different values of $Pe_f$.}
\label{orientn_profiles} 
\end{center}
\end{figure}

\subsection{Mean swimmer orientation profiles}
\label{mean-orientation}

Figure \ref{orientn_profiles} shows how ensemble averages of the parallel-to-flow and perpendicular-to-flow components of the swimmer orientation vector, $\langle \tilde p_x\rangle = \langle \cos \theta \rangle$ and $\langle \tilde p_y\rangle=\langle \sin \theta \rangle$, respectively, vary across the channel. Here, $\left \langle \cdots\right \rangle$ denotes the average 
\begin{equation}
\label{eq:av_def}
 \langle \cdots \rangle=\frac{1}{\tilde c(\tilde y)} \int_{-\pi/2}^{3\pi/2} (\cdots)\tilde\Psi(\tilde y, \theta)\,\mathrm{d}\theta,
\end{equation} 
where, with no loss of generality, we have chosen the $[-\pi/2,3\pi/2)$ range for $\theta$, so that the interval can be conveniently divided into a first $[-\pi/2,\pi/2)$ and a second $[\pi/2,3\pi/2)$ half, corresponding to {\em downstream} and {\em upstream} swimmers, respectively. 

When there is no imposed flow ($Pe_f=0$), the swimmer orientation is, on average, perpendicular to the direction of channel and toward the walls (i.e., $\langle \tilde p_x\rangle=0$ and $\langle \tilde p_y\rangle\neq 0$ as seen in Fig. \ref{orientn_profiles}). When Couette flow is imposed, the swimmers take a finite mean parallel-to-flow orientation (see, e.g., the curves for $Pe_f=5$ and 10 in Fig. \ref{orientn_profiles}). In this case,  negative (positive) values of $\langle \tilde p_x\rangle$ for $\tilde y<0$ ($\tilde y>0$) are representative of the fact that,  on average, swimmers orient in the  upstream (downstream) direction in the lower (upper) half of the channel. As noted before, this kind of up- or downstream swimming in imposed flows is already well established (see, e.g., Refs. \cite{son_review,ecoliupstream,ecoliupstream2,ecoliupstream3,upstreamold,upstream3,rusconi,ezhilan}) and can be understood in terms of the shear-induced torques acting on swimmers near the lower or upper wall of the channel (see Section \ref{popsplitting}). 

\begin{figure}[t!]
\begin{center}
	\begin{minipage}[t]{0.238\textwidth}\begin{center}
		\includegraphics[width=\textwidth]{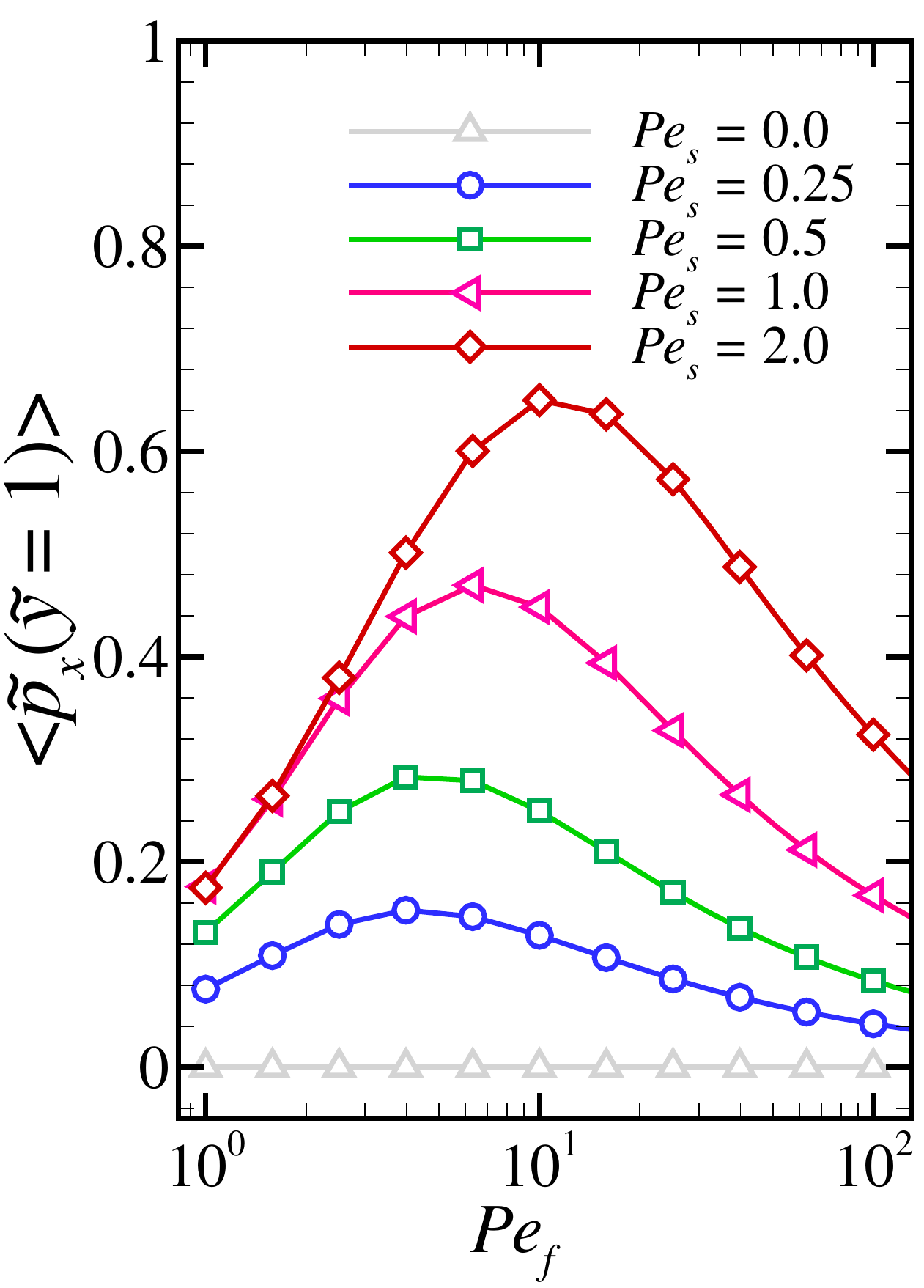} (a)
		\label{wallpy}
	\end{center}\end{minipage} \hskip0.cm	
	\begin{minipage}[t]{0.238\textwidth}\begin{center}
		\includegraphics[width=\textwidth]{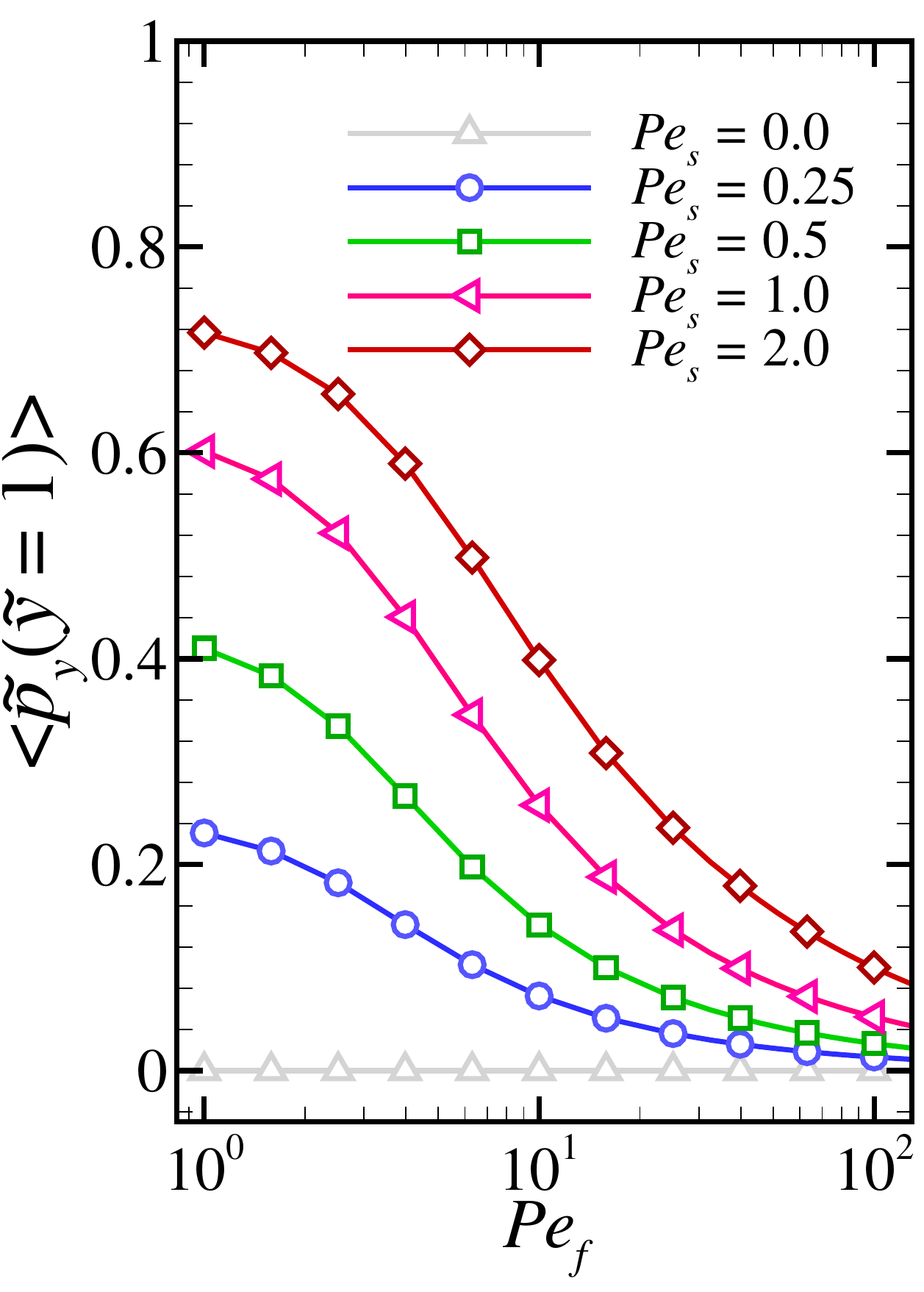} (b)
		\label{wallpz}
	\end{center}\end{minipage} 
\caption{(a) Parallel-to-flow and (b) perpendicular-to-flow components of the mean swimmer orientation vector, $\langle \tilde p_x\rangle$ and $\langle \tilde p_y\rangle$, respectively, on the top wall of the channel ($\tilde y=1$) plotted as a function of $Pe_f$, for fixed $\xi=0.82$, and different values of $Pe_s$ as shown on the graphs. Symbols show the computed results and curves are plotted as guides to the eye.
} 
\label{pxy_log} 
\end{center}
\end{figure}

It is interesting to note that, once the shear flow is established and gradually strengthened, both components of the mean orientation vector become smaller. The reduction in $\langle \tilde p_y\rangle$ with increasing shear strength (or flow P\'eclet number) is a natural consequence of the horizontally directed Couette flow barring swimmer tendency to approach the walls. However, the reduction in $\langle \tilde p_x\rangle$ with increasing shear strength (see, e.g., the curves for $Pe_f>10$ in Fig. \ref{psi_a}) is less trivial. A more clear view of the non-monotonic (monotonic) behavior of the mean parallel-to-flow (perpendicular-to-flow) orientation component is given in Fig. \ref{pxy_log}, where $\langle \tilde p_x\rangle$ and $\langle \tilde p_y\rangle$ at the {\em top} wall are plotted as a function of the flow P\'eclet number and for different swim P\'eclet numbers. As seen, increasing $Pe_f$ leads to a monotonic decrease in $\langle \tilde p_y\rangle$ and a corresponding increase in $\langle \tilde p_x\rangle$, but only up to a certain threshold (the peaks of the plots in Fig. \ref{pxy_log}a), beyond which $\langle \tilde p_x\rangle$ {\em decreases} with increasing $Pe_f$. Such a behavior has also been found,  but remained unexplained, for swimmers in a Poiseuille flow \cite{ezhilan}. This behavior in fact portrays the onset of a population splitting phenomenon in swimmer orientation that shall be discussed in the following section (see also Appendix \ref{mean_nearwall_orientation} for more details).  

\begin{figure*}[t!]
\begin{center}
	\begin{minipage}[t]{0.33\textwidth}\begin{center}
		\includegraphics[width=\textwidth]{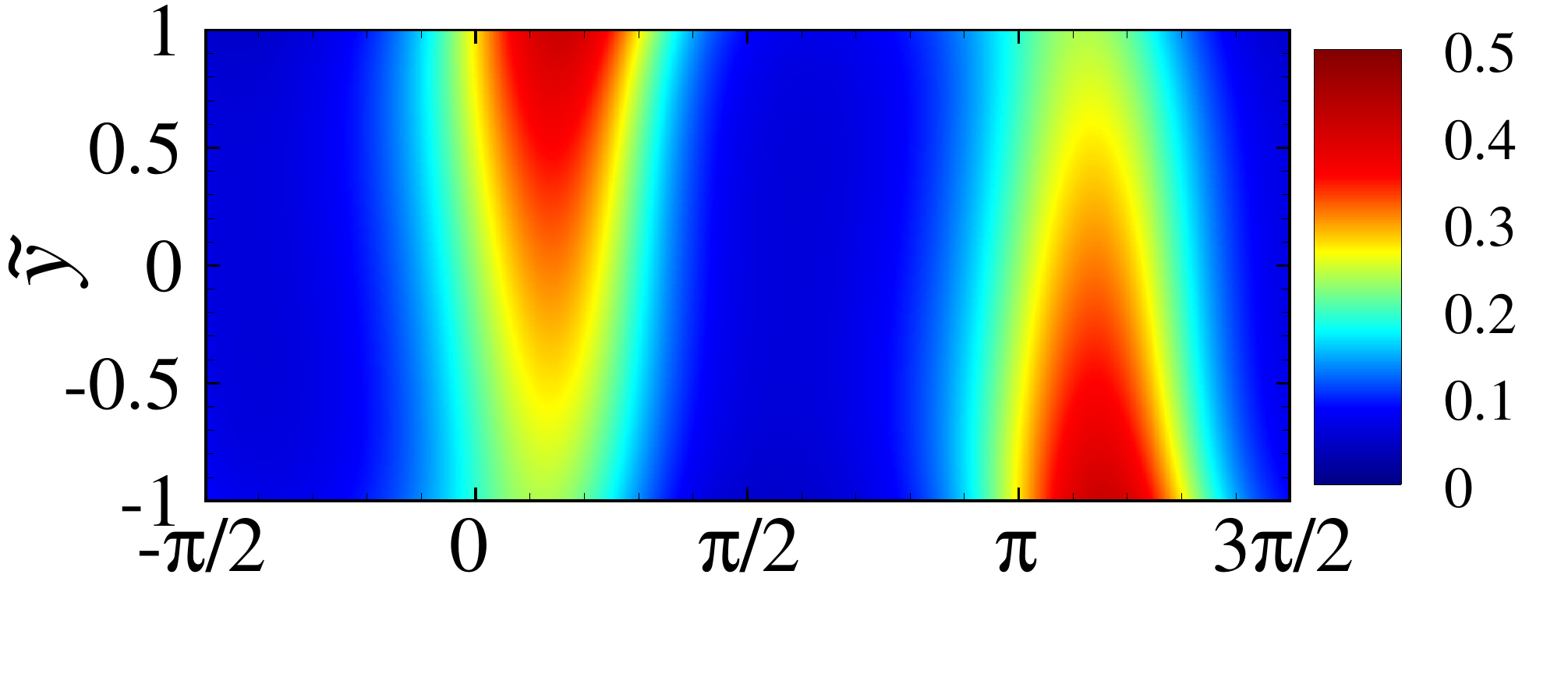} \vskip-2mm (a) $Pe_s=0.25$, $Pe_f=5$
	\end{center}\end{minipage} \hskip-0.2cm	
	\begin{minipage}[t]{0.33\textwidth}\begin{center}
		\includegraphics[width=\textwidth]{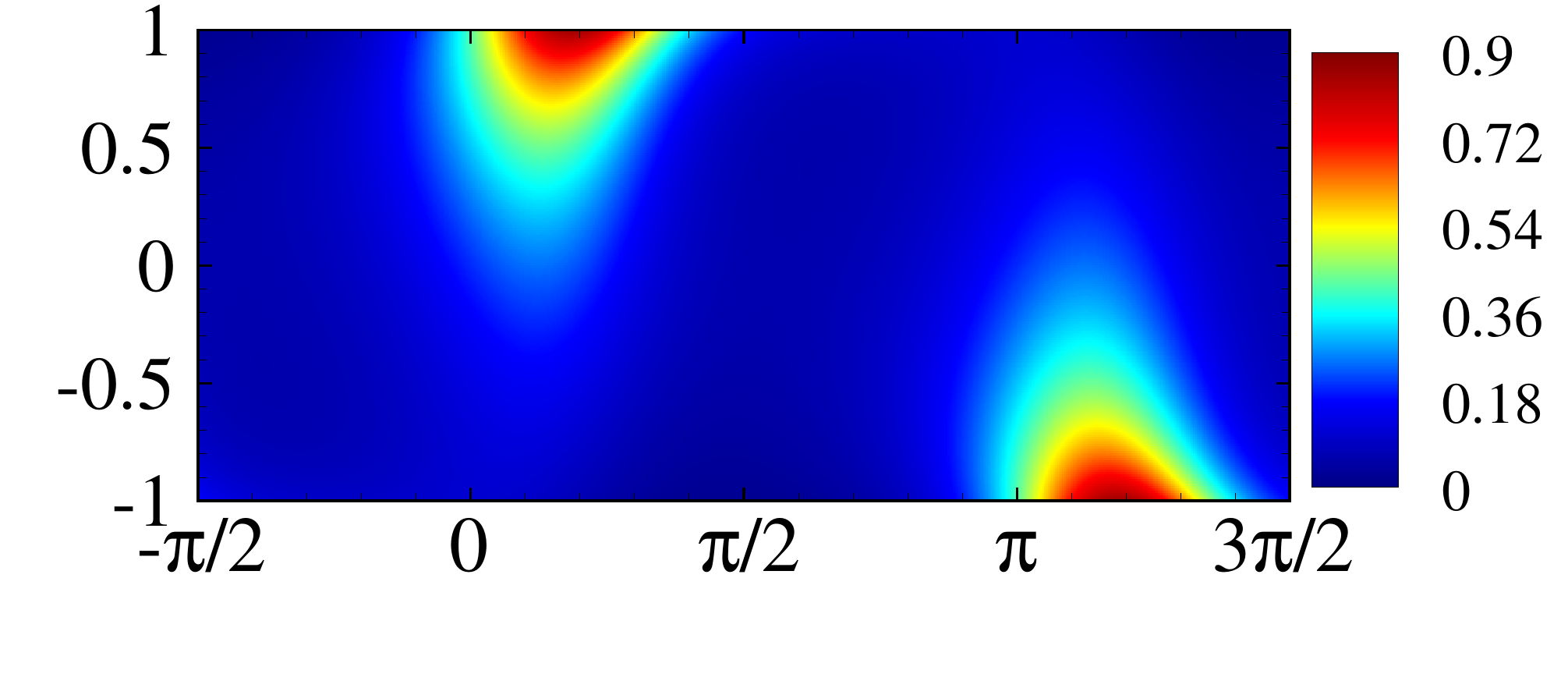} \vskip-2mm (b)   $Pe_s=1$, $Pe_f=5$
	\end{center}\end{minipage} \hskip-0.2cm	
	\begin{minipage}[t]{0.33\textwidth}\begin{center}
		\includegraphics[width=\textwidth]{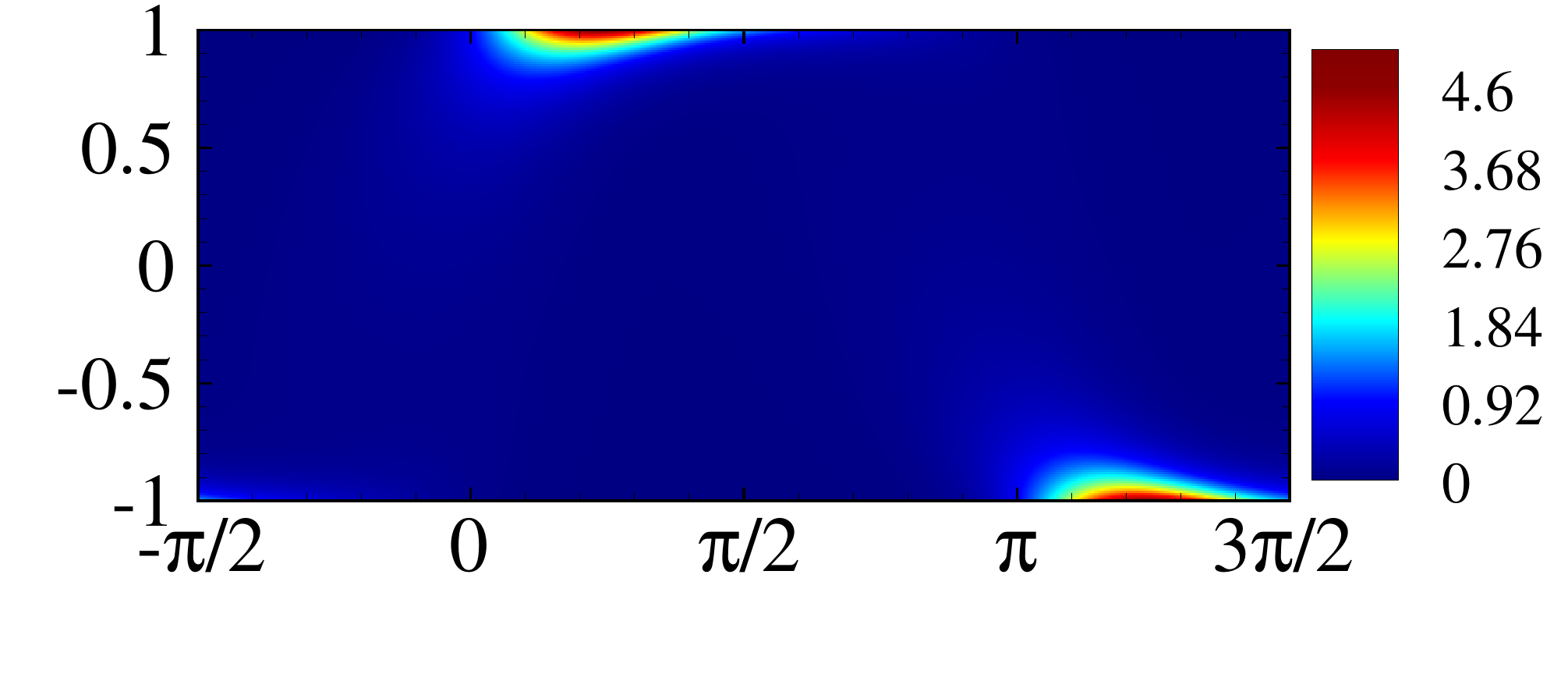} \vskip-2mm (c)  $Pe_s=5$, $Pe_f=5$
	\end{center}\end{minipage} \vskip0.2cm	
	\begin{minipage}[t]{0.33\textwidth}\begin{center}
		\includegraphics[width=\textwidth]{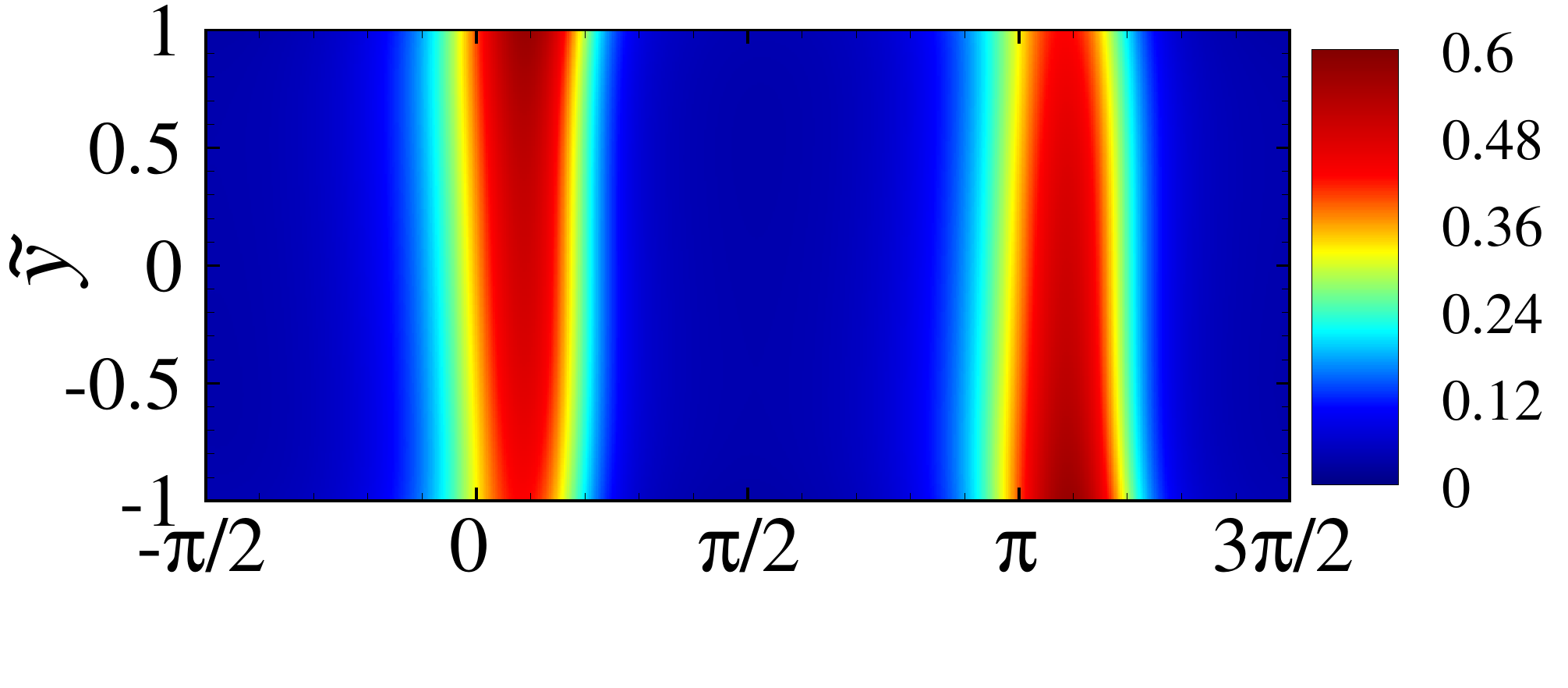} \vskip-2mm (d) $Pe_s=0.25$, $Pe_f=20$
	\end{center}\end{minipage} \hskip-0.2cm	
	\begin{minipage}[t]{0.33\textwidth}\begin{center}
		\includegraphics[width=\textwidth]{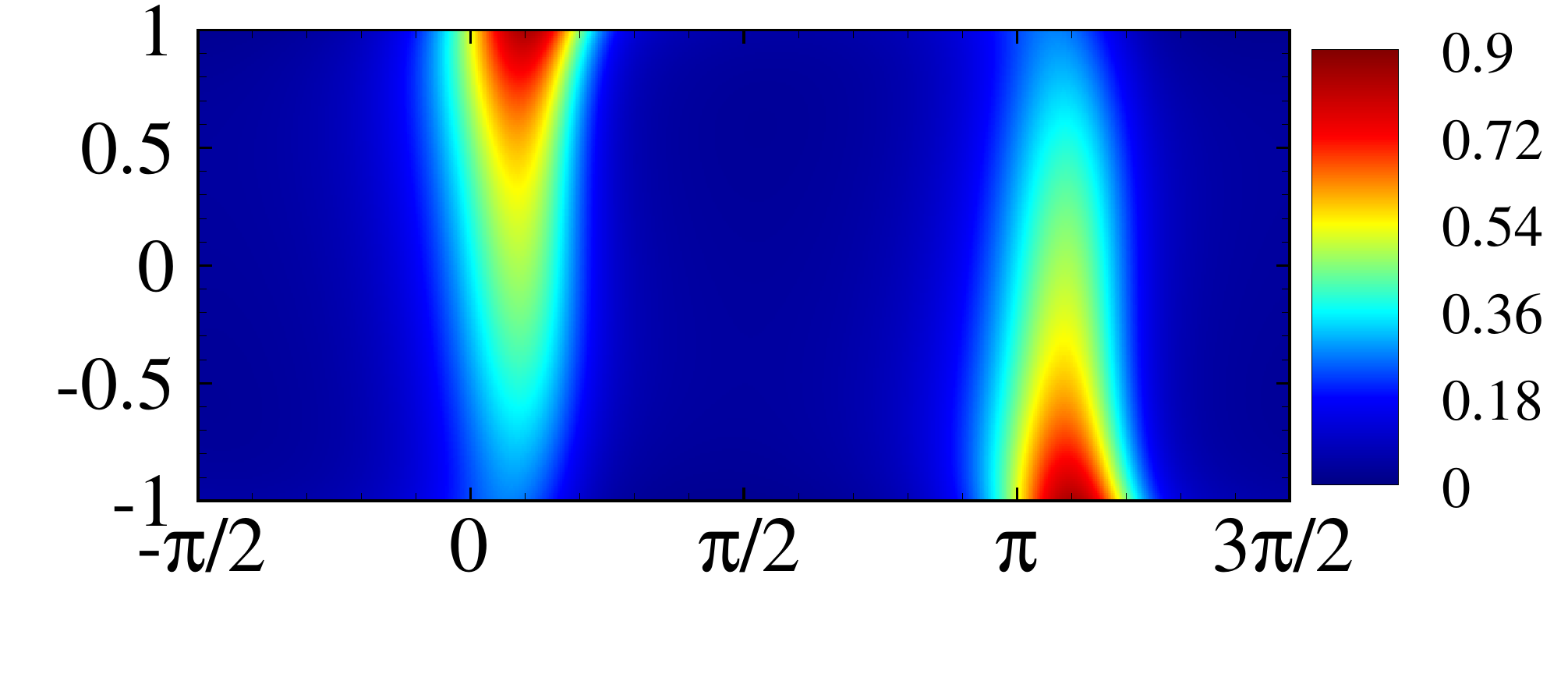} \vskip-2mm (e) $Pe_s=1$, $Pe_f=20$
	\end{center}\end{minipage} \hskip-0.2cm		
	\begin{minipage}[t]{0.33\textwidth}\begin{center}
		\includegraphics[width=\textwidth]{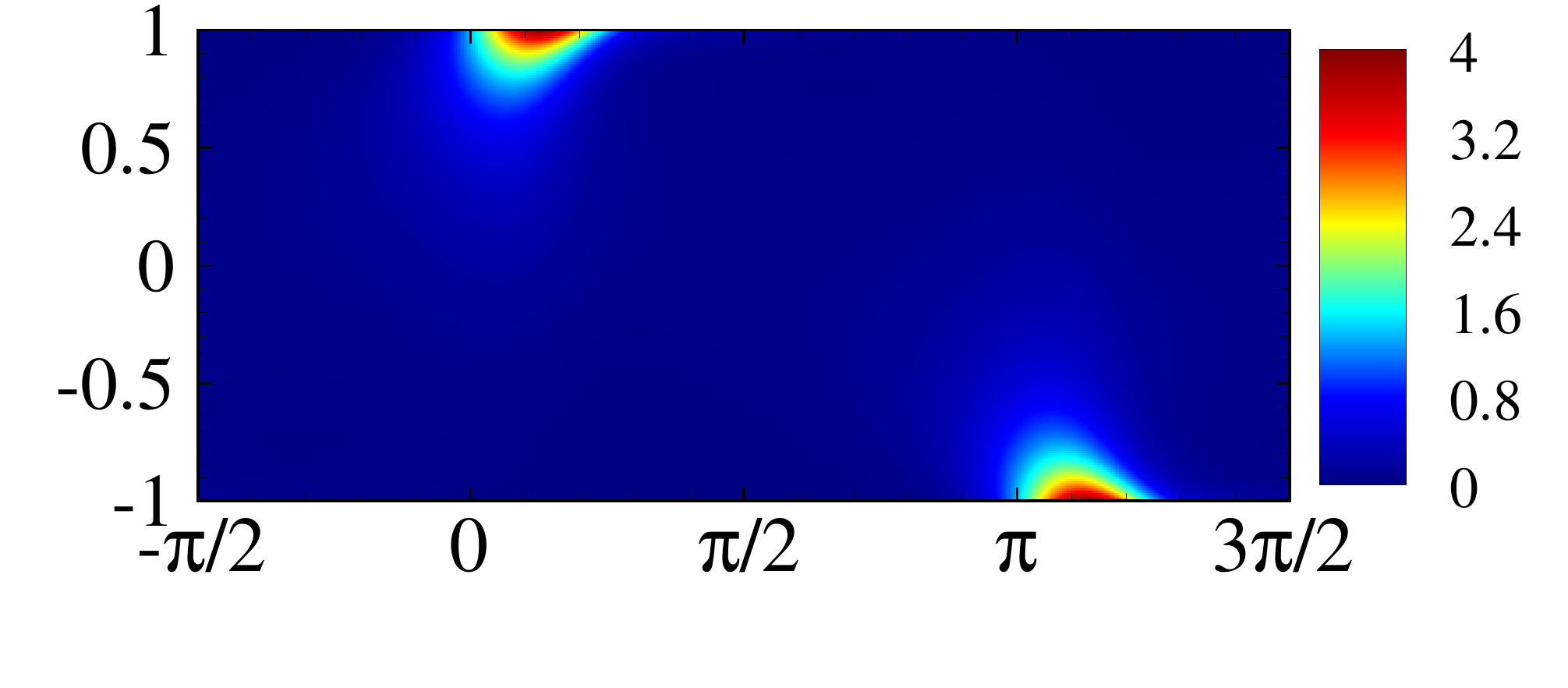} \vskip-2mm (f) $Pe_s=5$, $Pe_f=20$
	\end{center}\end{minipage} \vskip0.2cm
	\begin{minipage}[t]{0.33\textwidth}\begin{center}
		\includegraphics[width=\textwidth]{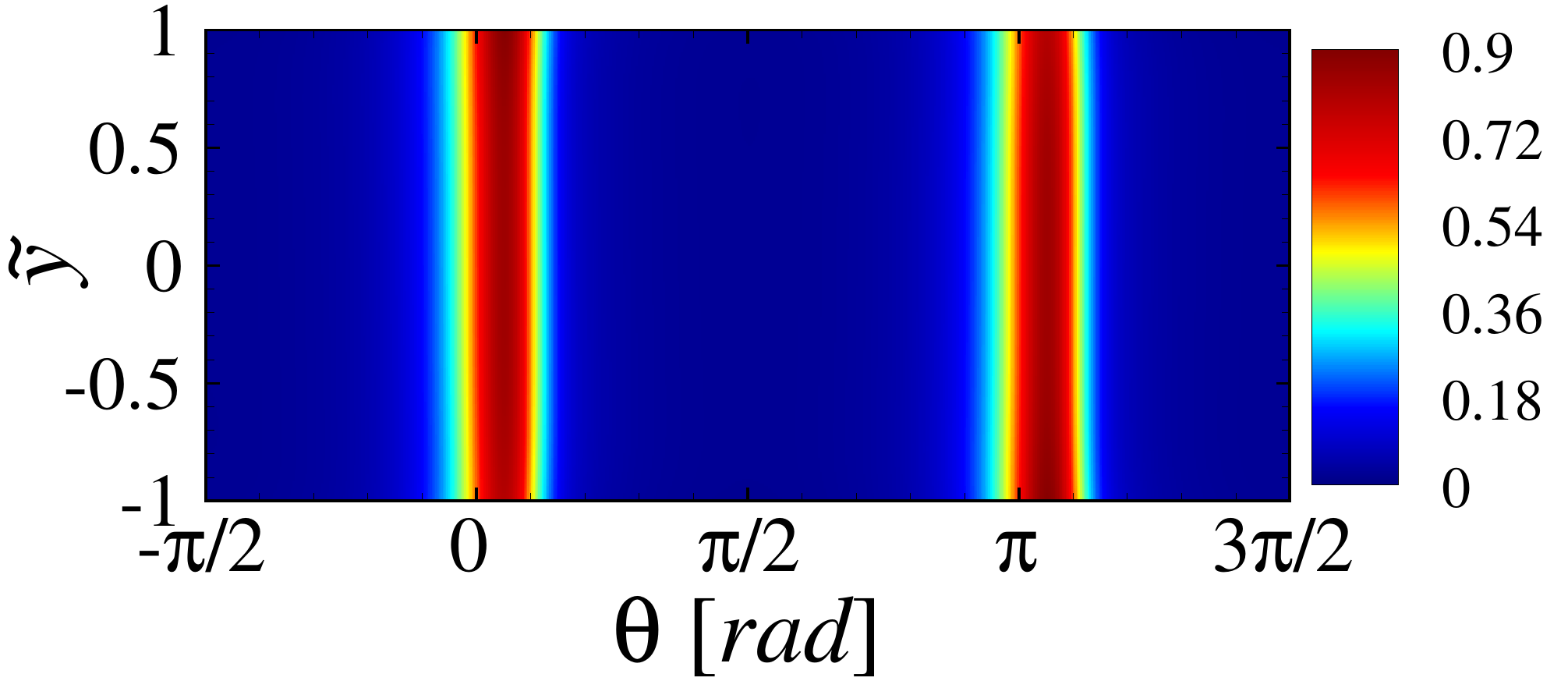} \vskip0mm (g) $Pe_s=0.25$, $Pe_f=100$
	\end{center}\end{minipage} \hskip-0.2cm	
	\begin{minipage}[t]{0.33\textwidth}\begin{center}
		\includegraphics[width=\textwidth]{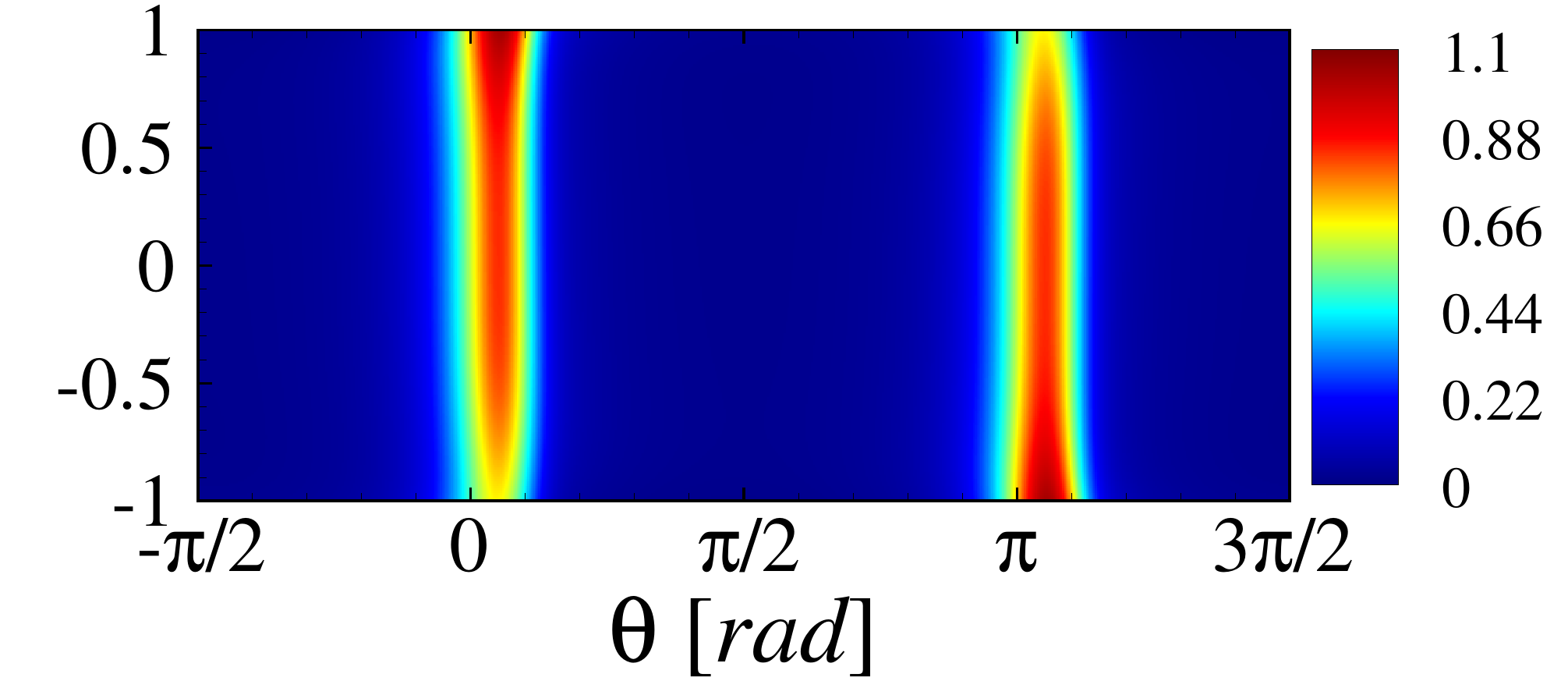} \vskip0mm (h) $Pe_s=1$, $Pe_f=100$
	\end{center}\end{minipage} \hskip-0.2cm
	\begin{minipage}[t]{0.33\textwidth}\begin{center}
		\includegraphics[width=\textwidth]{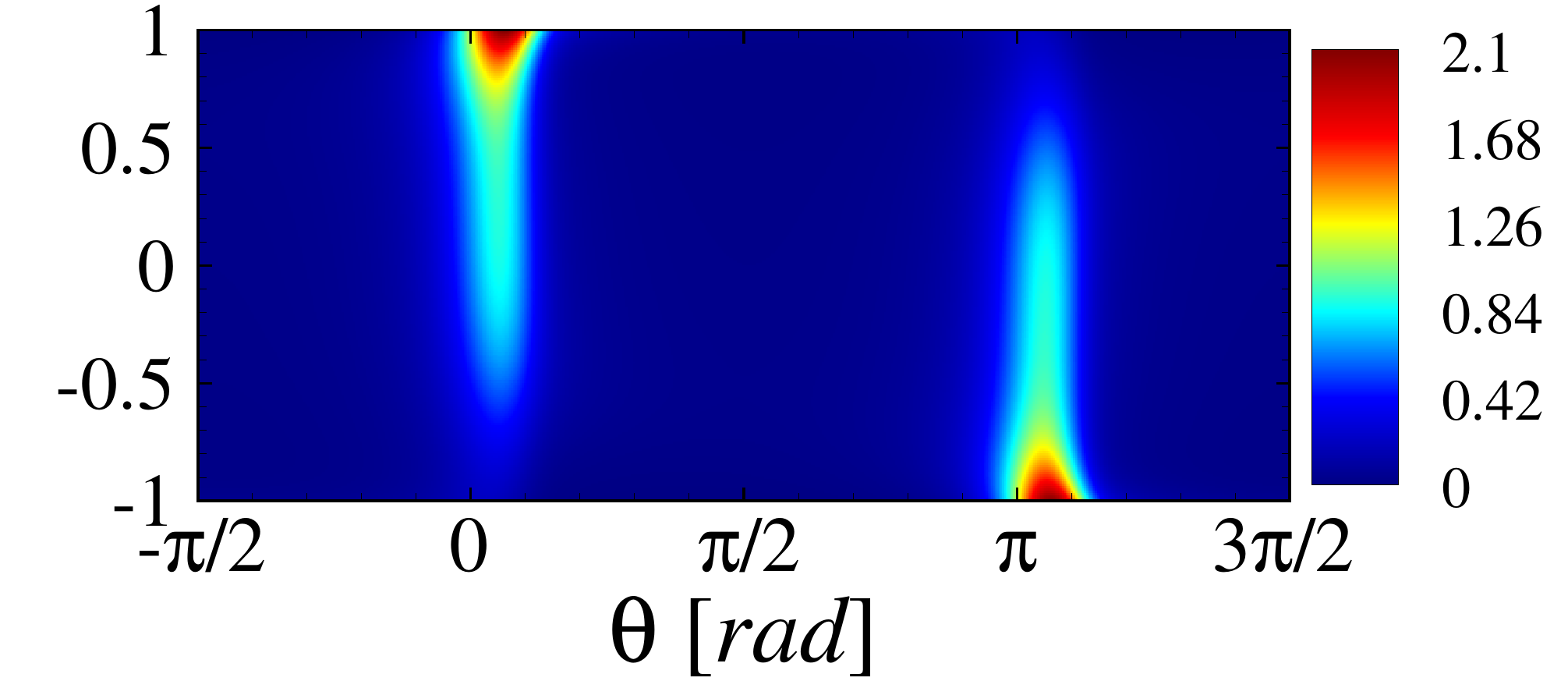} \vskip0mm (i)  $Pe_s=5$, $Pe_f=100$
	\end{center}\end{minipage} 
\caption{Density plots of the rescaled swimmer PDF, $\tilde{\Psi}(\tilde y,\theta)$, within the {\em  computational} domain, where the vertical and horizontal axes represent the position $\tilde y$ across the channel and the swimmer orientation angle $\theta$ (see Fig. \ref{figure_no1}). Here we show typical cases of weak, intermediate, and strong self-propulsion/shear rate as denoted by the different values of swim and flow P\'eclet numbers in the frame captions. In all cases, we have  $\xi=0.82$. 
}
\label{fig_contours_full}
\end{center}
\end{figure*}

\subsection{The population splitting phenomenon} 
\label{popsplitting}

Figure \ref{fig_contours_full} shows the profiles of the PDF, $\tilde\Psi(\tilde y, \theta)$, within the computational domain $-\pi/2\leq \theta< 3\pi/2$ and $-1\leq \tilde y \leq 1$. As can be seen in the captions, we have chosen $Pe_{s}=\left \{ 0.25,1,5 \right \}$ and $Pe_{f}=\left \{ 5,20,100 \right \}$ as sample parameter values representing (in the same order as they appear in the brackets) active self-propulsion and shear flow of relatively low, medium, and high strength, respectively. As the color bars show, the warmer (cooler) colors represent higher (lower) swimmer density. The symmetry in all profiles arises from the linear structure of Couette flow, resulting in $(\tilde y, \theta)\rightarrow (-\tilde y, \pi+\theta)$ symmetry for the solution of the Smoluchowski equation, Eq. (\ref{pde_nondim}). 

The tendency of swimmers to accumulate at confining boundaries is evident in all frames of Fig. \ref{fig_contours_full}: There are always two areas of high swimmer density near the top and bottom wall of the channel. Recalling the definition of $\theta$ discussed after Eq. (\ref{eq:av_def}), the first and second halves of the horizontal axes in the profiles of Fig. \ref{fig_contours_full} correspond to down- and upstream swimming, respectively. Hence, the high  density region near the top and bottom wall of the channel corresponds to active particles swimming in the downstream and upstream direction, respectively. 

It is interesting to note that, in the {\em self-propulsion-dominated regime} (represented by typically large values of $Pe_s$ such as in Fig. \ref{fig_contours_full}c),  there is a {\em single} population of swimmers on {\em each} of the channel walls, swimming in the up- or downstream direction depending on whether they are closer to the bottom or the top wall, respectively (on the top wall in panel c, this single population can be identified as the high-density region around an orientation angle between 0 and $\pi /2$, representing a tilting of swimmer orientation toward the wall). In the {\em shear-dominated regime} (represented by typically large values of $Pe_f$ such as in Fig. \ref{fig_contours_full}g), there are {\em two} distinct and well-separated populations of swimmers on {\em each} of the walls, swimming in  opposite down- or upstream directions (on the top wall in panel g, these populations can be identified as the two high-density regions around the orientation angles 0 and $\pi$, corresponding to down- and upstream swimmers, respectively). Similar features can be seen in other cases (e.g., in panels b and f that appear closer to the self-propulsion-dominated behavior in panel c,  and in panels d and h that appear closer to the shear-dominated behavior in panel g). However, the situation becomes more intriguing in the intermediate regime, where neither of the mechanisms are completely dominant (e.g., in panels a, e, i). In these cases, there still appear to be two distinct populations of swimmers on each of the walls but with very different densities. Therefore, the near-wall swimmers are still divided into two distinct populations (with angular separation of around $\pi$), representing a {\em majority} and a {\em minority} population, which, on the top wall, swim in the down- and upstream directions, respectively. The exact opposite of this situation occurs on the bottom wall. Therefore, the interplay between self-propulsion and imposed shear appears to lead to a discernible change in swimmer distribution from a single population to two separate populations with different up- and downstreaming behavior near the walls. This phenomenon can also occur in other positions across the channel width, depending on the system parameters (compare, e.g., panels c, f and i). We shall refer to this behavior as the {\em population splitting phenomenon}, whose characteristic features will be discussed in more detail below.

\begin{figure}[t!]
\begin{center}
	\begin{minipage}[t]{0.41\textwidth}\begin{center}
		\includegraphics[width=\textwidth]{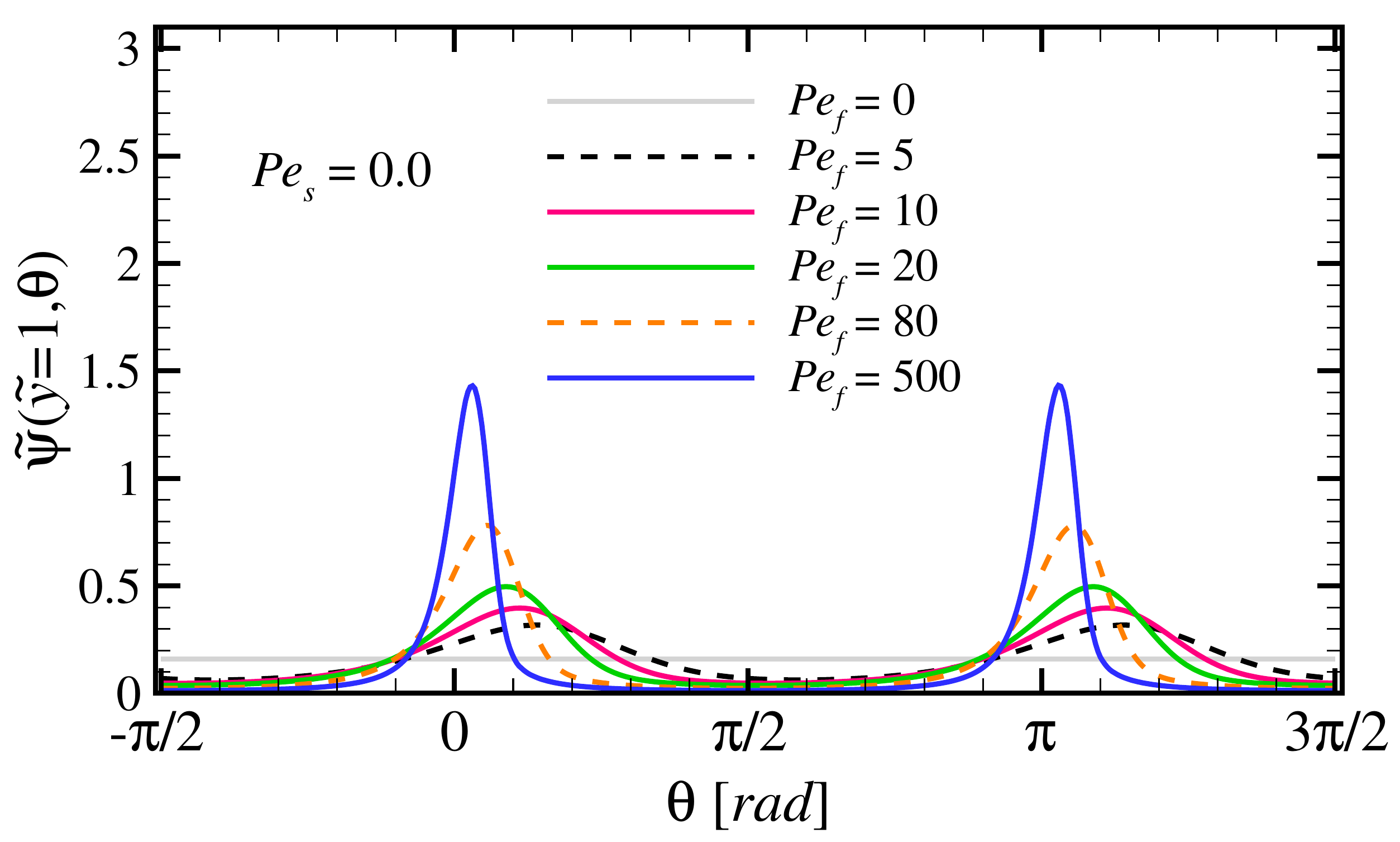} (a)
	\end{center}\end{minipage}\\ 
	\begin{minipage}[t]{0.41\textwidth}\begin{center}
		\includegraphics[width=\textwidth]{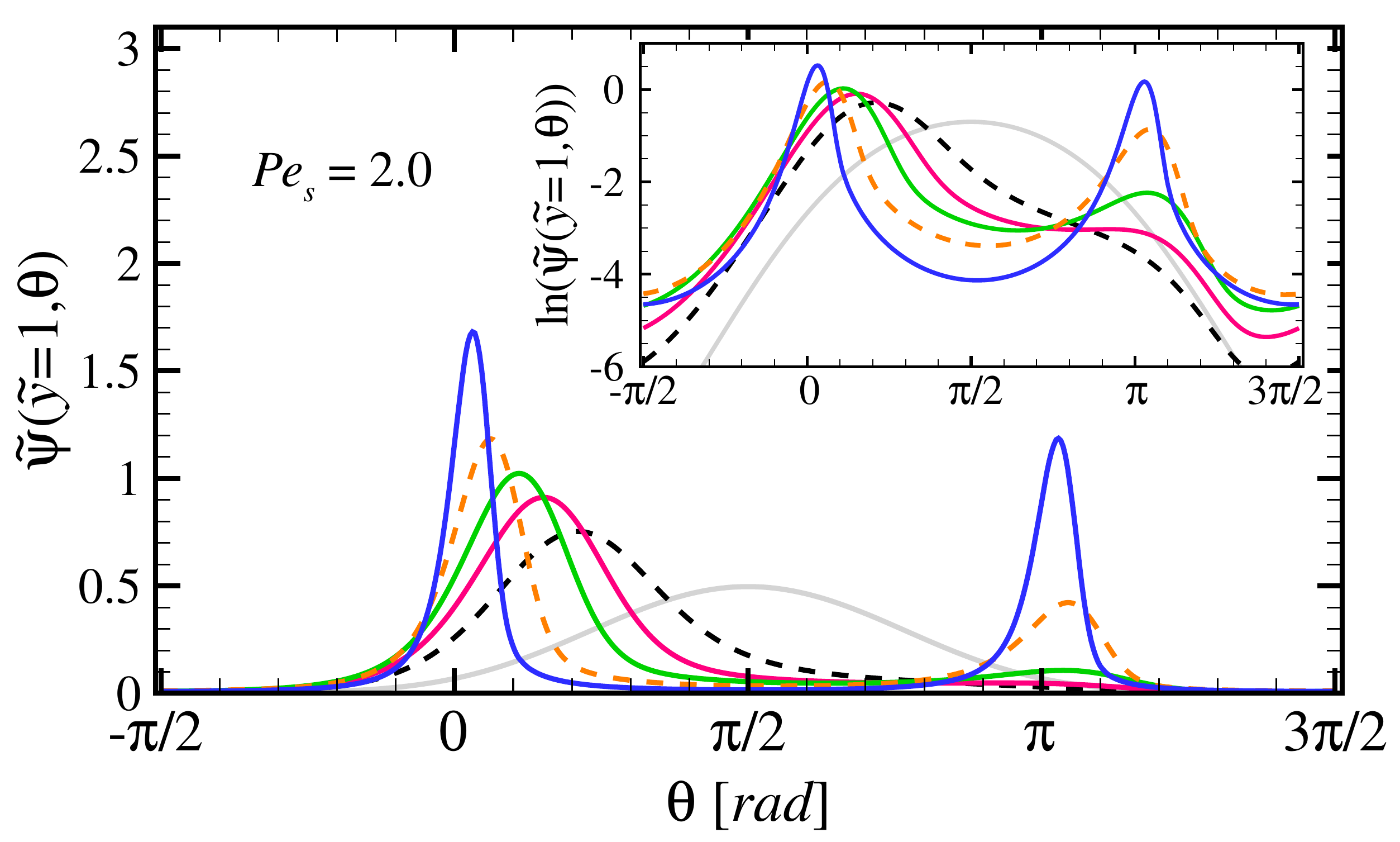} (b) 
	\end{center}\end{minipage}\\ 
	\begin{minipage}[t]{0.41\textwidth}\begin{center}
		\includegraphics[width=\textwidth]{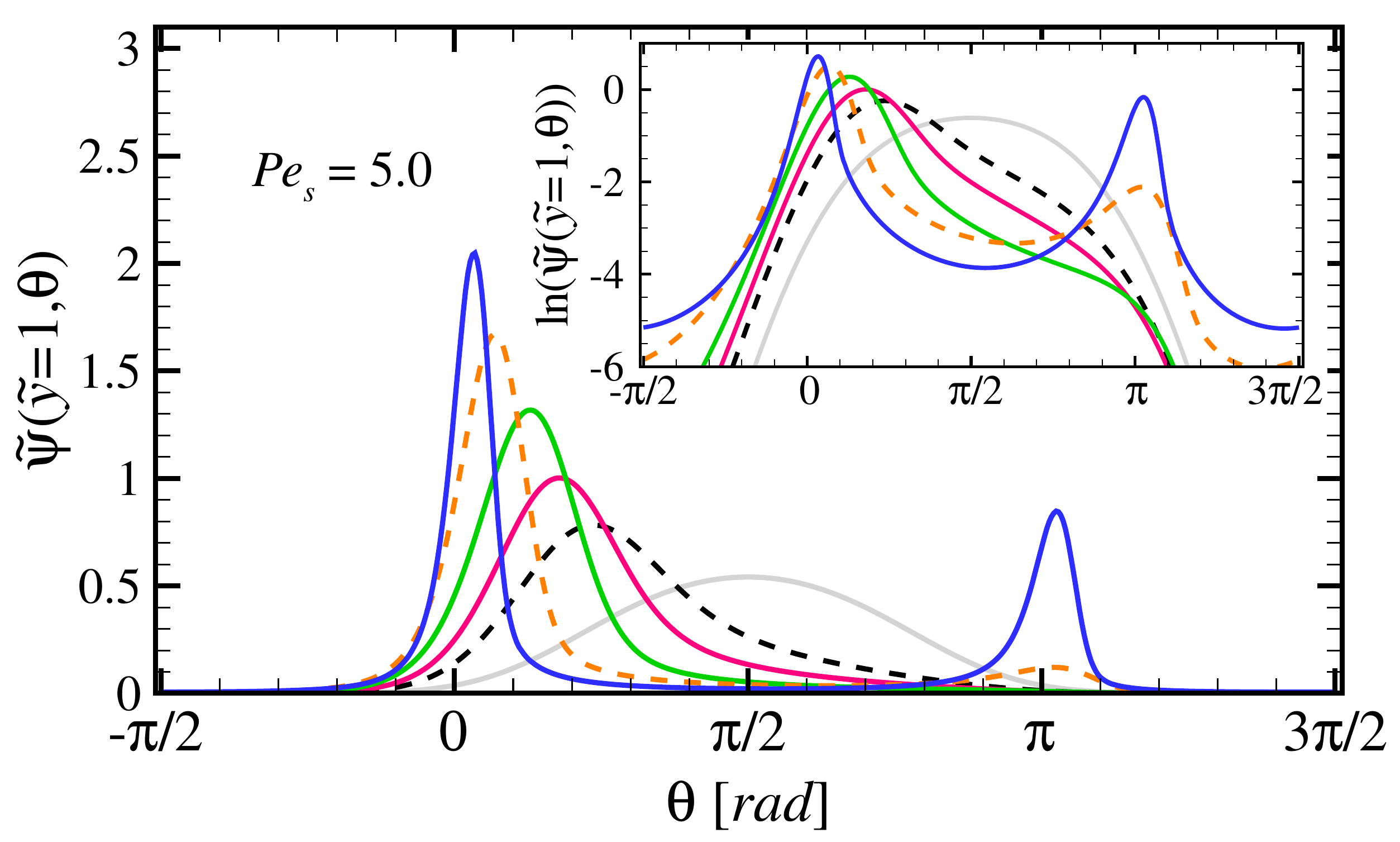} (c) 
	\end{center}\end{minipage} 
\caption{Rescaled top-wall PDF, $\tilde{\Psi}(\tilde y=1,\theta)$, of swimmer orientation is plotted in the range $-\pi/2<\theta<3\pi/2$  (see Fig. \ref{figure_no1}) for different values of the flow P\'eclet number as given in the legend and for different swim P\'eclet numbers (a) $Pe_s=0$, (b) $Pe_s=1$, and (c) $Pe_s=5$ at fixed $\xi=0.82$. Left ($-\pi/2<\theta<\pi/2$) and right ($\pi/2<\theta<3\pi/2$) halves of the horizontal axis are indicative of down- and upstream swimming directions, respectively. Insets show the log-linear view of the same profiles as shown in the main sets.
} 
\label{psi_profiles}
\end{center}
\end{figure}

\begin{figure}[t!]
\begin{center}
\includegraphics[width=6.cm]{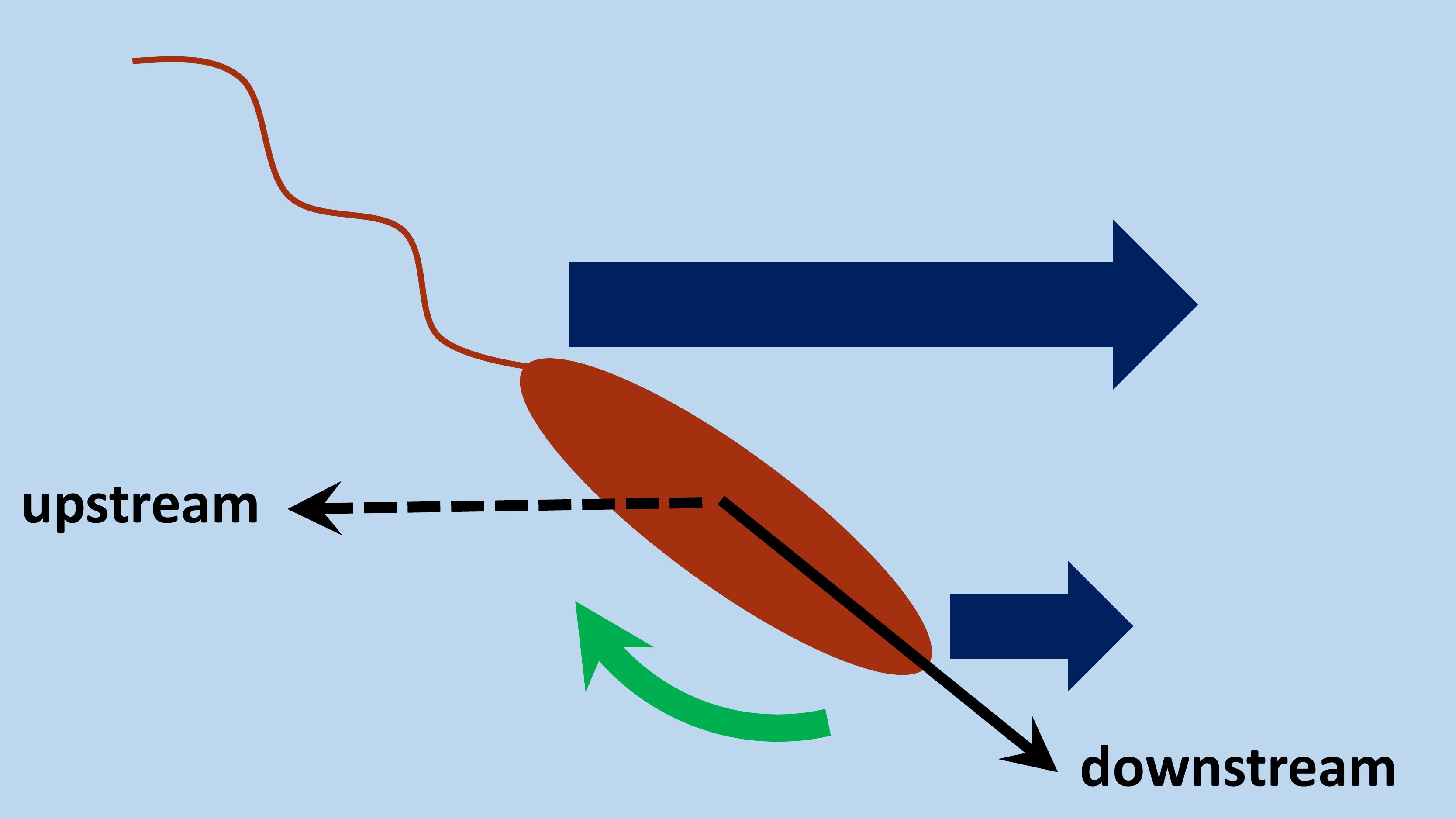}
\caption{Schematic view of a rodlike swimmer near the top channel wall  (Fig. \ref{figure_no1}), subjected to an external Couette flow, whose strength is denoted  by thicker arrows at the two ends of the sample swimmer. The tail-head direction is that of particle swimming. The sample swimmer is representative of a fraction of particles swimming {\em downstream}, while pointing {\em away} from the top wall. This fraction of particles reverse their swimming (tail-head) direction to {\em upstream}, thanks to the shear-induced torque (curved green arrow) in a flow of sufficient strength. The mentioned fraction  thus form a {\em minority} upstream population as opposed to the {\em majority} population of active particles that keep swimming downstream.  
} 
\label{downupstream}
\end{center}
\end{figure}

\subsection{Bimodality of the near-wall swimming}
\label{bimodal}

The population splitting phenomenon, and how it emerges and develops upon varying the swim and flow P\'eclet numbers can be elucidated further by focusing on the PDF of swimmers on the channel walls. As before, we concentrate on the top wall at $\tilde y=1$ (all that follows is true for the bottom wall at $\tilde y=-1$, if upstream and downstream change places in the wording). 

The  PDF, $\tilde{\Psi}(\tilde y=1, \theta)$, has been plotted in Fig. \ref{psi_profiles} for four different $Pe_s$ and a wide range of $Pe_f$. Figure \ref{psi_profiles}a shows the case with non-active rodlike particles ($Pe_s=0$). With no driving force for the particles to tend toward the walls, they would exhibit equal chance of aligning in or against the direction of flow (corresponding to the left or right peaks, respectively). This results in a symmetric bimodal distribution (with equal-height peaks) for the particles and vanishing components for their mean orientation vector (see Fig. \ref{pxy_log}). Note that, in this case, the particles are passive thin rods  \cite{doiedwards} and, as such, they can only nematically align with the flow, regardless of their positions in the channel. 

The situation is different for active swimmers, in which case the near-wall behavior of the particles plays a central role and their population splitting leads to two {\em oppositely swimming} majority and minority populations. In the absence of flow ($Pe_f=0$), the active particles will, on average, swim straight up toward the (top) channel wall, with no hindering effect from an imposed flow. This results in symmetric {\em unimodal} $\tilde{\Psi}$-profiles about the orientation angle $\theta_\ast=\pi/2$  (light-gray solid curves in Figs. \ref{psi_profiles}b and c). When the shear is imposed and strengthened, it will attempt to align the swimmers in the direction of flow, leading to {\em shift} of this single peak from $\theta_\ast=\pi/2$ toward smaller angles (see, e.g., the black dashed curves for $Pe_f=5$ in panels b and c). In this regime, self-propulsion is still the dominant mechanism driving the active particles, which form a single statistical population with a {\em unimodal} (but not necessarily symmetric) PDF peaked around the mean swimmer orientation, pointing in the downstream direction near the top wall. When $Pe_f$ is further increased at a given $Pe_s$, we find a second peak, whose height increases and gradually tends  toward that of the first peak, as the imposed flow becomes stronger (this is seen more clearly in log-linear scale in the insets of panels b and c). The emergence of this smaller second peak marks the onset of the population splitting phenomenon, as from there onwards, with $Pe_f$ increasing further, a growing number of swimmers from the {\em majority} population of {\em downstream} swimmers will be converting to develop the {\em minority} population of {\em upstream} swimmers (see the schematic illustration in Fig. \ref{downupstream}). The exact opposite of this conversion occurs on the bottom wall. It is thus clear that the PDF of swimmer orientation changes from {\em unimodal} to {\em bimodal}, exhibiting two peaks that remain well separated (by an angular separation of almost $\pi$) and represent two distinct macroscopic populations of swimmers, as the population splitting takes place and develops further as the flow P\'eclet number, $Pe_f$, is increased. Note also that larger values of $Pe_f$ lead generally to narrower PDF peak(s) as stronger shear is more effective in barring particle tilting due to rotational diffusion.

Going back to the results for the parallel-to-flow component of the mean swimmer orientation vector  $\langle \tilde p_x\rangle$ (Section \ref{mean-orientation}), we found that this quantity exhibits a non-monotonic behavior, eventually, tending to zero across the channel (Fig. \ref{orientn_profiles}a) and, in particular, near the channel walls (Fig. \ref{pxy_log}a) as $Pe_f$ is increased. Our foregoing analysis thus clearly shows that describing the near-wall behavior of swimmers in an imposed flow merely based on their mean parallel-to-flow orientation, in which swimmers are perceived to swim either up- or downstream near the walls, may be inaccurate and may even lead to the misleading interpretation that the near-wall, up- or downstream swimming trends are weakened at large shear rates, where $\langle \tilde p_x\rangle$ tends to zero. This is in fact where the swimmers split into two nearly equal but oppositely swimming populations. Thus, in order to capture the bimodal nature of the swimmer orientation, one needs to consider statistical quantities other than the mean orientation, including higher-order moments of swimmer orientation angle as shall be discussed later.

\subsection{The bimodal ratio}
\label{subsec:bi_ratio}

\begin{figure}[t!]
 \includegraphics[width=7.5cm]{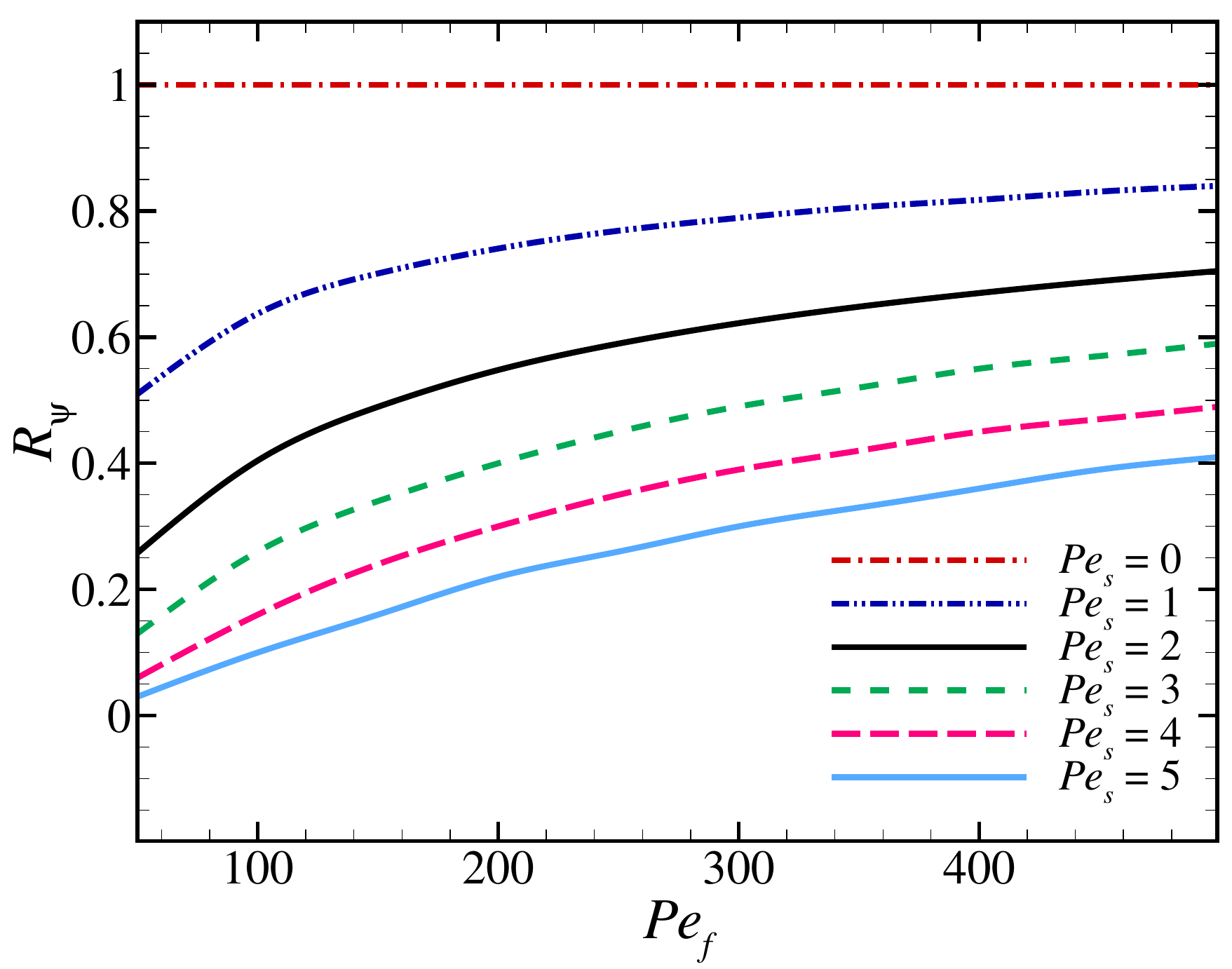}
 \caption{The bimodal ratio, as defined in the text, is plotted as a function of the flow P\'eclet number $Pe_f$ for $\xi=0.82$ and different values of $Pe_s$ as shown on the graph.}
\label{popratio}
 \end{figure}

The flow strength ($Pe_f$) required to generate a certain degree of population splitting would be larger under stronger swimmer self-propulsion (larger $Pe_s$). This can be seen from Fig. \ref{popratio}, which shows the bimodal ratio  $R_\Psi$, defined as the ratio of the minority population of upstream swimmers on the top channel wall (right peak of the profiles in Fig. \ref{psi_profiles}) and the majority population of downstream swimmers (left peak of the profiles in Fig. \ref{psi_profiles}). When there is no active self-propulsion, the two (non-swimming) populations are exactly equal and the ratio is identical to 1, regardless of the flow strength. For finite $Pe_s$, it can be seen that the ratio is smaller, for all flow strengths, when self-propulsion is stronger (larger $Pe_s$). 

Under imposed flow of (theoretically) infinite strength, the two populations will become equal (in this sense like the non-active situation) and in the near-wall $\tilde{\Psi}$-profiles, we would have two Dirac delta functions at $\theta=0$ and $\theta=\pi$. The curves in Fig. \ref{popratio} are accordingly seen to verge toward $R_\Psi=1$ as the flow P\'eclet number is increased. 

It is important to note that, at experimentally feasible values of $Pe_s$, like those shown on the graph, the situation is still far from the theoretical limits of zero self-propulsion or infinite flow strength. This has resulted in $R_\Psi$ falling well below the limiting value $R_\Psi\rightarrow 1$.  

\subsection{Onset of population splitting: Binder cumulant} 
\label{subsec:binder}

As noted in Sections \ref{popsplitting}-\ref{subsec:bi_ratio}, the onset and the extent of population splitting and bimodality in swimmer orientation in a Couette flow vary significantly depending on the flow and swim P\'eclet numbers as well as  the  vertical position of the swimmers across the channel. 

In general, when the PDF takes a bimodal shape, the first-order moments cease to act as suitable quantities to characterize the {\em typical} behavior of swimmers. One can instead turn to quantities  expressed in terms of the higher-order moments that may reflect the bimodal nature of the PDF. We make use of the notion of the Binder cumulant, commonly used in the context of symmetry-breaking equilibrium phase transitions (as, e.g., occurring within Ising models \cite{binder} with a characteristic uni- to bimodal shape change in the corresponding Boltzmann weight of the mean spin on a lattice). It is defined as 
\begin{equation}
	U_4 = 1 - \frac{\mu_{4}}{3\mu_{2}^{2}}, 
\end{equation}
where $\mu_{2}$ and $\mu_{4}$ are the second- and fourth-order central moments, respectively, of the swimmer orientation angle, $\theta$, at the walls.

 \begin{figure}[t!]
 \begin{center}
 \includegraphics[width=7.5cm]{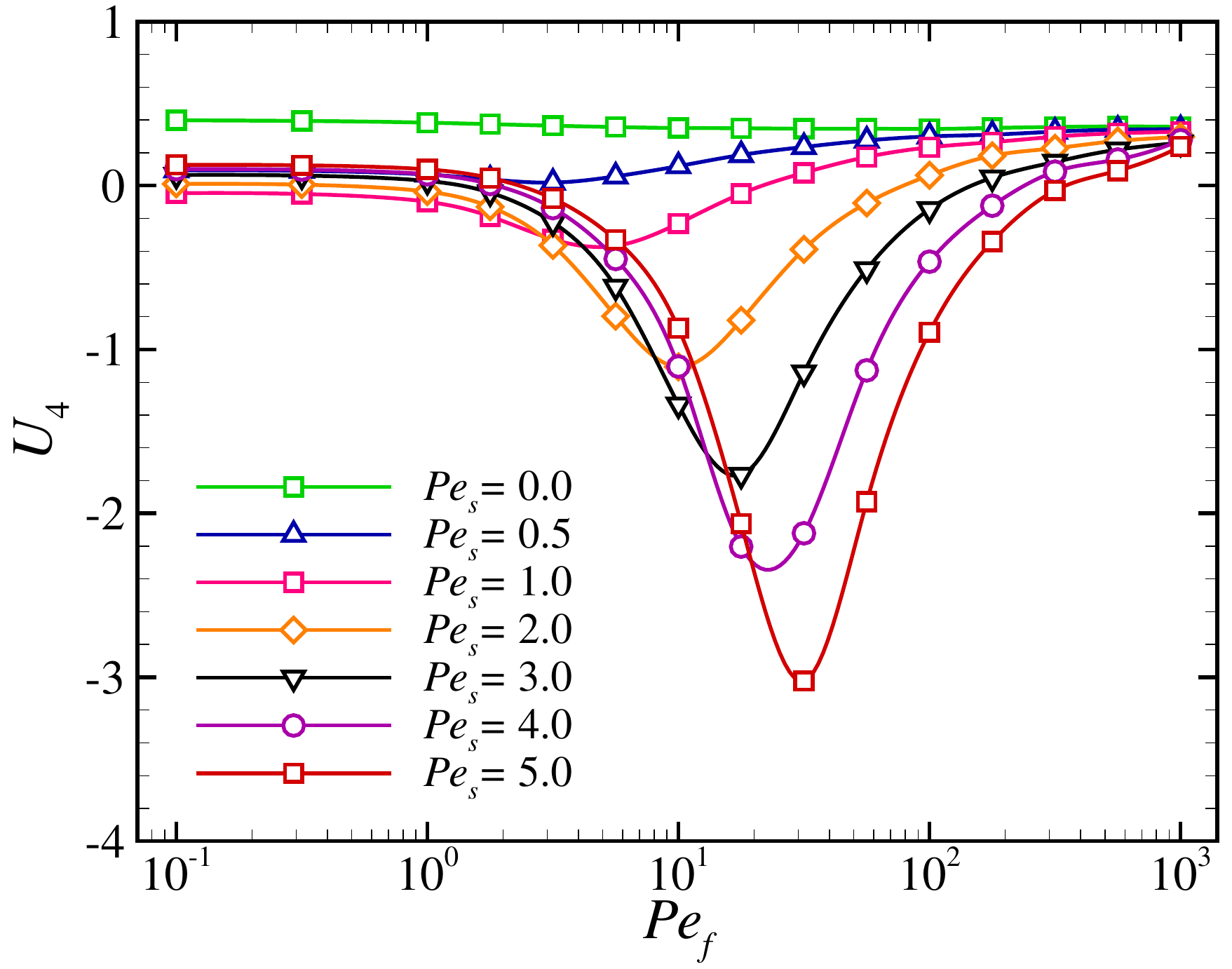}
 \caption{\label{binder} The Binder cumulant computed for the swimmer orientation angle on the top wall as a function of the flow P\'eclet number for $\xi=0.82$ and different values of $Pe_s$ as shown on the graph. The curves are guides to the eye. 
 }
 \end{center}
 \end{figure}
 
For a non-periodic random variable running over the whole real axis, the central moments can be calculated using standard textbook definitions; hence, the ratio $\mu_4/\mu_2^2$ will coincide with the standard kurtosis $\kappa$, and the Binder cumulant will be expressed in terms of the excess kurtosis $\kappa_e= \kappa  - 3$ as $U_4 = -\kappa_e/3$. In this case, the Binder cumulant vanishes ($U_4=0$) for a unimodal Gaussian PDF. It takes non-zero values when the PDF is bimodal and, in the limiting case of a distribution consisting of two delta-function peaks, it gives $U_4=2/3$.  

In the present case, however, the angular random variable $\theta$ is defined over a {\em circle} and the PDF is periodic in $\theta$, i.e., $\tilde{\Psi}(\tilde y,\theta)=\tilde{\Psi}(\tilde y,\theta+2\pi)$. Naively applying the standard definitions to calculate the central moments over a full period leads to results that generally depend on the choice of the period. This problem can be circumvented systematically by resorting to directional statistics \cite{mardia-jupp}. However, for the sake of simplicity,  we take a heuristic and more intuitive route to construct an interval-independent Binder cumulant as follows. For an arbitrary choice of angular interval $[\theta_1,\theta_1+2\pi)$ on the top or bottom wall, we determine the majority peak location $\theta_{\mathrm{max}}$ inside the interval. We evaluate the mean orientation angle, $\overline{\theta}$, by taking a symmetric interval equal to a full period around $\theta_{\mathrm{max}}$. This step is performed because the location of the peak varies and shifts toward  smaller (larger) angles on the top (bottom) wall upon increasing the flow P\'eclet number. The result for  $\overline{\theta}$  is then used to evaluate the central moments  $\mu_{n}=\overline{(\theta- \overline{\theta})^n}$ by taking a symmetric interval equal to a full period around $\overline{\theta}$. The resulting central moments and  the Binder cumulant will be independent of $\theta_1$. Using this procedure, the Binder cumulant can be shown to have the following limiting values: $U_4=2/5$ for $Pe_f=Pe_s=0$, where the distribution function is  uniform, $U_4\rightarrow 0$ for $Pe_f=0$ and $Pe_s\gg 1$, where the distribution takes a sharp unimodal shape, $U_4\rightarrow 1/3$ for $Pe_s=0$ and $Pe_f\gg 1$, where the distribution takes a bimodal shape with two sharp peaks at an angular separation equal to half the period (note that $U_4$ in this latter case is half the limiting value it takes for a non-periodic random variable mentioned above).  

The computed Binder cumulant is shown in Fig. \ref{binder} as a function of the flow P\'eclet number for different values of the swim P\'eclet number. In the absence of self-propulsion ($Pe_s=0$, green symbols), $U_4$ decreases monotonically with $Pe_f$ from a value close to 2/5 (uniform PDF) on the one end to 1/3 in the other (two sharp peaks), in accordance with the expected limiting values. For a finite $Pe_s$, the morphological variations of the PDF from a unimodal  to a bimodal shape is clearly indicated by a drop in $U_4$ at intermediate $Pe_f$, while $U_4$ stays close to its limiting values on the two ends (zero for unimodal PDF at small $Pe_f$ and 1/3 for two sharp peaks at large $Pe_f$). The larger the swim P\'eclet number, the larger is the drop in the Binder cumulant at its minimum. The latter occurs at a swim P\'eclet number nearly equal (within our margin of error)  to the value representing the onset of population splitting,  indicating that this onset can also be identified by a large negative Binder cumulant. 

We note that excess kurtosis, $\kappa_e$ (which, as noted previously, is proportional to the Binder cumulant, $U_4 = -\kappa_e/3$, for non-periodic random variables) has long been considered as a measure for unimodality versus bimodality in statistics literature, where its usefulness and limitations have also been discussed extensively \cite{Darlington1970,Moors1986,Macgillivray1988}. In the case of two-point distributions (equivalent to the two-state model introduced in Appendix \ref{mean_nearwall_orientation}), the Binder cumulant can be shown to take large negative values as the distribution becomes less symmetric (i.e., as the probability density at one of the points or, equivalently, the bimodal ratio, $R_\Psi$, is decreased), and, eventually, tend to negative infinity as the onset of bimodality is approached ($R_\Psi\rightarrow 0$) \cite{Darlington1970}. Although such a behavior is consistent with the trend seen in Fig. \ref{binder}, further attempt to analyze the behavior of the shown Binder cumulant requires an account of the cyclic nature of the orientation angle, which goes beyond the scope of the present work. 


 \begin{figure}[t!]
 \begin{center}
 \includegraphics[width=7.5cm]{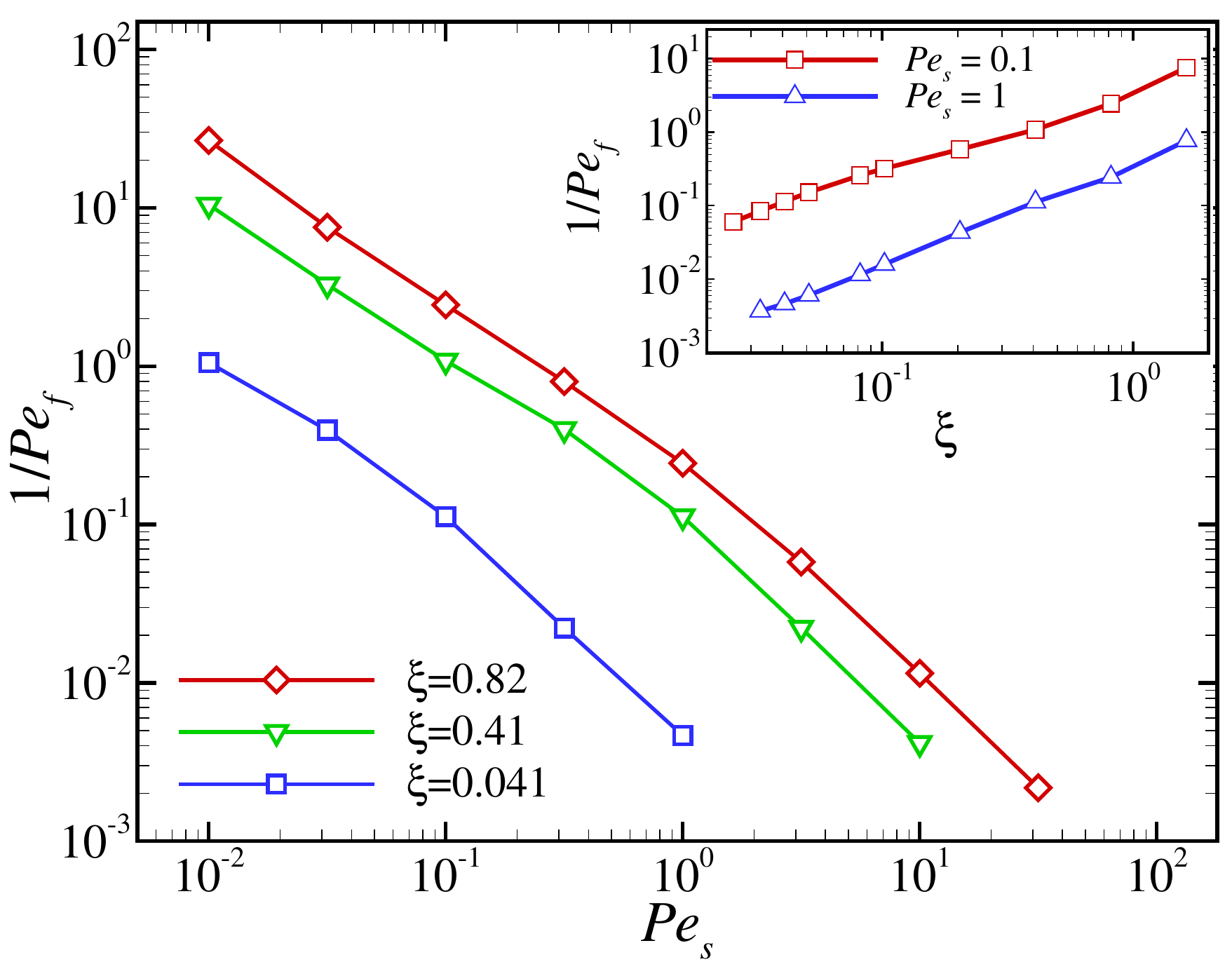}
 \caption{\label{pashetrans} Phase diagram in terms of the inverse flow P\'eclet number $1/Pe_f$, and the swim P\'eclet number $Pe_s$, for $\xi = 0.041, 0.41$ and 0.82. The bimodal (unimodal) regime is the region below (above) each curve. Symbols show the values of the inverse flow P\'eclet number at the onset of population splitting $1/Pe_f^\ast$. Inset shows the same quantity as a function of $\xi$ for  $Pe_s=0.1$ and 1. The lines are guides to the eye.
 }
 \end{center}
 \end{figure}
 
\subsection{Phase diagram} 
\label{phasediag}

In Fig. \ref{pashetrans}, we present a phase diagram in terms of the inverse flow P\'eclet number $1/Pe_f$, the swim P\'eclet number $Pe_s$, and the third dimensionless parameter of the system $\xi$ (see Eq. (\ref{xi_def})). The plots in this figure are shown in log-log scale and symbols represent the inverse values of the flow P\'eclet number at the onset of population splitting, $Pe_f^\ast$ (computed either directly by studying the $\tilde \Psi$-profiles or by examining the Binder cumulant plots, with closely matching results), as a function of $Pe_s$ for fixed $\xi = 0.041, 0.41, 0.82$ in the main set, and as a function of $\xi$ for fixed $Pe_s=0.1$ and 1 in the inset. Therefore, the area below (above) each curve represents the bimodal (unimodal) regime. As noted in Section \ref{non-dim}, the base value of $\xi=0.82$ was used to enable a quantitive comparison between our results and those in the case of an imposed  Poiseuille flow \cite{ezhilan}, as discussed throughout the previous sections. Figure \ref{pashetrans} presents our results for the population-splitting behavior of swimmers over a much wider range of values for $\xi$. The curves that indicate the boundary between the uni- and bimodal regimes appear to take approximately linear forms in the log-log representation of Fig. \ref{pashetrans}. This can suggest possible scaling-like behavior for $Pe_f^\ast$, especially in the regime $Pe_s < 1$; in the regime of parameters explored here, the exponent,
\begin{equation}
\alpha=\frac{{\mathrm d}\ln Pe_f^\ast}{{\mathrm d}\ln Pe_s}>0,
\end{equation}
appears to be weakly dependent on $\xi$ and given roughly as $\alpha\simeq 1$ for $Pe_s < 1$.

We also note that the lines separating the unimodal and bimodal regimes in the phase diagram can be identified as lines of {\em discontinuous}  transitions in analogy with the standard terminology in the context of equilibrium phase transitions \cite{kardar}. 
To further elucidate this point, we consider the behavior of the loci $\theta_\ast$ of the peaks of the PDF (see Fig. \ref{psi_profiles}), corresponding to the most probable orientations of swimmers near the walls. We also consider both positive and negative shear rates and, thus, allow $Pe_f$ to take both positive and negative values, with the negative values corresponding to a situation where the direction of the fluid flow in the channel is reversed. 
 
\begin{figure}[t!]
 \begin{center}
 \includegraphics[width=7.5cm]{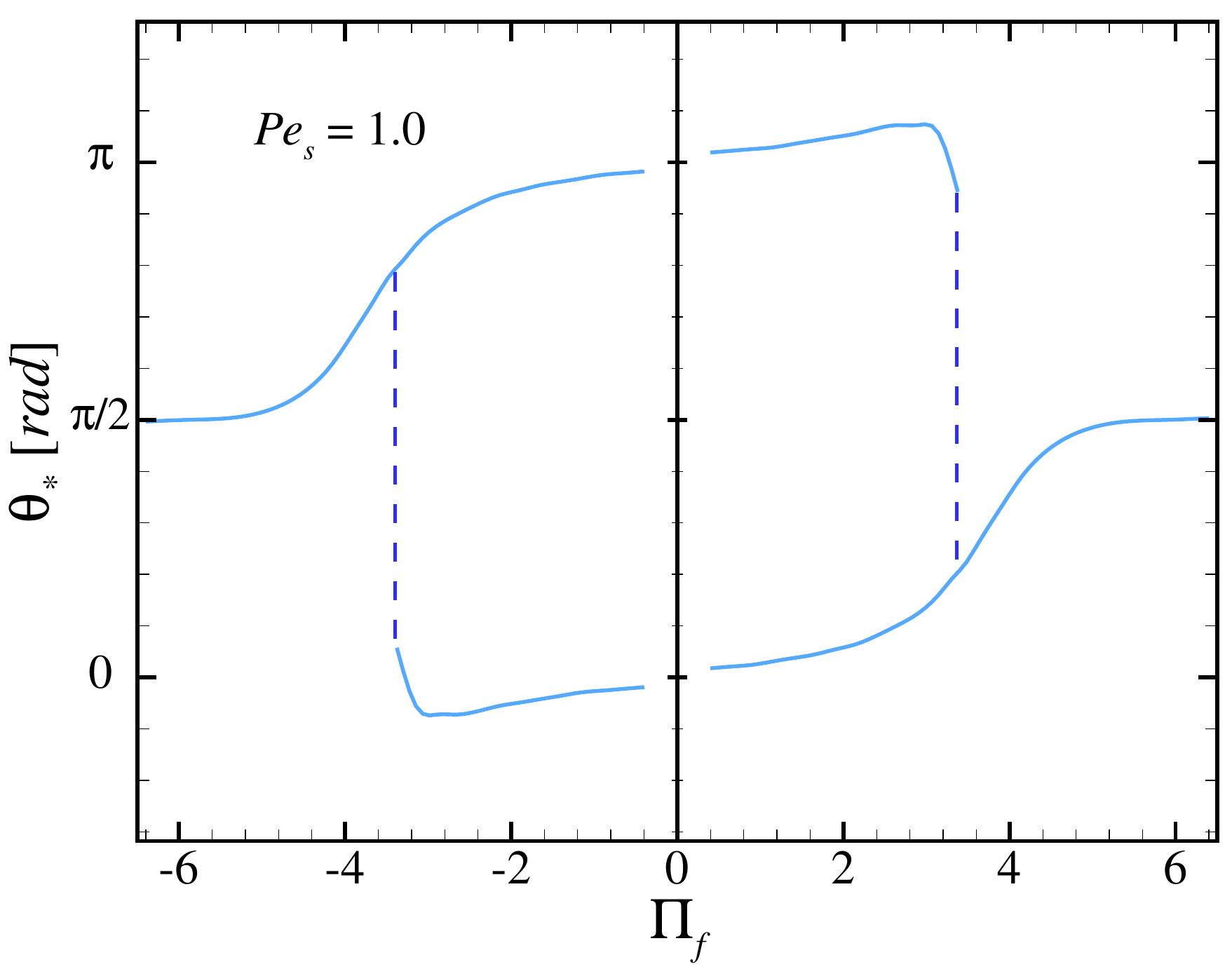}
 \caption{\label{hysteresis} Most probable swimmer orientation angle (at the peaks of the PDF) on the top wall, $\theta_\ast$, plotted as a function of  the redefined variable $\Pi_f = {\mathrm{sgn}}(Pe_f)\left(\log_{10}|Pe_f|^{-1}+C\right)$  with $C=4$. This change of variable is used to show the wide range of negative and positive values for $Pe_f$ on a logarithmic scale (see the text). Large values of $|Pe_f|$ are mapped to small values for $|\Pi_f|$, and vice versa.
 The lower (upper) branches for positive shear rates (with $\Pi_f>0$)  correspond  to the downstream majority (upstream minority) population, respectively. The situation is the exact opposite for $\Pi_f<0$. Here, we have set $Pe_s=1$ and $\xi=0.82$. 
 }
 \end{center}
 \end{figure}

In order to show a wide range of negative and positive values for $Pe_f$ in a single frame, we plot $\theta_\ast$ on the top wall as a function of  the redefined variable $\Pi_f = {\mathrm{sgn}}(Pe_f)\left(\log_{10}|Pe_f|^{-1}+C\right)$ with $C=4$ in Fig. \ref{hysteresis}. This change of (abscissa) variable enables a plot on a logarithmic scale with the branches, representing negative (positive) values of $Pe_f$, appearing separately on the left (right) half of the frame. Note that,  for this latter purpose,  any positive value of $C$ can be used, provided that $\log_{10}|Pe_f|^{-1}+C$ remains positive over the range plotted for $Pe_f$; in the figure, we vary $|Pe_f|$ from $4\times 10^{-3}$ up to $4\times 10^3$ and we have thus conveniently set $C=4$. Also, in the mentioned representation, large values of $|Pe_f|$ are mapped to small values for $|\Pi_f|$, and vice versa. The narrow region around $\Pi_f=0$ corresponds to excessively high shear rates $|Pe_f|\gg 1$ that cannot be treated numerically. In spite of this, and given that the states with $Pe_f=\pm \infty$ would be indistinguishable, one can formally  extrapolate the curves shown in Fig. \ref{hysteresis} down to $\Pi_f=0$ from both sides, in which case the right and left branches will be joined. It should be pointed out that the current approach based on low-Reynolds-number hydrodynamics \cite{purcellmain,happelbook} breaks down at very large $Pe_f$, and the region around $\Pi_f=0$ is expected to remain outside the regime of applicability of our results.

Figure \ref{hysteresis} is obtained for fixed  $Pe_s=1$ and $\xi=0.82$ but it clearly shows the characteristic features discussed in the preceding sections: $\theta_\ast$ is single-valued and tends to $\pi/2$ when $|\Pi_f|\rightarrow \infty$ or, equivalently, when $|Pe_f|\rightarrow 0$; $\theta_\ast$ remains single-valued in the unimodal regime above (below) a threshold value for $|\Pi_f|$ ($|Pe_f|$), and becomes multi-valued with two separate branches in the bimodal regime below (above) that threshold.  The threshold value of $|\Pi_f|$, corresponding to the {\em transition point} separating the uni- and bimodal regimes, is indicated by a dashed vertical line in Fig. \ref{hysteresis} (occurring at $Pe_f\simeq 4.1$ for the given parameter values). The {\em discontinuous} nature of the transition is thus  evident from this graph. 

The lower and upper branches for positive shear rate, corresponding to $\Pi_f>0$ (right half of the plot), give the most probable swimmer orientation near the (top) wall. The lower (upper) branch corresponds to the majority (minority) population and can be interpreted as the globally (locally) most probable state.
The situation is the exact opposite for negative shear rate, corresponding to $\Pi_f<0$ (left half of the plot). 

The results in Fig. \ref{hysteresis} may thus appear as indicating the possibility of a hysteresis-like behavior for the most probable orientation of swimmers with respect to the flow P\'eclet number in a channel, where the direction of the Couette flow can be reversed. This kind of behavior might indeed be possible in reality, that is, swimmers in a majority population, oriented in the direction of a sufficiently large (positive) shear rate ($\theta_\ast\simeq 0$), may resist flipping at once to form a majority population in the opposite direction ($\theta_\ast\simeq \pi$) in accordance with the reversed flow.  It is important to note, however, that interpretation of our results in Fig. \ref{hysteresis} in terms of a hysteresis-like effect and, in general,  any analogies with equilibrium phase transitions, should  be done with caution (see Section \ref{conclusion}). 

\section{Conclusion and discussion}
\label{conclusion}

We present a thorough study of the effects of an externally imposed Couette flow on the steady-state behavior of a dilute  suspension of active  (self-propelled) rodlike swimmers in a planar channel with rigid no-slip walls. Using a continuum description for non-interacting active rods \cite{ezhilan}, we present a detailed investigation of the interplay between the imposed shear and the intrinsic self-propulsion of the  swimmers in the channel. 

In the self-propulsion-dominated regime, the swimmers form a single population, strongly accumulated near the channel walls and swimming in the down- and upstream directions near the top and bottom walls, respectively, in accordance with the directions intuitively implied from the linear structure of the Couette flow. Upon increasing the shear rate, the swimmers become more strongly aligned with the flow, resulting in larger magnitudes for the mean parallel-to-flow component of the orientation vector of the down- (up-) streaming swimmers near the top (bottom) wall. This behavior, however, holds only up to a threshold shear rate (or flow P\'eclet number, $Pe_f$), beyond which the shear-induced torque acting on the particles becomes strong enough to reverse the swimming direction of a finite fraction of swimmers from the direction dictated by the flow; this fraction of swimmers thus forms a minority, but statistically significant and macroscopic, population of particles swimming in the opposite direction to the majority population. Near the top wall, the conversion is from the downstream (majority) to upstream (minority), and vice versa near the bottom wall.   This behavior, which we refer to as the {\em population splitting phenomenon}, corresponds to a  morphological {\em uni-} to  {\em bimodal} change in the joint (position-orientation) probability distribution function (PDF) of the swimmers. 

We thus argue that the standard picture for the near-wall behavior of swimmers in a shear flow, in which the swimmers are perceived, based on their {\em mean} parallel-to-flow orientation, to swim up- or downstream may be inaccurate and misleading, as it implies overall up- or downstream swimming near the walls, missing the fact that beyond the onset of population splitting, swimmers form two discernible, oppositely swimming, statistical populations. Consequently, while one might expect stronger imposed flow (larger $Pe_f$) to lead to larger magnitudes for the parallel-to-flow component of the swimmer orientation, this latter quantity increases only up to the threshold value of $Pe_f$ and takes the reverse path of {\em decreasing} in magnitude as the population splitting takes place and develops further with increasing $Pe_f$. The inadequacy of the mean swimmer orientation to describe the `typical' behavior of swimmers is nothing but a direct consequence of the bimodal nature of the swimmer orientation, necessitating the knowledge of higher-order moments of the PDF of swimmer orientation. We show that a conveniently defined Binder cumulant (used commonly in the context of equilibrium phase transitions  \cite{binder}) can identify the uni- to bimodal transition of the PDF (or, the population splitting of swimmers) by exhibiting a global minimum of large negative value at the flow P\'eclet number corresponding to the onset of this transition. It is also worthwhile to note that in the limits, where the imposed shear (self-propulsion) effects are completely dominant, the PDF tends to the one found in the non-active (no-shear) case with two  equal peaks (one symmetric peak); hence, assuming these two limiting cases per se, a uni- to bimodal transition, going through a population-splitting threshold for swimmer orientation, could be envisaged. Our analysis places this intuitive picture (which, however and to the best of our knowledge, remained unexplored) in a systematic framework and elucidates its nature.

We map out a phase diagram in terms of the flow and swim P\'eclet numbers and show that the uni- and bimodal regimes are separated by a {\em discontinuous} transition line. Our results also suggest the possibility of a hysteresis-like behavior for the most probable orientation of swimmers with the  flow P\'eclet number in a channel, where the direction of the  flow can be reversed. It is important to note, however, that a hysteresis-like behavior would be possible in situations where the time-scale of swimmer response is much longer than that of flow reversal (in equilibrium phase transitions, such a mechanism is provided by the inter-particle couplings that lead to free energy barriers between stable and/or meta-stable states of the system). Addressing this issue thus requires a more sophisticated modeling approach to incorporate several key factors such as (time-dependent) swimmer response to variations in the external flow, interactions between swimmers and interactions between swimmers and the walls, which have been neglected in our current analysis. Hence, interpretation of our results in terms of a hysteresis-like effect and making any analogies with equilibrium phase transitions should be done with caution. 

Our analysis is based on a few simplifying assumptions such as  neglect of hydrodynamic interactions, either between the active particles themselves, or between the particles and the channel walls. 
In dense swimmer suspensions, such inter-particle interactions, including also steric interactions, would have to be included (see, e.g., Refs. \cite{wioland2013confinement,lushi2014fluid,pof2008,prl2008,comptes_rendus,ezhilan,ardekani,Hernandez-Ortiz1,Hernandez-Ortiz2,Hernandez-Ortiz3,zottl_Poiseuille}). Wall hydrodynamic interactions have been suggested to have a prominent role in the well-established migration of active particles toward walls (see, e.g., Refs. \cite{wallattraction,wallattraction2,wallattraction3,ardekani,Mathijssen:2016a,Mathijssen:2016b,Mathijssen:2016c,zottl_Poiseuille}), while there are several studies that suggest otherwise (see, e.g., Refs. \cite{elgeti2015run,costanzo,ezhilan,li2011accumulation,li2011accumulation2} and references therein). This work is thus an example among many simulation works that demonstrate wall accumulation of swimmers, while wall hydrodynamic interactions are neglected. 

Another interesting problem is the role of particle shape and its finite aspect ratio. Here, we have modeled the swimmers as thin rods (needles) and, as such, our analysis ignores hydrodynamic effects, such as oscillatory motion of particle orientation in Jeffery orbits \cite{Jeffery1922}, arising as a result of finite particle aspect ratio. This and other factors such as swimmer chirality are among the directions we plan to pursue in our future research.


\begin{acknowledgments}
M.K., J.A. and A.F. acknowledge the School of Physics, IPM, Tehran, for hospitality and support during the completion of this work. A.N. acknowledges Iran Science Elites Federation for partial support. We thank M.R. Shabanniya for useful discussions. 
\end{acknowledgments}

\appendix 

\section{Numerical methods} 
\label{methods}

The steady-state properties of swimmers in the channel follow from the solution of Eq. (\ref{pde_nondim}) that can be evaluated numerically in the rescaled  domain $-1\leq \tilde y \leq 1$ and $0\leq \theta < 2\pi$ 
using the boundary conditions (\ref{bc_nondim}), the constraint (\ref{c_int_cons}) and the fact that  $\tilde{\Psi}(\tilde y,\theta=0)=\tilde{\Psi}(\tilde y,\theta=2\pi)$. 
All numerical simulations in this work have been performed using COMSOL Multiphysics v5.2. In all cases reported here, the relative numerical error was determined by direct validation of the numerical solutions to be less than $1\%$. 
We used structured (square-shaped) mesh elements and cubic discretization in all simulations. A mesh-independency study was performed for the test case of swimmer density on the top wall ($\tilde{y}=1$). We observed the results converging, with an error margin below $1\%$, with quadratic discretization. For our case (cubic), the convergence rate was, expectedly, even better.

\section{Mean near-wall swimmer orientation} 
\label{mean_nearwall_orientation}

Here, we show how the non-monotonic (monotonic) behavior of the mean parallel-to-flow (perpendicular-to-flow) orientation component $\langle \tilde p_x\rangle$ ($\langle \tilde p_y\rangle$) as a function of $Pe_f$ (Fig. \ref{pxy_log}) can be understood on a semi-quantitative basis using the population spitting phenomenon. We focus on the top channel wall at $\tilde y =1$ and fisrt consider the unimodal regime of small to intermediate $Pe_f$ at a fixed  $Pe_s>0$. When there is no imposed shear ($Pe_f=0$), the near-wall PDF $\tilde{\Psi}(\tilde y=1,\theta)$ of swimmer orientation is unimodal and symmetric, being centered at $\theta_\ast\simeq \pi/2$. 
It thus follows that $\langle \tilde p_x\rangle= \langle \cos \theta \rangle= 0$, while  $\langle \tilde p_y\rangle=\langle \sin \theta \rangle$ is finite and depends  on the exact value of $Pe_s$ (see Figs. \ref{orientn_profiles} and \ref{pxy_log}). As noted in Section \ref{bimodal}, upon increasing $Pe_f$ and before the population splitting sets in, the unimodal distribution for swimmer orientation becomes increasingly more asymmetric and its peak location shifts to smaller angles, i.e., $\theta_\ast$ tends to zero. Hence, $\langle \tilde p_x\rangle$ increases in magnitude, while $\langle \tilde p_y\rangle$ decreases. 

We now turn to the bimodal regime, where swimmer orientation is sharply peaked around the orientation angles $ \theta_\ast = \theta_u$ (for upstream minority) and $\theta_d$ (for downstream majority) with an angular separation of $\theta_u-\theta_d\simeq \pi$ (in fact, $\theta_u\simeq \pi$ and  $\theta_d\simeq 0$). A rough estimate for $\langle \tilde p_x\rangle$ and $\langle \tilde p_y\rangle$ can be obtained by adopting an idealized two-state model, in which the swimmer orientation angle on the top wall can only take one of the two values $\{\theta_u, \theta_d\}$, representing the minority and majority populations, respectively, with no angular dispersion around these peak values. 
Hence, we find
\begin{subequations} 
\begin{align}
 \langle \tilde p_x  \rangle &= \frac{\cos\theta_d + R_\Psi \cos\theta_u}{1+R_\Psi} \simeq \frac{1-R_\Psi}{1+R_\Psi}\cos\theta_d,
\label{px_delta}
\\
 \langle \tilde p_y  \rangle &= \frac{\sin\theta_d + R_\Psi \sin\theta_u}{1+R_\Psi}\simeq \frac{1-R_\Psi}{1+R_\Psi}\sin\theta_d,
\label{py_delta}
\end{align}
\end{subequations}
where $R_\Psi$ is the bimodal ratio at the top wall (Section \ref{subsec:bi_ratio}). Thus, within the mentioned two-state model, $\langle \tilde p_y \rangle$ vanishes as $\theta_d$ tends to zero, which, together with what we found above for the unimodal regime, suggests that $\langle \tilde p_y \rangle$ should monotonically decrease to zero as $Pe_f$ is increased in accord with our numerical results (Fig. \ref{pxy_log}b). The situation is found to be different for $\langle \tilde p_x \rangle$; since $\theta_d$ is close to zero in the bimodal regime, $\cos\theta_d\simeq 1$ and  $\langle \tilde p_x \rangle$ can be further approximated as $ \langle \tilde p_x  \rangle \simeq (1-R_\Psi)/(1+R_\Psi)$, 
which finds its maximum near the onset of population splitting ($R_\Psi=0$) and decreases to zero as the minority population is strengthened ($R_\Psi\rightarrow 1$). 

Our foregoing analysis clearly demonstrates that the increasing (decreasing) regime for $\langle \tilde p_x \rangle$ as a function of $Pe_f$, as shown in Fig. \ref{pxy_log}a, is associated with the unimodal (bimodal) regime, respectively. Therefore, the non-monotonic behavior of $\langle \tilde p_x \rangle$ signifies the underlying population splitting phenomenon. This same mechanism can explain the non-monotonic near-wall behavior of $\langle \tilde p_x \rangle$ found in the case of an imposed Poiseuille flow \cite{ezhilan}, as the latter represents a linear flow profile near the walls. The foregoing argument can be used for swimmers on the bottom wall by replacing minority with majority, upstream with downstream, and $R_\Psi$ with $1/R_\Psi$.

\bibliography{refs_draft_rv}

\begin{thebibliography}{108}%
\makeatletter
\providecommand \@ifxundefined [1]{%
 \@ifx{#1\undefined}
}%
\providecommand \@ifnum [1]{%
 \ifnum #1\expandafter \@firstoftwo
 \else \expandafter \@secondoftwo
 \fi
}%
\providecommand \@ifx [1]{%
 \ifx #1\expandafter \@firstoftwo
 \else \expandafter \@secondoftwo
 \fi
}%
\providecommand \natexlab [1]{#1}%
\providecommand \enquote  [1]{``#1''}%
\providecommand \bibnamefont  [1]{#1}%
\providecommand \bibfnamefont [1]{#1}%
\providecommand \citenamefont [1]{#1}%
\providecommand \href@noop [0]{\@secondoftwo}%
\providecommand \href [0]{\begingroup \@sanitize@url \@href}%
\providecommand \@href[1]{\@@startlink{#1}\@@href}%
\providecommand \@@href[1]{\endgroup#1\@@endlink}%
\providecommand \@sanitize@url [0]{\catcode `\\12\catcode `\$12\catcode
  `\&12\catcode `\#12\catcode `\^12\catcode `\_12\catcode `\%12\relax}%
\providecommand \@@startlink[1]{}%
\providecommand \@@endlink[0]{}%
\providecommand \url  [0]{\begingroup\@sanitize@url \@url }%
\providecommand \@url [1]{\endgroup\@href {#1}{\urlprefix }}%
\providecommand \urlprefix  [0]{URL }%
\providecommand \Eprint [0]{\href }%
\providecommand \doibase [0]{http://dx.doi.org/}%
\providecommand \selectlanguage [0]{\@gobble}%
\providecommand \bibinfo  [0]{\@secondoftwo}%
\providecommand \bibfield  [0]{\@secondoftwo}%
\providecommand \translation [1]{[#1]}%
\providecommand \BibitemOpen [0]{}%
\providecommand \bibitemStop [0]{}%
\providecommand \bibitemNoStop [0]{.\EOS\space}%
\providecommand \EOS [0]{\spacefactor3000\relax}%
\providecommand \BibitemShut  [1]{\csname bibitem#1\endcsname}%
\let\auto@bib@innerbib\@empty
\bibitem [{\citenamefont {Ramaswamy}(2010)}]{ramaswamyreview}%
  \BibitemOpen
  \bibfield  {author} {\bibinfo {author} {\bibfnamefont {S.}~\bibnamefont
  {Ramaswamy}},\ }\href@noop {} {\bibfield  {journal} {\bibinfo  {journal}
  {Annu. Rev. Condens. Matter Phys.}\ }\textbf {\bibinfo {volume} {1}},\
  \bibinfo {pages} {323} (\bibinfo {year} {2010})}\BibitemShut {NoStop}%
\bibitem [{\citenamefont {Marchetti}\ \emph {et~al.}(2013)\citenamefont
  {Marchetti}, \citenamefont {Joanny}, \citenamefont {Ramaswamy}, \citenamefont
  {Liverpool}, \citenamefont {Prost}, \citenamefont {Rao},\ and\ \citenamefont
  {Simha}}]{marchetti_review}%
  \BibitemOpen
  \bibfield  {author} {\bibinfo {author} {\bibfnamefont {M.~C.}\ \bibnamefont
  {Marchetti}}, \bibinfo {author} {\bibfnamefont {J.~F.}\ \bibnamefont
  {Joanny}}, \bibinfo {author} {\bibfnamefont {S.}~\bibnamefont {Ramaswamy}},
  \bibinfo {author} {\bibfnamefont {T.~B.}\ \bibnamefont {Liverpool}}, \bibinfo
  {author} {\bibfnamefont {J.}~\bibnamefont {Prost}}, \bibinfo {author}
  {\bibfnamefont {M.}~\bibnamefont {Rao}}, \ and\ \bibinfo {author}
  {\bibfnamefont {R.~A.}\ \bibnamefont {Simha}},\ }\href
  {http://link.aps.org/doi/10.1103/RevModPhys.85.1143} {\bibfield  {journal}
  {\bibinfo  {journal} {Rev. Mod. Phys.}\ }\textbf {\bibinfo {volume} {85}},\
  \bibinfo {pages} {1143} (\bibinfo {year} {2013})}\BibitemShut {NoStop}%
\bibitem [{\citenamefont {Elgeti}\ \emph {et~al.}(2015)\citenamefont {Elgeti},
  \citenamefont {Winkler},\ and\ \citenamefont {Gompper}}]{gompper_review}%
  \BibitemOpen
  \bibfield  {author} {\bibinfo {author} {\bibfnamefont {J.}~\bibnamefont
  {Elgeti}}, \bibinfo {author} {\bibfnamefont {R.~G.}\ \bibnamefont {Winkler}},
  \ and\ \bibinfo {author} {\bibfnamefont {G.}~\bibnamefont {Gompper}},\ }\href
  {http://stacks.iop.org/0034-4885/78/i=5/a=056601} {\bibfield  {journal}
  {\bibinfo  {journal} {Rep. Prog. Phys.}\ }\textbf {\bibinfo {volume} {78}},\
  \bibinfo {pages} {056601} (\bibinfo {year} {2015})}\BibitemShut {NoStop}%
\bibitem [{\citenamefont {Goldstein}(2015)}]{goldstein_review}%
  \BibitemOpen
  \bibfield  {author} {\bibinfo {author} {\bibfnamefont {R.~E.}\ \bibnamefont
  {Goldstein}},\ }\href {https://doi.org/10.1146/annurev-fluid-010313-141426}
  {\bibfield  {journal} {\bibinfo  {journal} {Annu. Rev. Fluid Mech.}\ }\textbf
  {\bibinfo {volume} {47}},\ \bibinfo {pages} {343} (\bibinfo {year}
  {2015})}\BibitemShut {NoStop}%
\bibitem [{\citenamefont {Son}\ \emph {et~al.}(2015)\citenamefont {Son},
  \citenamefont {Brumley},\ and\ \citenamefont {Stocker}}]{son_review}%
  \BibitemOpen
  \bibfield  {author} {\bibinfo {author} {\bibfnamefont {K.}~\bibnamefont
  {Son}}, \bibinfo {author} {\bibfnamefont {D.~R.}\ \bibnamefont {Brumley}}, \
  and\ \bibinfo {author} {\bibfnamefont {R.}~\bibnamefont {Stocker}},\ }\href
  {https://doi.org/10.1038/nrmicro3567} {\bibfield  {journal} {\bibinfo
  {journal} {Nat. Rev. Microbiol.}\ }\textbf {\bibinfo {volume} {13}},\
  \bibinfo {pages} {761} (\bibinfo {year} {2015})}\BibitemShut {NoStop}%
\bibitem [{\citenamefont {Cates}\ and\ \citenamefont
  {Tailleur}(2015)}]{cates_review}%
  \BibitemOpen
  \bibfield  {author} {\bibinfo {author} {\bibfnamefont {M.~E.}\ \bibnamefont
  {Cates}}\ and\ \bibinfo {author} {\bibfnamefont {J.}~\bibnamefont
  {Tailleur}},\ }\href
  {https://doi.org/10.1146/annurev-conmatphys-031214-014710} {\bibfield
  {journal} {\bibinfo  {journal} {Annu. Rev. Condens. Matter Phys.}\ }\textbf
  {\bibinfo {volume} {6}},\ \bibinfo {pages} {219} (\bibinfo {year}
  {2015})}\BibitemShut {NoStop}%
\bibitem [{\citenamefont {Rusconi}\ and\ \citenamefont
  {Stocker}(2015)}]{rusconi-stocker-review}%
  \BibitemOpen
  \bibfield  {author} {\bibinfo {author} {\bibfnamefont {R.}~\bibnamefont
  {Rusconi}}\ and\ \bibinfo {author} {\bibfnamefont {R.}~\bibnamefont
  {Stocker}},\ }\href {https://doi.org/10.1016/j.mib.2015.03.003} {\bibfield
  {journal} {\bibinfo  {journal} {Curr. Opin. Microbiol.}\ }\textbf {\bibinfo
  {volume} {25}},\ \bibinfo {pages} {1} (\bibinfo {year} {2015})}\BibitemShut
  {NoStop}%
\bibitem [{\citenamefont {Golestanian}\ \emph {et~al.}(2011)\citenamefont
  {Golestanian}, \citenamefont {Yeomans},\ and\ \citenamefont
  {Uchida}}]{golestanian_review}%
  \BibitemOpen
  \bibfield  {author} {\bibinfo {author} {\bibfnamefont {R.}~\bibnamefont
  {Golestanian}}, \bibinfo {author} {\bibfnamefont {J.~M.}\ \bibnamefont
  {Yeomans}}, \ and\ \bibinfo {author} {\bibfnamefont {N.}~\bibnamefont
  {Uchida}},\ }\href {https://doi.org/10.1039/c0sm01121e} {\bibfield  {journal}
  {\bibinfo  {journal} {Soft Matter}\ }\textbf {\bibinfo {volume} {7}},\
  \bibinfo {pages} {3074} (\bibinfo {year} {2011})}\BibitemShut {NoStop}%
\bibitem [{\citenamefont {Paxton}\ \emph {et~al.}(2006)\citenamefont {Paxton},
  \citenamefont {Sundararajan}, \citenamefont {Mallouk},\ and\ \citenamefont
  {Sen}}]{paxton-review}%
  \BibitemOpen
  \bibfield  {author} {\bibinfo {author} {\bibfnamefont {W.~F.}\ \bibnamefont
  {Paxton}}, \bibinfo {author} {\bibfnamefont {S.}~\bibnamefont
  {Sundararajan}}, \bibinfo {author} {\bibfnamefont {T.~E.}\ \bibnamefont
  {Mallouk}}, \ and\ \bibinfo {author} {\bibfnamefont {A.}~\bibnamefont
  {Sen}},\ }\href {https://doi.org/10.1002/anie.200600060} {\bibfield
  {journal} {\bibinfo  {journal} {Angew. Chem.}\ }\textbf {\bibinfo {volume}
  {45}},\ \bibinfo {pages} {5420} (\bibinfo {year} {2006})}\BibitemShut
  {NoStop}%
\bibitem [{\citenamefont {Romanczuk}\ \emph {et~al.}(2012)\citenamefont
  {Romanczuk}, \citenamefont {B\"ar}, \citenamefont {Ebeling}, \citenamefont
  {Lindner},\ and\ \citenamefont {Schimansky-Geier}}]{Romanczuk:EPJ2012}%
  \BibitemOpen
  \bibfield  {author} {\bibinfo {author} {\bibfnamefont {P.}~\bibnamefont
  {Romanczuk}}, \bibinfo {author} {\bibfnamefont {M.}~\bibnamefont {B\"ar}},
  \bibinfo {author} {\bibfnamefont {W.}~\bibnamefont {Ebeling}}, \bibinfo
  {author} {\bibfnamefont {B.}~\bibnamefont {Lindner}}, \ and\ \bibinfo
  {author} {\bibfnamefont {L.}~\bibnamefont {Schimansky-Geier}},\ }\href
  {https://doi.org/10.1140/epjst/e2012-01529-y} {\bibfield  {journal} {\bibinfo
   {journal} {Eur. Phys. J. Special Topics}\ }\textbf {\bibinfo {volume}
  {202}},\ \bibinfo {pages} {1} (\bibinfo {year} {2012})}\BibitemShut {NoStop}%
\bibitem [{\citenamefont {Bechinger}\ \emph {et~al.}(2016)\citenamefont
  {Bechinger}, \citenamefont {Di~Leonardo}, \citenamefont {L\"owen},
  \citenamefont {Reichhardt}, \citenamefont {Volpe},\ and\ \citenamefont
  {Volpe}}]{bechinger_review}%
  \BibitemOpen
  \bibfield  {author} {\bibinfo {author} {\bibfnamefont {C.}~\bibnamefont
  {Bechinger}}, \bibinfo {author} {\bibfnamefont {R.}~\bibnamefont
  {Di~Leonardo}}, \bibinfo {author} {\bibfnamefont {H.}~\bibnamefont
  {L\"owen}}, \bibinfo {author} {\bibfnamefont {C.}~\bibnamefont {Reichhardt}},
  \bibinfo {author} {\bibfnamefont {G.}~\bibnamefont {Volpe}}, \ and\ \bibinfo
  {author} {\bibfnamefont {G.}~\bibnamefont {Volpe}},\ }\href
  {https://doi.org/10.1103/RevModPhys.88.045006} {\bibfield  {journal}
  {\bibinfo  {journal} {Rev. Mod. Phys.}\ }\textbf {\bibinfo {volume} {88}},\
  \bibinfo {pages} {045006} (\bibinfo {year} {2016})}\BibitemShut {NoStop}%
\bibitem [{\citenamefont {Koumakis}\ \emph {et~al.}(2013)\citenamefont
  {Koumakis}, \citenamefont {Lepore}, \citenamefont {Maggi},\ and\
  \citenamefont {Di~Leonardo}}]{cargo}%
  \BibitemOpen
  \bibfield  {author} {\bibinfo {author} {\bibfnamefont {N.}~\bibnamefont
  {Koumakis}}, \bibinfo {author} {\bibfnamefont {A.}~\bibnamefont {Lepore}},
  \bibinfo {author} {\bibfnamefont {C.}~\bibnamefont {Maggi}}, \ and\ \bibinfo
  {author} {\bibfnamefont {R.}~\bibnamefont {Di~Leonardo}},\ }\href
  {https://doi.org/10.1038/ncomms3588} {\bibfield  {journal} {\bibinfo
  {journal} {Nat. Commun.}\ }\textbf {\bibinfo {volume} {4}},\ \bibinfo {pages}
  {2588} (\bibinfo {year} {2013})}\BibitemShut {NoStop}%
\bibitem [{\citenamefont {Dreyfus}\ \emph {et~al.}(2005)\citenamefont
  {Dreyfus}, \citenamefont {Baudry}, \citenamefont {Roper}, \citenamefont
  {Fermigier}, \citenamefont {Stone},\ and\ \citenamefont {Bibette}}]{dreyfus}%
  \BibitemOpen
  \bibfield  {author} {\bibinfo {author} {\bibfnamefont {R.}~\bibnamefont
  {Dreyfus}}, \bibinfo {author} {\bibfnamefont {J.}~\bibnamefont {Baudry}},
  \bibinfo {author} {\bibfnamefont {M.~L.}\ \bibnamefont {Roper}}, \bibinfo
  {author} {\bibfnamefont {M.}~\bibnamefont {Fermigier}}, \bibinfo {author}
  {\bibfnamefont {H.~A.}\ \bibnamefont {Stone}}, \ and\ \bibinfo {author}
  {\bibfnamefont {J.}~\bibnamefont {Bibette}},\ }\href
  {https://doi.org/10.1038/nature04090} {\bibfield  {journal} {\bibinfo
  {journal} {Nature}\ }\textbf {\bibinfo {volume} {437}},\ \bibinfo {pages}
  {862} (\bibinfo {year} {2005})}\BibitemShut {NoStop}%
\bibitem [{\citenamefont {Paxton}\ \emph {et~al.}(2004)\citenamefont {Paxton},
  \citenamefont {Kistler}, \citenamefont {Olmeda}, \citenamefont {Sen},
  \citenamefont {St.~Angelo}, \citenamefont {Cao}, \citenamefont {Mallouk},
  \citenamefont {Lammert},\ and\ \citenamefont {Crespi}}]{paxton}%
  \BibitemOpen
  \bibfield  {author} {\bibinfo {author} {\bibfnamefont {W.~F.}\ \bibnamefont
  {Paxton}}, \bibinfo {author} {\bibfnamefont {K.~C.}\ \bibnamefont {Kistler}},
  \bibinfo {author} {\bibfnamefont {C.~C.}\ \bibnamefont {Olmeda}}, \bibinfo
  {author} {\bibfnamefont {A.}~\bibnamefont {Sen}}, \bibinfo {author}
  {\bibfnamefont {S.~K.}\ \bibnamefont {St.~Angelo}}, \bibinfo {author}
  {\bibfnamefont {Y.}~\bibnamefont {Cao}}, \bibinfo {author} {\bibfnamefont
  {T.~E.}\ \bibnamefont {Mallouk}}, \bibinfo {author} {\bibfnamefont {P.~E.}\
  \bibnamefont {Lammert}}, \ and\ \bibinfo {author} {\bibfnamefont {V.~H.}\
  \bibnamefont {Crespi}},\ }\href {https://doi.org/10.1021/ja047697z}
  {\bibfield  {journal} {\bibinfo  {journal} {J. Am. Chem. Soc.}\ }\textbf
  {\bibinfo {volume} {126}},\ \bibinfo {pages} {13424} (\bibinfo {year}
  {2004})}\BibitemShut {NoStop}%
\bibitem [{\citenamefont {Medina-S\'anchez}\ \emph {et~al.}(2016)\citenamefont
  {Medina-S\'anchez}, \citenamefont {Schwarz}, \citenamefont {Meyer},
  \citenamefont {Hebenstreit},\ and\ \citenamefont {Schmidt}}]{sperm-carrying}%
  \BibitemOpen
  \bibfield  {author} {\bibinfo {author} {\bibfnamefont {M.}~\bibnamefont
  {Medina-S\'anchez}}, \bibinfo {author} {\bibfnamefont {L.}~\bibnamefont
  {Schwarz}}, \bibinfo {author} {\bibfnamefont {A.~K.}\ \bibnamefont {Meyer}},
  \bibinfo {author} {\bibfnamefont {F.}~\bibnamefont {Hebenstreit}}, \ and\
  \bibinfo {author} {\bibfnamefont {O.~G.}\ \bibnamefont {Schmidt}},\ }\href
  {https://doi.org/10.1021/acs.nanolett.5b04221} {\bibfield  {journal}
  {\bibinfo  {journal} {Nano Lett.}\ }\textbf {\bibinfo {volume} {16}},\
  \bibinfo {pages} {555} (\bibinfo {year} {2016})}\BibitemShut {NoStop}%
\bibitem [{\citenamefont {Cheang}\ \emph {et~al.}(2014)\citenamefont {Cheang},
  \citenamefont {Lee}, \citenamefont {Julius},\ and\ \citenamefont
  {Kim}}]{robotic}%
  \BibitemOpen
  \bibfield  {author} {\bibinfo {author} {\bibfnamefont {U.~K.}\ \bibnamefont
  {Cheang}}, \bibinfo {author} {\bibfnamefont {K.}~\bibnamefont {Lee}},
  \bibinfo {author} {\bibfnamefont {A.~A.}\ \bibnamefont {Julius}}, \ and\
  \bibinfo {author} {\bibfnamefont {M.~J.}\ \bibnamefont {Kim}},\ }\href
  {https://doi.org/10.1063/1.4893695} {\bibfield  {journal} {\bibinfo
  {journal} {Appl. Phys. Lett.}\ }\textbf {\bibinfo {volume} {105}},\ \bibinfo
  {pages} {083705} (\bibinfo {year} {2014})}\BibitemShut {NoStop}%
\bibitem [{\citenamefont {Khalil}\ \emph {et~al.}(2014)\citenamefont {Khalil},
  \citenamefont {Magdanz}, \citenamefont {Sanchez}, \citenamefont {Schmidt},\
  and\ \citenamefont {Misra}}]{micro-bio-robot}%
  \BibitemOpen
  \bibfield  {author} {\bibinfo {author} {\bibfnamefont {I.~S.~M.}\
  \bibnamefont {Khalil}}, \bibinfo {author} {\bibfnamefont {V.}~\bibnamefont
  {Magdanz}}, \bibinfo {author} {\bibfnamefont {S.}~\bibnamefont {Sanchez}},
  \bibinfo {author} {\bibfnamefont {O.~G.}\ \bibnamefont {Schmidt}}, \ and\
  \bibinfo {author} {\bibfnamefont {S.}~\bibnamefont {Misra}},\ }\href
  {https://doi.org/10.1007/s12213-014-0077-9} {\bibfield  {journal} {\bibinfo
  {journal} {J. Micro-Bio Robot.}\ }\textbf {\bibinfo {volume} {9}},\ \bibinfo
  {pages} {79} (\bibinfo {year} {2014})}\BibitemShut {NoStop}%
\bibitem [{\citenamefont {Vach}\ \emph {et~al.}(2013)\citenamefont {Vach},
  \citenamefont {Brun}, \citenamefont {Bennet}, \citenamefont {Bertinetti},
  \citenamefont {Widdrat}, \citenamefont {Baumgartner}, \citenamefont {Klumpp},
  \citenamefont {Fratzl},\ and\ \citenamefont {Faivre}}]{klumpp1}%
  \BibitemOpen
  \bibfield  {author} {\bibinfo {author} {\bibfnamefont {P.~J.}\ \bibnamefont
  {Vach}}, \bibinfo {author} {\bibfnamefont {N.}~\bibnamefont {Brun}}, \bibinfo
  {author} {\bibfnamefont {M.}~\bibnamefont {Bennet}}, \bibinfo {author}
  {\bibfnamefont {L.}~\bibnamefont {Bertinetti}}, \bibinfo {author}
  {\bibfnamefont {M.}~\bibnamefont {Widdrat}}, \bibinfo {author} {\bibfnamefont
  {J.}~\bibnamefont {Baumgartner}}, \bibinfo {author} {\bibfnamefont
  {S.}~\bibnamefont {Klumpp}}, \bibinfo {author} {\bibfnamefont
  {P.}~\bibnamefont {Fratzl}}, \ and\ \bibinfo {author} {\bibfnamefont
  {D.}~\bibnamefont {Faivre}},\ }\href {https://doi.org/10.1021/nl402897x}
  {\bibfield  {journal} {\bibinfo  {journal} {Nano Lett.}\ }\textbf {\bibinfo
  {volume} {13}},\ \bibinfo {pages} {5373} (\bibinfo {year}
  {2013})}\BibitemShut {NoStop}%
\bibitem [{\citenamefont {Vach}\ \emph {et~al.}(2015)\citenamefont {Vach},
  \citenamefont {Fratzl}, \citenamefont {Klumpp}, \citenamefont {Fratzl},\ and\
  \citenamefont {Faivre}}]{klumpp2}%
  \BibitemOpen
  \bibfield  {author} {\bibinfo {author} {\bibfnamefont {P.~J.}\ \bibnamefont
  {Vach}}, \bibinfo {author} {\bibfnamefont {P.}~\bibnamefont {Fratzl}},
  \bibinfo {author} {\bibfnamefont {S.}~\bibnamefont {Klumpp}}, \bibinfo
  {author} {\bibfnamefont {P.}~\bibnamefont {Fratzl}}, \ and\ \bibinfo {author}
  {\bibfnamefont {D.}~\bibnamefont {Faivre}},\ }\href
  {https://doi.org/10.1021/acs.nanolett.5b03131} {\bibfield  {journal}
  {\bibinfo  {journal} {Nano Lett.}\ }\textbf {\bibinfo {volume} {15}},\
  \bibinfo {pages} {7064} (\bibinfo {year} {2015})}\BibitemShut {NoStop}%
\bibitem [{\citenamefont {Purcell}(1977)}]{purcellmain}%
  \BibitemOpen
  \bibfield  {author} {\bibinfo {author} {\bibfnamefont {E.~M.}\ \bibnamefont
  {Purcell}},\ }\href {https://doi.org/10.1119/1.10903} {\bibfield  {journal}
  {\bibinfo  {journal} {Am. J. Phys.}\ }\textbf {\bibinfo {volume} {45}},\
  \bibinfo {pages} {3} (\bibinfo {year} {1977})}\BibitemShut {NoStop}%
\bibitem [{\citenamefont {Najafi}\ and\ \citenamefont
  {Golestanian}(2004)}]{najafiramin}%
  \BibitemOpen
  \bibfield  {author} {\bibinfo {author} {\bibfnamefont {A.}~\bibnamefont
  {Najafi}}\ and\ \bibinfo {author} {\bibfnamefont {R.}~\bibnamefont
  {Golestanian}},\ }\href {https://doi.org/10.1103/PhysRevE.69.062901}
  {\bibfield  {journal} {\bibinfo  {journal} {Phys. Rev. E}\ }\textbf {\bibinfo
  {volume} {69}},\ \bibinfo {pages} {062901} (\bibinfo {year}
  {2004})}\BibitemShut {NoStop}%
\bibitem [{\citenamefont {Lauga}(2011)}]{laugascallop1}%
  \BibitemOpen
  \bibfield  {author} {\bibinfo {author} {\bibfnamefont {E.}~\bibnamefont
  {Lauga}},\ }\href {https://doi.org/10.1039/C0SM00953A} {\bibfield  {journal}
  {\bibinfo  {journal} {Soft Matter}\ }\textbf {\bibinfo {volume} {7}},\
  \bibinfo {pages} {3060} (\bibinfo {year} {2011})}\BibitemShut {NoStop}%
\bibitem [{\citenamefont {Z\"ottl}\ and\ \citenamefont
  {Stark}(2014)}]{zottlcollective}%
  \BibitemOpen
  \bibfield  {author} {\bibinfo {author} {\bibfnamefont {A.}~\bibnamefont
  {Z\"ottl}}\ and\ \bibinfo {author} {\bibfnamefont {H.}~\bibnamefont
  {Stark}},\ }\href {https://doi.org/10.1103/PhysRevLett.112.118101} {\bibfield
   {journal} {\bibinfo  {journal} {Phys. Rev. Lett.}\ }\textbf {\bibinfo
  {volume} {112}},\ \bibinfo {pages} {118101} (\bibinfo {year}
  {2014})}\BibitemShut {NoStop}%
\bibitem [{\citenamefont {Sokolov}\ and\ \citenamefont
  {Aranson}(2012)}]{sokolovcollective}%
  \BibitemOpen
  \bibfield  {author} {\bibinfo {author} {\bibfnamefont {A.}~\bibnamefont
  {Sokolov}}\ and\ \bibinfo {author} {\bibfnamefont {I.~S.}\ \bibnamefont
  {Aranson}},\ }\href {https://doi.org/10.1103/PhysRevLett.109.248109}
  {\bibfield  {journal} {\bibinfo  {journal} {Phys. Rev. Lett.}\ }\textbf
  {\bibinfo {volume} {109}},\ \bibinfo {pages} {248109} (\bibinfo {year}
  {2012})}\BibitemShut {NoStop}%
\bibitem [{\citenamefont {Golestanian}\ \emph {et~al.}(2007)\citenamefont
  {Golestanian}, \citenamefont {Liverpool},\ and\ \citenamefont
  {Ajdari}}]{golestanian_njp}%
  \BibitemOpen
  \bibfield  {author} {\bibinfo {author} {\bibfnamefont {R.}~\bibnamefont
  {Golestanian}}, \bibinfo {author} {\bibfnamefont {T.~B.}\ \bibnamefont
  {Liverpool}}, \ and\ \bibinfo {author} {\bibfnamefont {A.}~\bibnamefont
  {Ajdari}},\ }\href {http://stacks.iop.org/1367-2630/9/i=5/a=126} {\bibfield
  {journal} {\bibinfo  {journal} {New J. Phys.}\ }\textbf {\bibinfo {volume}
  {9}},\ \bibinfo {pages} {126} (\bibinfo {year} {2007})}\BibitemShut {NoStop}%
\bibitem [{\citenamefont {Becker}\ \emph {et~al.}(2003)\citenamefont {Becker},
  \citenamefont {Koehler},\ and\ \citenamefont {Stone}}]{becker}%
  \BibitemOpen
  \bibfield  {author} {\bibinfo {author} {\bibfnamefont {L.~E.}\ \bibnamefont
  {Becker}}, \bibinfo {author} {\bibfnamefont {S.~A.}\ \bibnamefont {Koehler}},
  \ and\ \bibinfo {author} {\bibfnamefont {H.~A.}\ \bibnamefont {Stone}},\
  }\href@noop {} {\bibfield  {journal} {\bibinfo  {journal} {J. Fluid Mech.}\
  }\textbf {\bibinfo {volume} {490}},\ \bibinfo {pages} {15} (\bibinfo {year}
  {2003})}\BibitemShut {NoStop}%
\bibitem [{\citenamefont {Najafi}\ and\ \citenamefont
  {Golestanian}(2005)}]{najafiramin05}%
  \BibitemOpen
  \bibfield  {author} {\bibinfo {author} {\bibfnamefont {A.}~\bibnamefont
  {Najafi}}\ and\ \bibinfo {author} {\bibfnamefont {R.}~\bibnamefont
  {Golestanian}},\ }\href {http://stacks.iop.org/0953-8984/17/i=14/a=009}
  {\bibfield  {journal} {\bibinfo  {journal} {J. Phys.: Condens. Matter}\
  }\textbf {\bibinfo {volume} {17}},\ \bibinfo {pages} {S1203} (\bibinfo {year}
  {2005})}\BibitemShut {NoStop}%
\bibitem [{\citenamefont {Berg}\ and\ \citenamefont
  {Anderson}(1973)}]{berg1973bacteria}%
  \BibitemOpen
  \bibfield  {author} {\bibinfo {author} {\bibfnamefont {H.~C.}\ \bibnamefont
  {Berg}}\ and\ \bibinfo {author} {\bibfnamefont {R.~A.}\ \bibnamefont
  {Anderson}},\ }\href {\doibase 10.1038/245380a0} {\bibfield  {journal}
  {\bibinfo  {journal} {Nature}\ }\textbf {\bibinfo {volume} {245}},\ \bibinfo
  {pages} {380} (\bibinfo {year} {1973})}\BibitemShut {NoStop}%
\bibitem [{\citenamefont {Berg}(2003)}]{bergbook}%
  \BibitemOpen
  \bibfield  {author} {\bibinfo {author} {\bibfnamefont {H.~C.}\ \bibnamefont
  {Berg}},\ }\href@noop {} {\emph {\bibinfo {title} {E. coli in Motion}}}\
  (\bibinfo  {publisher} {Springer, New York},\ \bibinfo {year}
  {2003})\BibitemShut {NoStop}%
\bibitem [{\citenamefont {Budrene}\ and\ \citenamefont {Berg}(1991)}]{budrene}%
  \BibitemOpen
  \bibfield  {author} {\bibinfo {author} {\bibfnamefont {E.~O.}\ \bibnamefont
  {Budrene}}\ and\ \bibinfo {author} {\bibfnamefont {H.~C.}\ \bibnamefont
  {Berg}},\ }\href {https://doi.org/10.1038/349630a0} {\bibfield  {journal}
  {\bibinfo  {journal} {Nature}\ }\textbf {\bibinfo {volume} {349}},\ \bibinfo
  {pages} {630} (\bibinfo {year} {1991})}\BibitemShut {NoStop}%
\bibitem [{\citenamefont {Guasto}\ \emph {et~al.}(2012)\citenamefont {Guasto},
  \citenamefont {Rusconi},\ and\ \citenamefont {Stocker}}]{planktonic}%
  \BibitemOpen
  \bibfield  {author} {\bibinfo {author} {\bibfnamefont {J.~S.}\ \bibnamefont
  {Guasto}}, \bibinfo {author} {\bibfnamefont {R.}~\bibnamefont {Rusconi}}, \
  and\ \bibinfo {author} {\bibfnamefont {R.}~\bibnamefont {Stocker}},\ }\href
  {https://doi.org/10.1146/annurev-fluid-120710-101156} {\bibfield  {journal}
  {\bibinfo  {journal} {Annu. Rev. Fluid Mech.}\ }\textbf {\bibinfo {volume}
  {44}},\ \bibinfo {pages} {373} (\bibinfo {year} {2012})}\BibitemShut
  {NoStop}%
\bibitem [{\citenamefont {Golestanian}\ \emph {et~al.}(2005)\citenamefont
  {Golestanian}, \citenamefont {Liverpool},\ and\ \citenamefont
  {Ajdari}}]{ramin_molmach}%
  \BibitemOpen
  \bibfield  {author} {\bibinfo {author} {\bibfnamefont {R.}~\bibnamefont
  {Golestanian}}, \bibinfo {author} {\bibfnamefont {T.~B.}\ \bibnamefont
  {Liverpool}}, \ and\ \bibinfo {author} {\bibfnamefont {A.}~\bibnamefont
  {Ajdari}},\ }\href@noop {} {\bibfield  {journal} {\bibinfo  {journal} {Phys.
  Rev. Lett.}\ }\textbf {\bibinfo {volume} {94}},\ \bibinfo {pages} {220801}
  (\bibinfo {year} {2005})}\BibitemShut {NoStop}%
\bibitem [{\citenamefont {Blair}(1995)}]{blair1995bacteria}%
  \BibitemOpen
  \bibfield  {author} {\bibinfo {author} {\bibfnamefont {D.~F.}\ \bibnamefont
  {Blair}},\ }\href@noop {} {\bibfield  {journal} {\bibinfo  {journal} {Annu.
  Rev. Microbiol.}\ }\textbf {\bibinfo {volume} {49}},\ \bibinfo {pages} {489}
  (\bibinfo {year} {1995})}\BibitemShut {NoStop}%
\bibitem [{\citenamefont {Gray}(1955)}]{graysperm}%
  \BibitemOpen
  \bibfield  {author} {\bibinfo {author} {\bibfnamefont {J.}~\bibnamefont
  {Gray}},\ }\href@noop {} {\bibfield  {journal} {\bibinfo  {journal} {J. Exp.
  Biol.}\ }\textbf {\bibinfo {volume} {32}},\ \bibinfo {pages} {775} (\bibinfo
  {year} {1955})}\BibitemShut {NoStop}%
\bibitem [{\citenamefont {Shack}\ \emph {et~al.}(1974)\citenamefont {Shack},
  \citenamefont {Fray},\ and\ \citenamefont {Lardner}}]{shacksperm}%
  \BibitemOpen
  \bibfield  {author} {\bibinfo {author} {\bibfnamefont {W.~J.}\ \bibnamefont
  {Shack}}, \bibinfo {author} {\bibfnamefont {C.~S.}\ \bibnamefont {Fray}}, \
  and\ \bibinfo {author} {\bibfnamefont {T.~J.}\ \bibnamefont {Lardner}},\
  }\href {https://doi.org/10.1007/BF02463267} {\bibfield  {journal} {\bibinfo
  {journal} {Bull. Math. Biol}\ }\textbf {\bibinfo {volume} {36}},\ \bibinfo
  {pages} {555} (\bibinfo {year} {1974})}\BibitemShut {NoStop}%
\bibitem [{\citenamefont {Woolley}(2003)}]{woolleysperm}%
  \BibitemOpen
  \bibfield  {author} {\bibinfo {author} {\bibfnamefont {D.~M.}\ \bibnamefont
  {Woolley}},\ }\href {https://doi.org/10.1530/rep.0.1260259} {\bibfield
  {journal} {\bibinfo  {journal} {Reproduction}\ }\textbf {\bibinfo {volume}
  {126}},\ \bibinfo {pages} {259} (\bibinfo {year} {2003})}\BibitemShut
  {NoStop}%
\bibitem [{\citenamefont {Friedrich}\ \emph {et~al.}(2010)\citenamefont
  {Friedrich}, \citenamefont {Riedel-Kruse}, \citenamefont {Howard},\ and\
  \citenamefont {J{\"u}licher}}]{julichersperm}%
  \BibitemOpen
  \bibfield  {author} {\bibinfo {author} {\bibfnamefont {B.~M.}\ \bibnamefont
  {Friedrich}}, \bibinfo {author} {\bibfnamefont {I.~H.}\ \bibnamefont
  {Riedel-Kruse}}, \bibinfo {author} {\bibfnamefont {J.}~\bibnamefont
  {Howard}}, \ and\ \bibinfo {author} {\bibfnamefont {F.}~\bibnamefont
  {J{\"u}licher}},\ }\href {https://doi.org/10.1242/jeb.039800} {\bibfield
  {journal} {\bibinfo  {journal} {J. Exp. Biol.}\ }\textbf {\bibinfo {volume}
  {213}},\ \bibinfo {pages} {1226} (\bibinfo {year} {2010})}\BibitemShut
  {NoStop}%
\bibitem [{\citenamefont {Friedrich}\ and\ \citenamefont
  {J\"ulicher}(2007)}]{julicherspermchemx}%
  \BibitemOpen
  \bibfield  {author} {\bibinfo {author} {\bibfnamefont {B.~M.}\ \bibnamefont
  {Friedrich}}\ and\ \bibinfo {author} {\bibfnamefont {F.}~\bibnamefont
  {J\"ulicher}},\ }\href {https://doi.org/10.1073/pnas.0703530104} {\bibfield
  {journal} {\bibinfo  {journal} {Proc. Natl. Acad. Sci. U.S.A.}\ }\textbf
  {\bibinfo {volume} {104}},\ \bibinfo {pages} {13256} (\bibinfo {year}
  {2007})}\BibitemShut {NoStop}%
\bibitem [{\citenamefont {Ringo}(1967)}]{ringochlamy}%
  \BibitemOpen
  \bibfield  {author} {\bibinfo {author} {\bibfnamefont {D.~L.}\ \bibnamefont
  {Ringo}},\ }\href {https://doi.org/10.1083/jcb.33.3.543} {\bibfield
  {journal} {\bibinfo  {journal} {J. Cell Biol.}\ }\textbf {\bibinfo {volume}
  {33}},\ \bibinfo {pages} {543} (\bibinfo {year} {1967})}\BibitemShut
  {NoStop}%
\bibitem [{\citenamefont {Polin}\ \emph {et~al.}(2009)\citenamefont {Polin},
  \citenamefont {Tuval}, \citenamefont {Drescher}, \citenamefont {Gollub},\
  and\ \citenamefont {Goldstein}}]{goldstein_science}%
  \BibitemOpen
  \bibfield  {author} {\bibinfo {author} {\bibfnamefont {M.}~\bibnamefont
  {Polin}}, \bibinfo {author} {\bibfnamefont {I.}~\bibnamefont {Tuval}},
  \bibinfo {author} {\bibfnamefont {K.}~\bibnamefont {Drescher}}, \bibinfo
  {author} {\bibfnamefont {J.~P.}\ \bibnamefont {Gollub}}, \ and\ \bibinfo
  {author} {\bibfnamefont {R.~E.}\ \bibnamefont {Goldstein}},\ }\href
  {https://doi.org/10.1126/science.1172667} {\bibfield  {journal} {\bibinfo
  {journal} {Science}\ }\textbf {\bibinfo {volume} {325}},\ \bibinfo {pages}
  {487} (\bibinfo {year} {2009})}\BibitemShut {NoStop}%
\bibitem [{\citenamefont {Drescher}\ \emph {et~al.}(2009)\citenamefont
  {Drescher}, \citenamefont {Leptos}, \citenamefont {Tuval}, \citenamefont
  {Ishikawa}, \citenamefont {Pedley},\ and\ \citenamefont
  {Goldstein}}]{volvox1}%
  \BibitemOpen
  \bibfield  {author} {\bibinfo {author} {\bibfnamefont {K.}~\bibnamefont
  {Drescher}}, \bibinfo {author} {\bibfnamefont {K.~C.}\ \bibnamefont
  {Leptos}}, \bibinfo {author} {\bibfnamefont {I.}~\bibnamefont {Tuval}},
  \bibinfo {author} {\bibfnamefont {T.}~\bibnamefont {Ishikawa}}, \bibinfo
  {author} {\bibfnamefont {T.~J.}\ \bibnamefont {Pedley}}, \ and\ \bibinfo
  {author} {\bibfnamefont {R.~E.}\ \bibnamefont {Goldstein}},\ }\href
  {https://doi.org/10.1103/PhysRevLett.102.168101} {\bibfield  {journal}
  {\bibinfo  {journal} {Phys. Rev. Lett.}\ }\textbf {\bibinfo {volume} {102}},\
  \bibinfo {pages} {168101} (\bibinfo {year} {2009})}\BibitemShut {NoStop}%
\bibitem [{\citenamefont {Brumley}\ \emph {et~al.}(2015)\citenamefont
  {Brumley}, \citenamefont {Polin}, \citenamefont {Pedley},\ and\ \citenamefont
  {Goldstein}}]{volvox2}%
  \BibitemOpen
  \bibfield  {author} {\bibinfo {author} {\bibfnamefont {D.~R.}\ \bibnamefont
  {Brumley}}, \bibinfo {author} {\bibfnamefont {M.}~\bibnamefont {Polin}},
  \bibinfo {author} {\bibfnamefont {T.~J.}\ \bibnamefont {Pedley}}, \ and\
  \bibinfo {author} {\bibfnamefont {R.~E.}\ \bibnamefont {Goldstein}},\ }\href
  {https://doi.org/10.1098/rsif.2014.1358} {\bibfield  {journal} {\bibinfo
  {journal} {J. R. Soc. Interface}\ }\textbf {\bibinfo {volume} {12}},\
  \bibinfo {pages} {20141358} (\bibinfo {year} {2015})}\BibitemShut {NoStop}%
\bibitem [{\citenamefont {Walther}\ and\ \citenamefont
  {Muller}(2013)}]{janusmain}%
  \BibitemOpen
  \bibfield  {author} {\bibinfo {author} {\bibfnamefont {A.}~\bibnamefont
  {Walther}}\ and\ \bibinfo {author} {\bibfnamefont {A.~H.}\ \bibnamefont
  {Muller}},\ }\href@noop {} {\bibfield  {journal} {\bibinfo  {journal} {Chem.
  Rev.}\ }\textbf {\bibinfo {volume} {113}},\ \bibinfo {pages} {5194} (\bibinfo
  {year} {2013})}\BibitemShut {NoStop}%
\bibitem [{\citenamefont {Valadares}\ \emph {et~al.}(2010)\citenamefont
  {Valadares}, \citenamefont {Tao}, \citenamefont {Zacharia}, \citenamefont
  {Kitaev}, \citenamefont {Galembeck}, \citenamefont {Kapral},\ and\
  \citenamefont {Ozin}}]{valadares2010catalytic}%
  \BibitemOpen
  \bibfield  {author} {\bibinfo {author} {\bibfnamefont {L.~F.}\ \bibnamefont
  {Valadares}}, \bibinfo {author} {\bibfnamefont {Y.-G.}\ \bibnamefont {Tao}},
  \bibinfo {author} {\bibfnamefont {N.~S.}\ \bibnamefont {Zacharia}}, \bibinfo
  {author} {\bibfnamefont {V.}~\bibnamefont {Kitaev}}, \bibinfo {author}
  {\bibfnamefont {F.}~\bibnamefont {Galembeck}}, \bibinfo {author}
  {\bibfnamefont {R.}~\bibnamefont {Kapral}}, \ and\ \bibinfo {author}
  {\bibfnamefont {G.~A.}\ \bibnamefont {Ozin}},\ }\href@noop {} {\bibfield
  {journal} {\bibinfo  {journal} {Small}\ }\textbf {\bibinfo {volume} {6}},\
  \bibinfo {pages} {565} (\bibinfo {year} {2010})}\BibitemShut {NoStop}%
\bibitem [{\citenamefont {Jiang}\ \emph {et~al.}(2010)\citenamefont {Jiang},
  \citenamefont {Yoshinaga},\ and\ \citenamefont {Sano}}]{sanojanus}%
  \BibitemOpen
  \bibfield  {author} {\bibinfo {author} {\bibfnamefont {H.-R.}\ \bibnamefont
  {Jiang}}, \bibinfo {author} {\bibfnamefont {N.}~\bibnamefont {Yoshinaga}}, \
  and\ \bibinfo {author} {\bibfnamefont {M.}~\bibnamefont {Sano}},\ }\href
  {https://doi.org/10.1103/PhysRevLett.105.268302} {\bibfield  {journal}
  {\bibinfo  {journal} {Phys. Rev. Lett.}\ }\textbf {\bibinfo {volume} {105}},\
  \bibinfo {pages} {268302} (\bibinfo {year} {2010})}\BibitemShut {NoStop}%
\bibitem [{\citenamefont {Howse}\ \emph {et~al.}(2007)\citenamefont {Howse},
  \citenamefont {Jones}, \citenamefont {Ryan}, \citenamefont {Gough},
  \citenamefont {Vafabakhsh},\ and\ \citenamefont {Golestanian}}]{ramin07}%
  \BibitemOpen
  \bibfield  {author} {\bibinfo {author} {\bibfnamefont {J.~R.}\ \bibnamefont
  {Howse}}, \bibinfo {author} {\bibfnamefont {R.~A.~L.}\ \bibnamefont {Jones}},
  \bibinfo {author} {\bibfnamefont {A.~J.}\ \bibnamefont {Ryan}}, \bibinfo
  {author} {\bibfnamefont {T.}~\bibnamefont {Gough}}, \bibinfo {author}
  {\bibfnamefont {R.}~\bibnamefont {Vafabakhsh}}, \ and\ \bibinfo {author}
  {\bibfnamefont {R.}~\bibnamefont {Golestanian}},\ }\href
  {https://doi.org/10.1103/PhysRevLett.99.048102} {\bibfield  {journal}
  {\bibinfo  {journal} {Phys. Rev. Lett.}\ }\textbf {\bibinfo {volume} {99}},\
  \bibinfo {pages} {048102} (\bibinfo {year} {2007})}\BibitemShut {NoStop}%
\bibitem [{\citenamefont {Volpe}\ \emph {et~al.}(2011)\citenamefont {Volpe},
  \citenamefont {Buttinoni}, \citenamefont {Vogt}, \citenamefont {Kummerer},\
  and\ \citenamefont {Bechinger}}]{volpe1}%
  \BibitemOpen
  \bibfield  {author} {\bibinfo {author} {\bibfnamefont {G.}~\bibnamefont
  {Volpe}}, \bibinfo {author} {\bibfnamefont {I.}~\bibnamefont {Buttinoni}},
  \bibinfo {author} {\bibfnamefont {D.}~\bibnamefont {Vogt}}, \bibinfo {author}
  {\bibfnamefont {H.-J.}\ \bibnamefont {Kummerer}}, \ and\ \bibinfo {author}
  {\bibfnamefont {C.}~\bibnamefont {Bechinger}},\ }\href
  {https://doi.org/10.1039/C1SM05960B} {\bibfield  {journal} {\bibinfo
  {journal} {Soft Matter}\ }\textbf {\bibinfo {volume} {7}},\ \bibinfo {pages}
  {8810} (\bibinfo {year} {2011})}\BibitemShut {NoStop}%
\bibitem [{\citenamefont {Takagi}\ \emph {et~al.}(2013)\citenamefont {Takagi},
  \citenamefont {Braunschweig}, \citenamefont {Zhang},\ and\ \citenamefont
  {Shelley}}]{takagi2013}%
  \BibitemOpen
  \bibfield  {author} {\bibinfo {author} {\bibfnamefont {D.}~\bibnamefont
  {Takagi}}, \bibinfo {author} {\bibfnamefont {A.~B.}\ \bibnamefont
  {Braunschweig}}, \bibinfo {author} {\bibfnamefont {J.}~\bibnamefont {Zhang}},
  \ and\ \bibinfo {author} {\bibfnamefont {M.~J.}\ \bibnamefont {Shelley}},\
  }\href {https://doi.org/10.1103/PhysRevLett.110.038301} {\bibfield  {journal}
  {\bibinfo  {journal} {Phys. Rev. Lett.}\ }\textbf {\bibinfo {volume} {110}},\
  \bibinfo {pages} {038301} (\bibinfo {year} {2013})}\BibitemShut {NoStop}%
\bibitem [{\citenamefont {Saintillan}\ and\ \citenamefont
  {Shelley}(2007)}]{Saintillan2007}%
  \BibitemOpen
  \bibfield  {author} {\bibinfo {author} {\bibfnamefont {D.}~\bibnamefont
  {Saintillan}}\ and\ \bibinfo {author} {\bibfnamefont {M.~J.}\ \bibnamefont
  {Shelley}},\ }\href {https://doi.org/10.1103/PhysRevLett.99.058102}
  {\bibfield  {journal} {\bibinfo  {journal} {Phys. Rev. Lett.}\ }\textbf
  {\bibinfo {volume} {99}},\ \bibinfo {pages} {058102} (\bibinfo {year}
  {2007})}\BibitemShut {NoStop}%
\bibitem [{\citenamefont {Hong}\ \emph {et~al.}(2007)\citenamefont {Hong},
  \citenamefont {Blackman}, \citenamefont {Kopp}, \citenamefont {Sen},\ and\
  \citenamefont {Velegol}}]{Hong2007}%
  \BibitemOpen
  \bibfield  {author} {\bibinfo {author} {\bibfnamefont {Y.}~\bibnamefont
  {Hong}}, \bibinfo {author} {\bibfnamefont {N.~M.~K.}\ \bibnamefont
  {Blackman}}, \bibinfo {author} {\bibfnamefont {N.~D.}\ \bibnamefont {Kopp}},
  \bibinfo {author} {\bibfnamefont {A.}~\bibnamefont {Sen}}, \ and\ \bibinfo
  {author} {\bibfnamefont {D.}~\bibnamefont {Velegol}},\ }\href
  {https://doi.org/10.1103/PhysRevLett.99.178103} {\bibfield  {journal}
  {\bibinfo  {journal} {Phys. Rev. Lett.}\ }\textbf {\bibinfo {volume} {99}},\
  \bibinfo {pages} {178103} (\bibinfo {year} {2007})}\BibitemShut {NoStop}%
\bibitem [{\citenamefont {Zhang}\ \emph {et~al.}(2010)\citenamefont {Zhang},
  \citenamefont {Petit}, \citenamefont {Lu}, \citenamefont {Kratochvil},
  \citenamefont {Peyer}, \citenamefont {Pei}, \citenamefont {Lou},\ and\
  \citenamefont {Nelson}}]{nwirezhang}%
  \BibitemOpen
  \bibfield  {author} {\bibinfo {author} {\bibfnamefont {L.}~\bibnamefont
  {Zhang}}, \bibinfo {author} {\bibfnamefont {T.}~\bibnamefont {Petit}},
  \bibinfo {author} {\bibfnamefont {Y.}~\bibnamefont {Lu}}, \bibinfo {author}
  {\bibfnamefont {B.~E.}\ \bibnamefont {Kratochvil}}, \bibinfo {author}
  {\bibfnamefont {K.~E.}\ \bibnamefont {Peyer}}, \bibinfo {author}
  {\bibfnamefont {R.}~\bibnamefont {Pei}}, \bibinfo {author} {\bibfnamefont
  {J.}~\bibnamefont {Lou}}, \ and\ \bibinfo {author} {\bibfnamefont {B.~J.}\
  \bibnamefont {Nelson}},\ }\href@noop {} {\bibfield  {journal} {\bibinfo
  {journal} {ACS Nano}\ }\textbf {\bibinfo {volume} {4}},\ \bibinfo {pages}
  {6228} (\bibinfo {year} {2010})}\BibitemShut {NoStop}%
\bibitem [{\citenamefont {Sabass}\ and\ \citenamefont
  {Seifert}(2010)}]{sabass}%
  \BibitemOpen
  \bibfield  {author} {\bibinfo {author} {\bibfnamefont {B.}~\bibnamefont
  {Sabass}}\ and\ \bibinfo {author} {\bibfnamefont {U.}~\bibnamefont
  {Seifert}},\ }\href {https://doi.org/10.1103/PhysRevLett.105.218103}
  {\bibfield  {journal} {\bibinfo  {journal} {Phys. Rev. Lett.}\ }\textbf
  {\bibinfo {volume} {105}},\ \bibinfo {pages} {218103} (\bibinfo {year}
  {2010})}\BibitemShut {NoStop}%
\bibitem [{\citenamefont {Elgeti}\ and\ \citenamefont
  {Gompper}(2016)}]{elgeti2016}%
  \BibitemOpen
  \bibfield  {author} {\bibinfo {author} {\bibfnamefont {J.}~\bibnamefont
  {Elgeti}}\ and\ \bibinfo {author} {\bibfnamefont {G.}~\bibnamefont
  {Gompper}},\ }\href {https://doi.org/10.1140/epjst/e2016-60070-6} {\bibfield
  {journal} {\bibinfo  {journal} {Eur. Phys. J. Special Topics}\ }\textbf
  {\bibinfo {volume} {225}},\ \bibinfo {pages} {2333} (\bibinfo {year}
  {2016})}\BibitemShut {NoStop}%
\bibitem [{\citenamefont {Berke}\ \emph {et~al.}(2008)\citenamefont {Berke},
  \citenamefont {Turner}, \citenamefont {Berg},\ and\ \citenamefont
  {Lauga}}]{wallattraction}%
  \BibitemOpen
  \bibfield  {author} {\bibinfo {author} {\bibfnamefont {A.~P.}\ \bibnamefont
  {Berke}}, \bibinfo {author} {\bibfnamefont {L.}~\bibnamefont {Turner}},
  \bibinfo {author} {\bibfnamefont {H.~C.}\ \bibnamefont {Berg}}, \ and\
  \bibinfo {author} {\bibfnamefont {E.}~\bibnamefont {Lauga}},\ }\href@noop {}
  {\bibfield  {journal} {\bibinfo  {journal} {Phys. Rev. Lett.}\ }\textbf
  {\bibinfo {volume} {101}},\ \bibinfo {pages} {038102} (\bibinfo {year}
  {2008})}\BibitemShut {NoStop}%
\bibitem [{\citenamefont {Lauga}\ \emph {et~al.}(2006)\citenamefont {Lauga},
  \citenamefont {DiLuzio}, \citenamefont {Whitesides},\ and\ \citenamefont
  {Stone}}]{wallattraction2}%
  \BibitemOpen
  \bibfield  {author} {\bibinfo {author} {\bibfnamefont {E.}~\bibnamefont
  {Lauga}}, \bibinfo {author} {\bibfnamefont {W.~R.}\ \bibnamefont {DiLuzio}},
  \bibinfo {author} {\bibfnamefont {G.~M.}\ \bibnamefont {Whitesides}}, \ and\
  \bibinfo {author} {\bibfnamefont {H.~A.}\ \bibnamefont {Stone}},\ }\href@noop
  {} {\bibfield  {journal} {\bibinfo  {journal} {Biophys. J.}\ }\textbf
  {\bibinfo {volume} {90}},\ \bibinfo {pages} {400} (\bibinfo {year}
  {2006})}\BibitemShut {NoStop}%
\bibitem [{\citenamefont {Spagnolie}\ and\ \citenamefont
  {Lauga}(2012)}]{wallattraction3}%
  \BibitemOpen
  \bibfield  {author} {\bibinfo {author} {\bibfnamefont {S.~E.}\ \bibnamefont
  {Spagnolie}}\ and\ \bibinfo {author} {\bibfnamefont {E.}~\bibnamefont
  {Lauga}},\ }\href@noop {} {\bibfield  {journal} {\bibinfo  {journal} {J.
  Fluid Mech.}\ }\textbf {\bibinfo {volume} {700}},\ \bibinfo {pages} {105}
  (\bibinfo {year} {2012})}\BibitemShut {NoStop}%
\bibitem [{\citenamefont {Li}\ \emph {et~al.}(2011)\citenamefont {Li},
  \citenamefont {Bensson}, \citenamefont {Nisimova}, \citenamefont {Munger},
  \citenamefont {Mahautmr}, \citenamefont {Tang}, \citenamefont {Maxey},\ and\
  \citenamefont {Brun}}]{li2011accumulation}%
  \BibitemOpen
  \bibfield  {author} {\bibinfo {author} {\bibfnamefont {G.}~\bibnamefont
  {Li}}, \bibinfo {author} {\bibfnamefont {J.}~\bibnamefont {Bensson}},
  \bibinfo {author} {\bibfnamefont {L.}~\bibnamefont {Nisimova}}, \bibinfo
  {author} {\bibfnamefont {D.}~\bibnamefont {Munger}}, \bibinfo {author}
  {\bibfnamefont {P.}~\bibnamefont {Mahautmr}}, \bibinfo {author}
  {\bibfnamefont {J.~X.}\ \bibnamefont {Tang}}, \bibinfo {author}
  {\bibfnamefont {M.~R.}\ \bibnamefont {Maxey}}, \ and\ \bibinfo {author}
  {\bibfnamefont {Y.~V.}\ \bibnamefont {Brun}},\ }\href@noop {} {\bibfield
  {journal} {\bibinfo  {journal} {Phys. Rev. E}\ }\textbf {\bibinfo {volume}
  {84}},\ \bibinfo {pages} {041932} (\bibinfo {year} {2011})}\BibitemShut
  {NoStop}%
\bibitem [{\citenamefont {Li}\ and\ \citenamefont
  {Tang}(2009)}]{li2011accumulation2}%
  \BibitemOpen
  \bibfield  {author} {\bibinfo {author} {\bibfnamefont {G.}~\bibnamefont
  {Li}}\ and\ \bibinfo {author} {\bibfnamefont {J.~X.}\ \bibnamefont {Tang}},\
  }\href@noop {} {\bibfield  {journal} {\bibinfo  {journal} {Phys. Rev. Lett.}\
  }\textbf {\bibinfo {volume} {103}},\ \bibinfo {pages} {078101} (\bibinfo
  {year} {2009})}\BibitemShut {NoStop}%
\bibitem [{\citenamefont {Li}\ and\ \citenamefont {Ardekani}(2014)}]{ardekani}%
  \BibitemOpen
  \bibfield  {author} {\bibinfo {author} {\bibfnamefont {G.-J.}\ \bibnamefont
  {Li}}\ and\ \bibinfo {author} {\bibfnamefont {A.~M.}\ \bibnamefont
  {Ardekani}},\ }\href@noop {} {\bibfield  {journal} {\bibinfo  {journal}
  {Phys. Rev. E}\ }\textbf {\bibinfo {volume} {90}},\ \bibinfo {pages} {013010}
  (\bibinfo {year} {2014})}\BibitemShut {NoStop}%
\bibitem [{\citenamefont {Hernandez-Ortiz}\ \emph {et~al.}(2005)\citenamefont
  {Hernandez-Ortiz}, \citenamefont {Stoltz},\ and\ \citenamefont
  {Graham}}]{Hernandez-Ortiz1}%
  \BibitemOpen
  \bibfield  {author} {\bibinfo {author} {\bibfnamefont {J.~P.}\ \bibnamefont
  {Hernandez-Ortiz}}, \bibinfo {author} {\bibfnamefont {C.~G.}\ \bibnamefont
  {Stoltz}}, \ and\ \bibinfo {author} {\bibfnamefont {M.~D.}\ \bibnamefont
  {Graham}},\ }\href@noop {} {\bibfield  {journal} {\bibinfo  {journal} {Phys.
  Rev. Lett.}\ }\textbf {\bibinfo {volume} {95}},\ \bibinfo {pages} {204501}
  (\bibinfo {year} {2005})}\BibitemShut {NoStop}%
\bibitem [{\citenamefont {Underhill}\ \emph {et~al.}(2008)\citenamefont
  {Underhill}, \citenamefont {Hernandez-Ortiz},\ and\ \citenamefont
  {Graham}}]{Hernandez-Ortiz2}%
  \BibitemOpen
  \bibfield  {author} {\bibinfo {author} {\bibfnamefont {P.~T.}\ \bibnamefont
  {Underhill}}, \bibinfo {author} {\bibfnamefont {J.~P.}\ \bibnamefont
  {Hernandez-Ortiz}}, \ and\ \bibinfo {author} {\bibfnamefont {M.~D.}\
  \bibnamefont {Graham}},\ }\href@noop {} {\bibfield  {journal} {\bibinfo
  {journal} {Phys. Rev. Lett.}\ }\textbf {\bibinfo {volume} {100}},\ \bibinfo
  {pages} {248101} (\bibinfo {year} {2008})}\BibitemShut {NoStop}%
\bibitem [{\citenamefont {Hernandez-Ortiz}\ \emph {et~al.}(2009)\citenamefont
  {Hernandez-Ortiz}, \citenamefont {Underhill},\ and\ \citenamefont
  {Graham}}]{Hernandez-Ortiz3}%
  \BibitemOpen
  \bibfield  {author} {\bibinfo {author} {\bibfnamefont {J.~P.}\ \bibnamefont
  {Hernandez-Ortiz}}, \bibinfo {author} {\bibfnamefont {P.~T.}\ \bibnamefont
  {Underhill}}, \ and\ \bibinfo {author} {\bibfnamefont {M.~D.}\ \bibnamefont
  {Graham}},\ }\href@noop {} {\bibfield  {journal} {\bibinfo  {journal} {J.
  Phys.: Condens. Matter}\ }\textbf {\bibinfo {volume} {21}},\ \bibinfo {pages}
  {204107} (\bibinfo {year} {2009})}\BibitemShut {NoStop}%
\bibitem [{\citenamefont {Schaar}\ \emph {et~al.}(2015)\citenamefont {Schaar},
  \citenamefont {Z\"ottl},\ and\ \citenamefont {Stark}}]{stark-wall}%
  \BibitemOpen
  \bibfield  {author} {\bibinfo {author} {\bibfnamefont {K.}~\bibnamefont
  {Schaar}}, \bibinfo {author} {\bibfnamefont {A.}~\bibnamefont {Z\"ottl}}, \
  and\ \bibinfo {author} {\bibfnamefont {H.}~\bibnamefont {Stark}},\
  }\href@noop {} {\bibfield  {journal} {\bibinfo  {journal} {Phys. Rev. Lett.}\
  }\textbf {\bibinfo {volume} {115}},\ \bibinfo {pages} {038101} (\bibinfo
  {year} {2015})}\BibitemShut {NoStop}%
\bibitem [{\citenamefont {Hu}\ \emph {et~al.}(2015)\citenamefont {Hu},
  \citenamefont {Wysocki}, \citenamefont {Winkler},\ and\ \citenamefont
  {Gompper}}]{gompper-wall}%
  \BibitemOpen
  \bibfield  {author} {\bibinfo {author} {\bibfnamefont {J.}~\bibnamefont
  {Hu}}, \bibinfo {author} {\bibfnamefont {A.}~\bibnamefont {Wysocki}},
  \bibinfo {author} {\bibfnamefont {R.~G.}\ \bibnamefont {Winkler}}, \ and\
  \bibinfo {author} {\bibfnamefont {G.}~\bibnamefont {Gompper}},\ }\href@noop
  {} {\bibfield  {journal} {\bibinfo  {journal} {Sci. Rep.}\ }\textbf {\bibinfo
  {volume} {5}},\ \bibinfo {pages} {9586} (\bibinfo {year} {2015})}\BibitemShut
  {NoStop}%
\bibitem [{\citenamefont {Elgeti}\ and\ \citenamefont
  {Gompper}(2009)}]{elgeti2009}%
  \BibitemOpen
  \bibfield  {author} {\bibinfo {author} {\bibfnamefont {J.}~\bibnamefont
  {Elgeti}}\ and\ \bibinfo {author} {\bibfnamefont {G.}~\bibnamefont
  {Gompper}},\ }\href {https://doi.org/10.1209/0295-5075/85/38002} {\bibfield
  {journal} {\bibinfo  {journal} {EPL}\ }\textbf {\bibinfo {volume} {85}},\
  \bibinfo {pages} {38002} (\bibinfo {year} {2009})}\BibitemShut {NoStop}%
\bibitem [{\citenamefont {Elgeti}\ and\ \citenamefont
  {Gompper}(2013)}]{elgeti2013}%
  \BibitemOpen
  \bibfield  {author} {\bibinfo {author} {\bibfnamefont {J.}~\bibnamefont
  {Elgeti}}\ and\ \bibinfo {author} {\bibfnamefont {G.}~\bibnamefont
  {Gompper}},\ }\href {https://doi.org/10.1209/0295-5075/101/48003} {\bibfield
  {journal} {\bibinfo  {journal} {EPL}\ }\textbf {\bibinfo {volume} {101}},\
  \bibinfo {pages} {48003} (\bibinfo {year} {2013})}\BibitemShut {NoStop}%
\bibitem [{\citenamefont {Kantsler}\ \emph {et~al.}(2014)\citenamefont
  {Kantsler}, \citenamefont {Dunkel}, \citenamefont {Blayney},\ and\
  \citenamefont {Goldstein}}]{sperm-rheotaxis}%
  \BibitemOpen
  \bibfield  {author} {\bibinfo {author} {\bibfnamefont {V.}~\bibnamefont
  {Kantsler}}, \bibinfo {author} {\bibfnamefont {J.}~\bibnamefont {Dunkel}},
  \bibinfo {author} {\bibfnamefont {M.}~\bibnamefont {Blayney}}, \ and\
  \bibinfo {author} {\bibfnamefont {R.~E.}\ \bibnamefont {Goldstein}},\
  }\href@noop {} {\bibfield  {journal} {\bibinfo  {journal} {eLife}\ }\textbf
  {\bibinfo {volume} {3}},\ \bibinfo {pages} {e02403} (\bibinfo {year}
  {2014})}\BibitemShut {NoStop}%
\bibitem [{\citenamefont {Kantsler}\ \emph {et~al.}(2005)\citenamefont
  {Kantsler}, \citenamefont {Dunkel}, \citenamefont {Polin},\ and\
  \citenamefont {Goldstein}}]{upstream-goldstein}%
  \BibitemOpen
  \bibfield  {author} {\bibinfo {author} {\bibfnamefont {V.}~\bibnamefont
  {Kantsler}}, \bibinfo {author} {\bibfnamefont {J.}~\bibnamefont {Dunkel}},
  \bibinfo {author} {\bibfnamefont {M.}~\bibnamefont {Polin}}, \ and\ \bibinfo
  {author} {\bibfnamefont {R.~E.}\ \bibnamefont {Goldstein}},\ }\href
  {https://doi.org/10.1073/pnas.1210548110} {\bibfield  {journal} {\bibinfo
  {journal} {Proc. Natl. Acad. Sci. U.S.A.}\ }\textbf {\bibinfo {volume}
  {110}},\ \bibinfo {pages} {1187} (\bibinfo {year} {2005})}\BibitemShut
  {NoStop}%
\bibitem [{\citenamefont {Rusconi}\ \emph {et~al.}(2014)\citenamefont
  {Rusconi}, \citenamefont {Guasto},\ and\ \citenamefont {Stocker}}]{rusconi}%
  \BibitemOpen
  \bibfield  {author} {\bibinfo {author} {\bibfnamefont {R.}~\bibnamefont
  {Rusconi}}, \bibinfo {author} {\bibfnamefont {J.~S.}\ \bibnamefont {Guasto}},
  \ and\ \bibinfo {author} {\bibfnamefont {R.}~\bibnamefont {Stocker}},\ }\href
  {https://doi.org/10.1038/nphys2883} {\bibfield  {journal} {\bibinfo
  {journal} {Nature Phys.}\ }\textbf {\bibinfo {volume} {10}},\ \bibinfo
  {pages} {212} (\bibinfo {year} {2014})}\BibitemShut {NoStop}%
\bibitem [{\citenamefont {Elgeti}\ and\ \citenamefont
  {Gompper}(2015)}]{elgeti2015run}%
  \BibitemOpen
  \bibfield  {author} {\bibinfo {author} {\bibfnamefont {J.}~\bibnamefont
  {Elgeti}}\ and\ \bibinfo {author} {\bibfnamefont {G.}~\bibnamefont
  {Gompper}},\ }\href@noop {} {\bibfield  {journal} {\bibinfo  {journal} {EPL}\
  }\textbf {\bibinfo {volume} {109}},\ \bibinfo {pages} {58003} (\bibinfo
  {year} {2015})}\BibitemShut {NoStop}%
\bibitem [{\citenamefont {Costanzo}\ \emph {et~al.}(2012)\citenamefont
  {Costanzo}, \citenamefont {Di~Leonardo}, \citenamefont {Ruocco},\ and\
  \citenamefont {Angelani}}]{costanzo}%
  \BibitemOpen
  \bibfield  {author} {\bibinfo {author} {\bibfnamefont {A.}~\bibnamefont
  {Costanzo}}, \bibinfo {author} {\bibfnamefont {R.}~\bibnamefont
  {Di~Leonardo}}, \bibinfo {author} {\bibfnamefont {G.}~\bibnamefont {Ruocco}},
  \ and\ \bibinfo {author} {\bibfnamefont {L.}~\bibnamefont {Angelani}},\
  }\href@noop {} {\bibfield  {journal} {\bibinfo  {journal} {J. Phys.: Condens.
  Matter}\ }\textbf {\bibinfo {volume} {24}},\ \bibinfo {pages} {065101}
  (\bibinfo {year} {2012})}\BibitemShut {NoStop}%
\bibitem [{\citenamefont {Nash}\ \emph {et~al.}(2010)\citenamefont {Nash},
  \citenamefont {Adhikari}, \citenamefont {Tailleur},\ and\ \citenamefont
  {Cates}}]{catesupstream}%
  \BibitemOpen
  \bibfield  {author} {\bibinfo {author} {\bibfnamefont {R.~W.}\ \bibnamefont
  {Nash}}, \bibinfo {author} {\bibfnamefont {R.}~\bibnamefont {Adhikari}},
  \bibinfo {author} {\bibfnamefont {J.}~\bibnamefont {Tailleur}}, \ and\
  \bibinfo {author} {\bibfnamefont {M.~E.}\ \bibnamefont {Cates}},\ }\href
  {https://doi.org/10.1103/PhysRevLett.104.258101} {\bibfield  {journal}
  {\bibinfo  {journal} {Phys. Rev. Lett.}\ }\textbf {\bibinfo {volume} {104}},\
  \bibinfo {pages} {258101} (\bibinfo {year} {2010})}\BibitemShut {NoStop}%
\bibitem [{\citenamefont {Mathijssen}\ \emph
  {et~al.}(2016{\natexlab{a}})\citenamefont {Mathijssen}, \citenamefont
  {Shendruk}, \citenamefont {Yeomans},\ and\ \citenamefont
  {Doostmohammadi}}]{Mathijssen:2016a}%
  \BibitemOpen
  \bibfield  {author} {\bibinfo {author} {\bibfnamefont {A.~J. T.~M.}\
  \bibnamefont {Mathijssen}}, \bibinfo {author} {\bibfnamefont {T.~N.}\
  \bibnamefont {Shendruk}}, \bibinfo {author} {\bibfnamefont {J.~M.}\
  \bibnamefont {Yeomans}}, \ and\ \bibinfo {author} {\bibfnamefont
  {A.}~\bibnamefont {Doostmohammadi}},\ }\href
  {https://doi.org/10.1103/PhysRevLett.116.028104} {\bibfield  {journal}
  {\bibinfo  {journal} {Phys. Rev. Lett.}\ }\textbf {\bibinfo {volume} {116}},\
  \bibinfo {pages} {028104} (\bibinfo {year} {2016}{\natexlab{a}})}\BibitemShut
  {NoStop}%
\bibitem [{\citenamefont {Mathijssen}\ \emph
  {et~al.}(2016{\natexlab{b}})\citenamefont {Mathijssen}, \citenamefont
  {Doostmohammadi}, \citenamefont {Yeomans},\ and\ \citenamefont
  {Shendruk}}]{Mathijssen:2016b}%
  \BibitemOpen
  \bibfield  {author} {\bibinfo {author} {\bibfnamefont {A.~J. T.~M.}\
  \bibnamefont {Mathijssen}}, \bibinfo {author} {\bibfnamefont
  {A.}~\bibnamefont {Doostmohammadi}}, \bibinfo {author} {\bibfnamefont
  {J.~M.}\ \bibnamefont {Yeomans}}, \ and\ \bibinfo {author} {\bibfnamefont
  {T.~N.}\ \bibnamefont {Shendruk}},\ }\href
  {https://doi.org/10.1017/jfm.2016.479} {\bibfield  {journal} {\bibinfo
  {journal} {J. Fluid Mech.}\ }\textbf {\bibinfo {volume} {806}},\ \bibinfo
  {pages} {35} (\bibinfo {year} {2016}{\natexlab{b}})}\BibitemShut {NoStop}%
\bibitem [{\citenamefont {Mathijssen}\ \emph
  {et~al.}(2016{\natexlab{c}})\citenamefont {Mathijssen}, \citenamefont
  {Doostmohammadi}, \citenamefont {Yeomans},\ and\ \citenamefont
  {Shendruk}}]{Mathijssen:2016c}%
  \BibitemOpen
  \bibfield  {author} {\bibinfo {author} {\bibfnamefont {A.~J. T.~M.}\
  \bibnamefont {Mathijssen}}, \bibinfo {author} {\bibfnamefont
  {A.}~\bibnamefont {Doostmohammadi}}, \bibinfo {author} {\bibfnamefont
  {J.~M.}\ \bibnamefont {Yeomans}}, \ and\ \bibinfo {author} {\bibfnamefont
  {T.~N.}\ \bibnamefont {Shendruk}},\ }\href
  {https://doi.org/10.1098/rsif.2015.0936} {\bibfield  {journal} {\bibinfo
  {journal} {J. R. Soc. Interface}\ }\textbf {\bibinfo {volume} {13}} (\bibinfo
  {year} {2016}{\natexlab{c}})}\BibitemShut {NoStop}%
\bibitem [{\citenamefont {Ezhilan}\ and\ \citenamefont
  {Saintillan}(2015)}]{ezhilan}%
  \BibitemOpen
  \bibfield  {author} {\bibinfo {author} {\bibfnamefont {B.}~\bibnamefont
  {Ezhilan}}\ and\ \bibinfo {author} {\bibfnamefont {D.}~\bibnamefont
  {Saintillan}},\ }\href {10.1017/jfm.2015.372} {\bibfield  {journal} {\bibinfo
   {journal} {J. Fluid Mech.}\ }\textbf {\bibinfo {volume} {777}},\ \bibinfo
  {pages} {482} (\bibinfo {year} {2015})}\BibitemShut {NoStop}%
\bibitem [{\citenamefont {Tung}\ \emph {et~al.}(2015)\citenamefont {Tung},
  \citenamefont {Ardon}, \citenamefont {Roy}, \citenamefont {Koch},
  \citenamefont {Suarez},\ and\ \citenamefont {Wu}}]{upstream2015prl}%
  \BibitemOpen
  \bibfield  {author} {\bibinfo {author} {\bibfnamefont {C.-k.}\ \bibnamefont
  {Tung}}, \bibinfo {author} {\bibfnamefont {F.}~\bibnamefont {Ardon}},
  \bibinfo {author} {\bibfnamefont {A.}~\bibnamefont {Roy}}, \bibinfo {author}
  {\bibfnamefont {D.~L.}\ \bibnamefont {Koch}}, \bibinfo {author}
  {\bibfnamefont {S.~S.}\ \bibnamefont {Suarez}}, \ and\ \bibinfo {author}
  {\bibfnamefont {M.}~\bibnamefont {Wu}},\ }\href
  {https://doi.org/10.1103/PhysRevLett.114.108102} {\bibfield  {journal}
  {\bibinfo  {journal} {Phys. Rev. Lett.}\ }\textbf {\bibinfo {volume} {114}},\
  \bibinfo {pages} {108102} (\bibinfo {year} {2015})}\BibitemShut {NoStop}%
\bibitem [{\citenamefont {Z{\"o}ttl}\ and\ \citenamefont
  {Stark}(2013)}]{zottl_Poiseuille}%
  \BibitemOpen
  \bibfield  {author} {\bibinfo {author} {\bibfnamefont {A.}~\bibnamefont
  {Z{\"o}ttl}}\ and\ \bibinfo {author} {\bibfnamefont {H.}~\bibnamefont
  {Stark}},\ }\href {https://doi.org/10.1140/epje/i2013-13004-5} {\bibfield
  {journal} {\bibinfo  {journal} {Eur. Phys. J. E}\ }\textbf {\bibinfo {volume}
  {36}},\ \bibinfo {pages} {4} (\bibinfo {year} {2013})}\BibitemShut {NoStop}%
\bibitem [{\citenamefont {Chilukuri}\ \emph {et~al.}(2015)\citenamefont
  {Chilukuri}, \citenamefont {Collins},\ and\ \citenamefont
  {Underhill}}]{Underhill2015}%
  \BibitemOpen
  \bibfield  {author} {\bibinfo {author} {\bibfnamefont {S.}~\bibnamefont
  {Chilukuri}}, \bibinfo {author} {\bibfnamefont {C.~H.}\ \bibnamefont
  {Collins}}, \ and\ \bibinfo {author} {\bibfnamefont {P.~T.}\ \bibnamefont
  {Underhill}},\ }\href {https://doi.org/10.1063/1.4914129} {\bibfield
  {journal} {\bibinfo  {journal} {Phys. Fluids}\ }\textbf {\bibinfo {volume}
  {27}},\ \bibinfo {pages} {031902} (\bibinfo {year} {2015})}\BibitemShut
  {NoStop}%
\bibitem [{\citenamefont {O'Toole}\ \emph {et~al.}(2000)\citenamefont
  {O'Toole}, \citenamefont {Kaplan},\ and\ \citenamefont
  {Kolter}}]{biofilmformation}%
  \BibitemOpen
  \bibfield  {author} {\bibinfo {author} {\bibfnamefont {G.}~\bibnamefont
  {O'Toole}}, \bibinfo {author} {\bibfnamefont {H.~B.}\ \bibnamefont {Kaplan}},
  \ and\ \bibinfo {author} {\bibfnamefont {R.}~\bibnamefont {Kolter}},\
  }\href@noop {} {\bibfield  {journal} {\bibinfo  {journal} {Annu. Rev.
  Microbiol.}\ }\textbf {\bibinfo {volume} {54}},\ \bibinfo {pages} {49}
  (\bibinfo {year} {2000})}\BibitemShut {NoStop}%
\bibitem [{\citenamefont {Hall-Stoodley}\ \emph {et~al.}(2004)\citenamefont
  {Hall-Stoodley}, \citenamefont {Costerton},\ and\ \citenamefont
  {Stoodley}}]{biofilmstoodley}%
  \BibitemOpen
  \bibfield  {author} {\bibinfo {author} {\bibfnamefont {L.}~\bibnamefont
  {Hall-Stoodley}}, \bibinfo {author} {\bibfnamefont {J.~W.}\ \bibnamefont
  {Costerton}}, \ and\ \bibinfo {author} {\bibfnamefont {P.}~\bibnamefont
  {Stoodley}},\ }\href@noop {} {\bibfield  {journal} {\bibinfo  {journal}
  {Nature Rev. Microbiol.}\ }\textbf {\bibinfo {volume} {2}},\ \bibinfo {pages}
  {95} (\bibinfo {year} {2004})}\BibitemShut {NoStop}%
\bibitem [{\citenamefont {Elgeti}\ \emph {et~al.}(2010)\citenamefont {Elgeti},
  \citenamefont {Kaupp},\ and\ \citenamefont {Gompper}}]{spermsurface1}%
  \BibitemOpen
  \bibfield  {author} {\bibinfo {author} {\bibfnamefont {J.}~\bibnamefont
  {Elgeti}}, \bibinfo {author} {\bibfnamefont {U.~B.}\ \bibnamefont {Kaupp}}, \
  and\ \bibinfo {author} {\bibfnamefont {G.}~\bibnamefont {Gompper}},\ }\href
  {https://doi.org/10.1016/j.bpj.2010.05.015} {\bibfield  {journal} {\bibinfo
  {journal} {Biophys. J.}\ }\textbf {\bibinfo {volume} {99}},\ \bibinfo {pages}
  {1018 } (\bibinfo {year} {2010})}\BibitemShut {NoStop}%
\bibitem [{\citenamefont {Denissenko}\ \emph {et~al.}(2012)\citenamefont
  {Denissenko}, \citenamefont {Kantsler}, \citenamefont {Smith},\ and\
  \citenamefont {Kirkman-Brown}}]{spermsurface2}%
  \BibitemOpen
  \bibfield  {author} {\bibinfo {author} {\bibfnamefont {P.}~\bibnamefont
  {Denissenko}}, \bibinfo {author} {\bibfnamefont {V.}~\bibnamefont
  {Kantsler}}, \bibinfo {author} {\bibfnamefont {D.~J.}\ \bibnamefont {Smith}},
  \ and\ \bibinfo {author} {\bibfnamefont {J.}~\bibnamefont {Kirkman-Brown}},\
  }\href {https://doi.org/10.1073/pnas.1202934109} {\bibfield  {journal}
  {\bibinfo  {journal} {Proc. Natl. Acad. Sci. U.S.A.}\ }\textbf {\bibinfo
  {volume} {109}},\ \bibinfo {pages} {8007} (\bibinfo {year}
  {2012})}\BibitemShut {NoStop}%
\bibitem [{\citenamefont {Crowdy}(2011)}]{spermsurface3}%
  \BibitemOpen
  \bibfield  {author} {\bibinfo {author} {\bibfnamefont {D.}~\bibnamefont
  {Crowdy}},\ }\href {https://doi.org/10.1016/j.ijnonlinmec.2010.12.010}
  {\bibfield  {journal} {\bibinfo  {journal} {Int. J. Nonlinear Mech.}\
  }\textbf {\bibinfo {volume} {46}},\ \bibinfo {pages} {577 } (\bibinfo {year}
  {2011})}\BibitemShut {NoStop}%
\bibitem [{\citenamefont {Wu}\ \emph {et~al.}(2009)\citenamefont {Wu},
  \citenamefont {Willing}, \citenamefont {Bjerketorp}, \citenamefont
  {Jansson},\ and\ \citenamefont {Hjort}}]{ufluidics1}%
  \BibitemOpen
  \bibfield  {author} {\bibinfo {author} {\bibfnamefont {Z.}~\bibnamefont
  {Wu}}, \bibinfo {author} {\bibfnamefont {B.}~\bibnamefont {Willing}},
  \bibinfo {author} {\bibfnamefont {J.}~\bibnamefont {Bjerketorp}}, \bibinfo
  {author} {\bibfnamefont {J.~K.}\ \bibnamefont {Jansson}}, \ and\ \bibinfo
  {author} {\bibfnamefont {K.}~\bibnamefont {Hjort}},\ }\href
  {https://doi.org/10.1039/B817611F} {\bibfield  {journal} {\bibinfo  {journal}
  {Lab Chip}\ }\textbf {\bibinfo {volume} {9}},\ \bibinfo {pages} {1193}
  (\bibinfo {year} {2009})}\BibitemShut {NoStop}%
\bibitem [{\citenamefont {Qiu}\ \emph {et~al.}(2009)\citenamefont {Qiu},
  \citenamefont {Zhou}, \citenamefont {Chen},\ and\ \citenamefont
  {Lin}}]{ufluidics2}%
  \BibitemOpen
  \bibfield  {author} {\bibinfo {author} {\bibfnamefont {J.}~\bibnamefont
  {Qiu}}, \bibinfo {author} {\bibfnamefont {Y.}~\bibnamefont {Zhou}}, \bibinfo
  {author} {\bibfnamefont {H.}~\bibnamefont {Chen}}, \ and\ \bibinfo {author}
  {\bibfnamefont {J.-M.}\ \bibnamefont {Lin}},\ }\href
  {https://doi.org/10.1016/j.talanta.2009.05.003} {\bibfield  {journal}
  {\bibinfo  {journal} {Talanta}\ }\textbf {\bibinfo {volume} {79}},\ \bibinfo
  {pages} {787 } (\bibinfo {year} {2009})}\BibitemShut {NoStop}%
\bibitem [{\citenamefont {Sternberg}\ \emph {et~al.}(1999)\citenamefont
  {Sternberg}, \citenamefont {Christensen}, \citenamefont {Johansen},
  \citenamefont {Nielsen}, \citenamefont {Andersen}, \citenamefont {Givskov},\
  and\ \citenamefont {Molin}}]{flowbiofilm1}%
  \BibitemOpen
  \bibfield  {author} {\bibinfo {author} {\bibfnamefont {C.}~\bibnamefont
  {Sternberg}}, \bibinfo {author} {\bibfnamefont {B.~B.}\ \bibnamefont
  {Christensen}}, \bibinfo {author} {\bibfnamefont {T.}~\bibnamefont
  {Johansen}}, \bibinfo {author} {\bibfnamefont {A.~T.}\ \bibnamefont
  {Nielsen}}, \bibinfo {author} {\bibfnamefont {J.~B.}\ \bibnamefont
  {Andersen}}, \bibinfo {author} {\bibfnamefont {M.}~\bibnamefont {Givskov}}, \
  and\ \bibinfo {author} {\bibfnamefont {S.}~\bibnamefont {Molin}},\
  }\href@noop {} {\bibfield  {journal} {\bibinfo  {journal} {Appl. Environ.
  Microbiol.}\ }\textbf {\bibinfo {volume} {65}},\ \bibinfo {pages} {4108}
  (\bibinfo {year} {1999})}\BibitemShut {NoStop}%
\bibitem [{\citenamefont {Pereira}\ \emph {et~al.}(2002)\citenamefont
  {Pereira}, \citenamefont {Kuehn}, \citenamefont {Wuertz}, \citenamefont
  {Neu},\ and\ \citenamefont {Melo}}]{flowbiofilm2}%
  \BibitemOpen
  \bibfield  {author} {\bibinfo {author} {\bibfnamefont {M.~O.}\ \bibnamefont
  {Pereira}}, \bibinfo {author} {\bibfnamefont {M.}~\bibnamefont {Kuehn}},
  \bibinfo {author} {\bibfnamefont {S.}~\bibnamefont {Wuertz}}, \bibinfo
  {author} {\bibfnamefont {T.}~\bibnamefont {Neu}}, \ and\ \bibinfo {author}
  {\bibfnamefont {L.~F.}\ \bibnamefont {Melo}},\ }\href@noop {} {\bibfield
  {journal} {\bibinfo  {journal} {Biotechnol. Bioeng.}\ }\textbf {\bibinfo
  {volume} {78}},\ \bibinfo {pages} {164} (\bibinfo {year} {2002})}\BibitemShut
  {NoStop}%
\bibitem [{\citenamefont {Hill}\ \emph {et~al.}(2007)\citenamefont {Hill},
  \citenamefont {Kalkanci}, \citenamefont {McMurry},\ and\ \citenamefont
  {Koser}}]{ecoliupstream}%
  \BibitemOpen
  \bibfield  {author} {\bibinfo {author} {\bibfnamefont {J.}~\bibnamefont
  {Hill}}, \bibinfo {author} {\bibfnamefont {O.}~\bibnamefont {Kalkanci}},
  \bibinfo {author} {\bibfnamefont {J.~L.}\ \bibnamefont {McMurry}}, \ and\
  \bibinfo {author} {\bibfnamefont {H.}~\bibnamefont {Koser}},\ }\href
  {https://doi.org/10.1103/PhysRevLett.98.068101} {\bibfield  {journal}
  {\bibinfo  {journal} {Phys. Rev. Lett.}\ }\textbf {\bibinfo {volume} {98}},\
  \bibinfo {pages} {068101} (\bibinfo {year} {2007})}\BibitemShut {NoStop}%
\bibitem [{\citenamefont {Kaya}\ and\ \citenamefont
  {Koser}(2012)}]{ecoliupstream2}%
  \BibitemOpen
  \bibfield  {author} {\bibinfo {author} {\bibfnamefont {T.}~\bibnamefont
  {Kaya}}\ and\ \bibinfo {author} {\bibfnamefont {H.}~\bibnamefont {Koser}},\
  }\href@noop {} {\bibfield  {journal} {\bibinfo  {journal} {Biophys. J.}\
  }\textbf {\bibinfo {volume} {102}},\ \bibinfo {pages} {1514} (\bibinfo {year}
  {2012})}\BibitemShut {NoStop}%
\bibitem [{\citenamefont {Kaya}\ and\ \citenamefont
  {Koser}(2009)}]{ecoliupstream3}%
  \BibitemOpen
  \bibfield  {author} {\bibinfo {author} {\bibfnamefont {T.}~\bibnamefont
  {Kaya}}\ and\ \bibinfo {author} {\bibfnamefont {H.}~\bibnamefont {Koser}},\
  }\href@noop {} {\bibfield  {journal} {\bibinfo  {journal} {Phys. Rev. Lett.}\
  }\textbf {\bibinfo {volume} {103}},\ \bibinfo {pages} {138103} (\bibinfo
  {year} {2009})}\BibitemShut {NoStop}%
\bibitem [{\citenamefont {Rosengarten}\ \emph {et~al.}(1988)\citenamefont
  {Rosengarten}, \citenamefont {Klein-Struckmeier},\ and\ \citenamefont
  {Kirchhoff}}]{upstreamold}%
  \BibitemOpen
  \bibfield  {author} {\bibinfo {author} {\bibfnamefont {R.}~\bibnamefont
  {Rosengarten}}, \bibinfo {author} {\bibfnamefont {A.}~\bibnamefont
  {Klein-Struckmeier}}, \ and\ \bibinfo {author} {\bibfnamefont
  {H.}~\bibnamefont {Kirchhoff}},\ }\href@noop {} {\bibfield  {journal}
  {\bibinfo  {journal} {J. Bacteriol.}\ }\textbf {\bibinfo {volume} {170}},\
  \bibinfo {pages} {989} (\bibinfo {year} {1988})}\BibitemShut {NoStop}%
\bibitem [{\citenamefont {Meng}\ \emph {et~al.}(2005)\citenamefont {Meng},
  \citenamefont {Li}, \citenamefont {Galvani}, \citenamefont {Hao},
  \citenamefont {Turner}, \citenamefont {Burr},\ and\ \citenamefont
  {Hoch}}]{upstream3}%
  \BibitemOpen
  \bibfield  {author} {\bibinfo {author} {\bibfnamefont {Y.}~\bibnamefont
  {Meng}}, \bibinfo {author} {\bibfnamefont {Y.}~\bibnamefont {Li}}, \bibinfo
  {author} {\bibfnamefont {C.~D.}\ \bibnamefont {Galvani}}, \bibinfo {author}
  {\bibfnamefont {G.}~\bibnamefont {Hao}}, \bibinfo {author} {\bibfnamefont
  {J.~N.}\ \bibnamefont {Turner}}, \bibinfo {author} {\bibfnamefont {T.~J.}\
  \bibnamefont {Burr}}, \ and\ \bibinfo {author} {\bibfnamefont {H.~C.}\
  \bibnamefont {Hoch}},\ }\href@noop {} {\bibfield  {journal} {\bibinfo
  {journal} {J. Bacteriol.}\ }\textbf {\bibinfo {volume} {187}},\ \bibinfo
  {pages} {5560} (\bibinfo {year} {2005})}\BibitemShut {NoStop}%
\bibitem [{\citenamefont {Bretherton}(1962)}]{bretherton}%
  \BibitemOpen
  \bibfield  {author} {\bibinfo {author} {\bibfnamefont {F.~P.}\ \bibnamefont
  {Bretherton}},\ }\href@noop {} {\bibfield  {journal} {\bibinfo  {journal} {J.
  Fluid Mech.}\ }\textbf {\bibinfo {volume} {14}},\ \bibinfo {pages} {284}
  (\bibinfo {year} {1962})}\BibitemShut {NoStop}%
\bibitem [{\citenamefont {Doi}\ and\ \citenamefont
  {Edwards}(1988)}]{doiedwards}%
  \BibitemOpen
  \bibfield  {author} {\bibinfo {author} {\bibfnamefont {M.}~\bibnamefont
  {Doi}}\ and\ \bibinfo {author} {\bibfnamefont {S.~F.}\ \bibnamefont
  {Edwards}},\ }\href@noop {} {\emph {\bibinfo {title} {The Theory of Polymer
  Dynamics}}}\ (\bibinfo  {publisher} {Oxford University Press, Oxford},\
  \bibinfo {year} {1988})\BibitemShut {NoStop}%
\bibitem [{\citenamefont {Binder}(1981)}]{binder}%
  \BibitemOpen
  \bibfield  {author} {\bibinfo {author} {\bibfnamefont {K.}~\bibnamefont
  {Binder}},\ }\href@noop {} {\bibfield  {journal} {\bibinfo  {journal} {Z.
  Phys. B}\ }\textbf {\bibinfo {volume} {43}},\ \bibinfo {pages} {119}
  (\bibinfo {year} {1981})}\BibitemShut {NoStop}%
\bibitem [{\citenamefont {Mardia}\ and\ \citenamefont
  {Jupp}(2000)}]{mardia-jupp}%
  \BibitemOpen
  \bibfield  {author} {\bibinfo {author} {\bibfnamefont {K.~V.}\ \bibnamefont
  {Mardia}}\ and\ \bibinfo {author} {\bibfnamefont {P.}~\bibnamefont {Jupp}},\
  }\href@noop {} {\emph {\bibinfo {title} {Directional Statistics}}}\ (\bibinfo
   {publisher} {John Wiley and Sons Ltd, Chichcster},\ \bibinfo {year}
  {2000})\BibitemShut {NoStop}%
\bibitem [{\citenamefont {Darlington}(1970)}]{Darlington1970}%
  \BibitemOpen
  \bibfield  {author} {\bibinfo {author} {\bibfnamefont {R.~B.}\ \bibnamefont
  {Darlington}},\ }\href {https://doi.org/10.1080/00031305.1970.10478885}
  {\bibfield  {journal} {\bibinfo  {journal} {Am. Stat.}\ }\textbf {\bibinfo
  {volume} {24}},\ \bibinfo {pages} {19} (\bibinfo {year} {1970})}\BibitemShut
  {NoStop}%
\bibitem [{\citenamefont {Moors}(1986)}]{Moors1986}%
  \BibitemOpen
  \bibfield  {author} {\bibinfo {author} {\bibfnamefont {J.~J.~A.}\
  \bibnamefont {Moors}},\ }\href
  {https://doi.org/10.1080/00031305.1986.10475415} {\bibfield  {journal}
  {\bibinfo  {journal} {Am. Stat.}\ }\textbf {\bibinfo {volume} {40}},\
  \bibinfo {pages} {283} (\bibinfo {year} {1986})}\BibitemShut {NoStop}%
\bibitem [{\citenamefont {Balanda}\ and\ \citenamefont
  {Macgillivray}(1988)}]{Macgillivray1988}%
  \BibitemOpen
  \bibfield  {author} {\bibinfo {author} {\bibfnamefont {K.~P.}\ \bibnamefont
  {Balanda}}\ and\ \bibinfo {author} {\bibfnamefont {H.~L.}\ \bibnamefont
  {Macgillivray}},\ }\href {https://doi.org/10.1080/00031305.1988.10475539}
  {\bibfield  {journal} {\bibinfo  {journal} {Am. Stat.}\ }\textbf {\bibinfo
  {volume} {42}},\ \bibinfo {pages} {111} (\bibinfo {year} {1988})}\BibitemShut
  {NoStop}%
\bibitem [{\citenamefont {Kardar}(2007)}]{kardar}%
  \BibitemOpen
  \bibfield  {author} {\bibinfo {author} {\bibfnamefont {M.}~\bibnamefont
  {Kardar}},\ }\href@noop {} {\emph {\bibinfo {title} {Statistical Physics of
  Field}}}\ (\bibinfo  {publisher} {Cambridge University Press, Cambridge},\
  \bibinfo {year} {2007})\BibitemShut {NoStop}%
\bibitem [{\citenamefont {Happel}\ and\ \citenamefont
  {Brenner}(1983)}]{happelbook}%
  \BibitemOpen
  \bibfield  {author} {\bibinfo {author} {\bibfnamefont {J.}~\bibnamefont
  {Happel}}\ and\ \bibinfo {author} {\bibfnamefont {H.}~\bibnamefont
  {Brenner}},\ }\href@noop {} {\emph {\bibinfo {title} {Low Reynolds Number
  Hydrodynamics: with special applications to particulate media}}}\ (\bibinfo
  {publisher} {Springer Netherlands \& Martinus Nijhoff Publishers, The Hague,
  The Netherlands},\ \bibinfo {year} {1983})\BibitemShut {NoStop}%
\bibitem [{\citenamefont {Wioland}\ \emph {et~al.}(2013)\citenamefont
  {Wioland}, \citenamefont {Woodhouse}, \citenamefont {Dunkel}, \citenamefont
  {Kessler},\ and\ \citenamefont {Goldstein}}]{wioland2013confinement}%
  \BibitemOpen
  \bibfield  {author} {\bibinfo {author} {\bibfnamefont {H.}~\bibnamefont
  {Wioland}}, \bibinfo {author} {\bibfnamefont {F.~G.}\ \bibnamefont
  {Woodhouse}}, \bibinfo {author} {\bibfnamefont {J.}~\bibnamefont {Dunkel}},
  \bibinfo {author} {\bibfnamefont {J.~O.}\ \bibnamefont {Kessler}}, \ and\
  \bibinfo {author} {\bibfnamefont {R.~E.}\ \bibnamefont {Goldstein}},\
  }\href@noop {} {\bibfield  {journal} {\bibinfo  {journal} {Phys. Rev. Lett.}\
  }\textbf {\bibinfo {volume} {110}},\ \bibinfo {pages} {268102} (\bibinfo
  {year} {2013})}\BibitemShut {NoStop}%
\bibitem [{\citenamefont {Lushi}\ \emph {et~al.}(2014)\citenamefont {Lushi},
  \citenamefont {Wioland},\ and\ \citenamefont {Goldstein}}]{lushi2014fluid}%
  \BibitemOpen
  \bibfield  {author} {\bibinfo {author} {\bibfnamefont {E.}~\bibnamefont
  {Lushi}}, \bibinfo {author} {\bibfnamefont {H.}~\bibnamefont {Wioland}}, \
  and\ \bibinfo {author} {\bibfnamefont {R.~E.}\ \bibnamefont {Goldstein}},\
  }\href@noop {} {\bibfield  {journal} {\bibinfo  {journal} {Proc. Natl. Acad.
  Sci. U.S.A.}\ }\textbf {\bibinfo {volume} {111}},\ \bibinfo {pages} {9733}
  (\bibinfo {year} {2014})}\BibitemShut {NoStop}%
\bibitem [{\citenamefont {Saintillan}\ and\ \citenamefont
  {Shelley}(2008{\natexlab{a}})}]{pof2008}%
  \BibitemOpen
  \bibfield  {author} {\bibinfo {author} {\bibfnamefont {D.}~\bibnamefont
  {Saintillan}}\ and\ \bibinfo {author} {\bibfnamefont {M.~J.}\ \bibnamefont
  {Shelley}},\ }\href {https://doi.org/10.1063/1.3041776} {\bibfield  {journal}
  {\bibinfo  {journal} {Phys. Fluids}\ }\textbf {\bibinfo {volume} {20}},\
  \bibinfo {pages} {123304} (\bibinfo {year} {2008}{\natexlab{a}})}\BibitemShut
  {NoStop}%
\bibitem [{\citenamefont {Saintillan}\ and\ \citenamefont
  {Shelley}(2008{\natexlab{b}})}]{prl2008}%
  \BibitemOpen
  \bibfield  {author} {\bibinfo {author} {\bibfnamefont {D.}~\bibnamefont
  {Saintillan}}\ and\ \bibinfo {author} {\bibfnamefont {M.~J.}\ \bibnamefont
  {Shelley}},\ }\href {https://doi.org/10.1103/PhysRevLett.100.178103}
  {\bibfield  {journal} {\bibinfo  {journal} {Phys. Rev. Lett.}\ }\textbf
  {\bibinfo {volume} {100}},\ \bibinfo {pages} {178103} (\bibinfo {year}
  {2008}{\natexlab{b}})}\BibitemShut {NoStop}%
\bibitem [{\citenamefont {Saintillan}\ and\ \citenamefont
  {Shelley}(2013)}]{comptes_rendus}%
  \BibitemOpen
  \bibfield  {author} {\bibinfo {author} {\bibfnamefont {D.}~\bibnamefont
  {Saintillan}}\ and\ \bibinfo {author} {\bibfnamefont {M.~J.}\ \bibnamefont
  {Shelley}},\ }\href {https://doi.org/10.1016/j.crhy.2013.04.001} {\bibfield
  {journal} {\bibinfo  {journal} {C. R. Physique}\ }\textbf {\bibinfo {volume}
  {14}},\ \bibinfo {pages} {497} (\bibinfo {year} {2013})}\BibitemShut
  {NoStop}%
\bibitem [{\citenamefont {Jeffery}(1922)}]{Jeffery1922}%
  \BibitemOpen
  \bibfield  {author} {\bibinfo {author} {\bibfnamefont {G.~B.}\ \bibnamefont
  {Jeffery}},\ }\href {http://doi.org/10.1098/rspa.1922.0078} {\bibfield
  {journal} {\bibinfo  {journal} {Proc. R. Soc. (London) Ser. A}\ }\textbf
  {\bibinfo {volume} {102}},\ \bibinfo {pages} {161} (\bibinfo {year}
  {1922})}\BibitemShut {NoStop}%
\end{thebibliography}%

\end{document}